\documentclass[trackchanges]{aastex701}
\usepackage{amsmath}
\usepackage{graphicx,color}
\usepackage{overpic}

\usepackage{multirow}

\newcommand{\HII}{H\,{\sc ii}}
\newcommand{\HI}{H\,{\sc i}}
\newcommand{\WHI}{W_\mathrm{HI}}
\newcommand{\WCO}{W_\mathrm{CO}}
\newcommand{\XCO}{X_\mathrm{CO}}
\newcommand{\Np}{N_\mathrm{p}}
\newcommand{\Htwo}{\mathrm{H_{2}}}

\begin{document}

%%\title{Search for the GeV Counterpart of PeVatron LHAASO~J1912+1014u using \textit{Fermi}-LAT $\gamma$-ray and FUGIN CO Data}
\title{Hadronic Scenario for Galactic PeVatron LHAASO~J1912+1014u Supported by \textit{Fermi}-LAT $\gamma$-ray Data and FUGIN CO Data}

\author[orcid=0000-0001-7263-0296,sname='Mizuno']{Tsunefumi Mizuno}
%%\altaffiliation{Kitt Peak National Observatory}
\affiliation{Hiroshima University}
\email[show]{mizuno@astro.hiroshima-u.ac.jp}  

\author[orcid=0000-0003-2062-5692,gname=Hidetoshi, sname='Sano']{Hidetoshi Sano} 
%%\altaffiliation{Las Campanas Observatory}
\affiliation{Gifu University}
\email{fakeemail2@google.com}

\author[orcid=0000-0002-9552-3570,gname=Takeru,sname=Murase]{Takeru Murase}
%%\affiliation{South African Astronomical Observatory}
\affiliation{Gifu University}
\email{fakeemail3@google.com}

\author[0000-0002-9924-9978,sname=Oka,gname=Tomohiko]{Tomohiko Oka}
\affiliation{Institute for Theoretical Physics and Astrophysics, Julius-Maximilians-Universit\"{a}t W\"{u}rzburg, Emil-Fischer-Str. 31, 97074 W\"{u}rzburg, Germany}
\email{oka.tomohiko.25n@gmail.com}

\author[0000-0002-8152-6172,sname=Suzuki,gname=Hiromasa]{Hiromasa Suzuki}
\affiliation{Miyazaki University}
\email{fakeemail5@google.com}

\author{Naohito Nakahara}
\affiliation{Hiroshima University}
\email{fakeemail6@google.com}

%%\collaboration{all}{The Terra Mater collaboration}

\begin{abstract}
LHAASO has reported 43 sub-PeV $\gamma$-ray sources, which are promising candidates for cosmic-ray (CR) accelerators above the PeV energy, commonly called as PeVatrons.
%either proton or electron PeVatrons,
Multi-wavelength observations are crucial for identifying the underlying particle species and estimating the CR energy content of these sources.
In this work we investigate the region around LHAASO~J1912+1014u (and HESS~J1912+101) 
using \textit{Fermi}-LAT $\gamma$-ray data and FUGIN CO data.
We analyzed 15 years of \textit{Fermi}-LAT data in the 0.4--409.6~GeV energy range. 
By improving the standard \textit{Fermi}-LAT diffuse emission model,
we significantly reduced the large residuals around the source in the 1.6-12.8~GeV band.
We detected a statistically significant excess above the diffuse background,
which likely represents {$\ge$}10~GeV emission associated with the LHAASO/H.E.S.S. source. 
The GeV excess exhibits a hard spectrum (photon index of about 2.1) and is well reproduced by interstellar gas templates with systemic velocities of about 25~$\mathrm{km~s^{-1}}$ or
60~$\mathrm{km~s^{-1}}$.
We performed a comprehensive fit to the GeV--TeV spectral energy distribution. 
Although a leptonic scenario can reproduce the observed spectrum, a hadronic scenario is favored once electron cooling is considered. The inferred CR proton spectrum has an index of {$\sim$}2.2, and the total CR proton energy above 1~GeV is (1--5)~$\times 10^{49}~\mathrm{erg}$, depending on the assumed velocity range of the associated interstellar gas.
A stringent upper limit on diffuse X-ray emission further supports the proton PeVatron scenario.
\end{abstract}




\section{Introduction}

Cosmic-ray (CR) hadrons below the knee energy (${\sim}3 \times 10^{15}~\mathrm{eV}$) are believed to be produced and confined in the Milky Way Galaxy.
Hadronic CRs with energies of petaelectronvolts (PeV) should have gained energies at powerful particle accelerators in our Galaxy,
and identifying such accelerators called ``PeVatrons'' is of significant interest in modern astrophysics.
A PeVatron will produce intense $\gamma$-ray emission above 100~TeV through nucleon–nucleon interactions at the nearby interstellar medium (ISM) clouds.
The Tibet AS $\gamma$ experiment, a high-altitude CR and $\gamma$-ray observatory located in Tibet, discovered a sub-PeV $\gamma$-ray source that is positionally coincident with the molecular clouds toward the supernova remnant (SNR) 
G106.3+2.7 \citep{Amenomori2021}. Subsequently, 
the Large High Altitude Air Shower Observatory \citep[LHAASO, ][]{LHAASO_Inst}
reported more than 40 sub-PeV sources in its first source catalog \citep{LHAASO_1stCat}.

One caveat is that while CR hadrons primarily produce
sub-PeV $\gamma$-rays via nucleon-nucleon interactions,
CR electrons and positrons (leptons) can also contribute to the
sub-PeV $\gamma$-rays flux through inverse-Compton (IC) scattering off ambient soft photons.
Indeed, the Crab Nebula, a bright pulsar wind nebula (PWN) and an established CR electron/positron accelerator, is one of the LHAASO sub-PeV sources. 
The spatial distribution of $\gamma$-ray emission is key to distinguishing proton PeVatrons from electron PeVatrons, but the moderate angular resolution of LHAASO 
\citep[$0\fdg3$--$0\fdg8$ in 10-100~TeV;][]{LHAASO_Inst} prevents us from making a firm identification in most cases.
Although atmospheric Cherenkov detectors have excellent angular resolution \citep[${\sim}0\fdg08$; e.g.,][]{HESS2018_GPS}, their small field of view 
often prevents the firm identification of LHAASO sub-PeV sources,
which typically have angular extents $\ge 0\fdg5$.

With these scientific motivations and technical considerations in mind, we searched for a GeV counterpart of the sub-PeV source LHAASO~J1912+1014u using
the \textit{Fermi} Large Area Telescope \citep[LAT;][]{Atwood2009}. 
The \textit{Fermi}-LAT, with its wide field-of-view (${\sim}2.4~\mathrm{sr}$), good angular resolution 
(${\sim}0\fdg1$ above 10~GeV; see official \textit{Fermi}-LAT Performance webpage\footnote{
\url{https://www.slac.stanford.edu/exp/glast/groups/canda/lat_Performance.htm}}), and substantial
photon statistics (more than 17 years of operation), is suitable for our study.
The source was found to be extended (39\% radii when modeled by a 2D Gaussian, $r_{39}$, are $0\fdg36$ and $0\fdg50$ in the WCDA band (1--25~TeV) and the KM2A band ($\ge 25~\mathrm{TeV}$)
of LHAASO, respectively) and was significantly detected above 100~TeV at %%a test statistic %%(TS) 
%% ${\sim}68$ or 
{$\ge$}8$\sigma$. 
There is an old, powerful pulsar PSR~J1913+1011 \citep{Morris2002MNRAS} inside the WCDA/KM2A Gaussians, and another sub-PeV source LHAASO~J1914+1150u about $1\fdg4$ away.
Our source of interest (LHAASO~J1912+1014u)

is positionally coincident (source separation of only ${\sim}0\fdg1$) with
the well-known TeV source HESS~J1912+101 \citep{HESS2008,HESS2018}, which exhibits a shell-like morphology. Considering a small angular separation and comparable
source sizes, we will assume that LHAASO~J1912+1014u and HESS~J1912+101 are the same object
and call them collectively the ``LHAASO/H.E.S.S. source''
whenever we describe both sources.
Although several studies have been done on HESS~J1912+101 in the GeV energy band \citep{Li2023,Sun2022,Zeng2021,Zhang2020}, the nature of the source
has not been settled yet. This is mainly due to the difficulty in modeling the background diffuse emission and the source spatial distribution.

In this new study of the HESS~J1912+101 (and LHAASO~J1912+1014u) region, we constructed a custom background model. % and assessed the
% influence of nearby catalog sources marked with a caution flag. %%with "c" (coincident with interstellar clump) flag.
We revealed an extended $\gamma$-ray excess over known sources and Galactic diffuse emission above 10~GeV, coincident in position with the LHAASO sub-PeV source.
To accurately trace the ISM gas toward the object, 
we prepared two $\WCO$ maps (maps of the integrated $^{12}$CO (J=1--0) 2.6-mm line intensity)
using high-resolution FUGIN CO data \citep{Umemoto2017}.
We also included the contribution of atomic gas around the molecular clouds.
We then used those ISM gas templates and fit the \textit{Fermi}-LAT data to characterize the spectrum and morphology of the extended emission 
in the GeV energy band.
This paper is organized as follows: 
in Section~2, we describe $\gamma$-ray observations and data selection, and the preparation of the ISM gas templates.
In Section~3, we present the results of the data analysis, which confirm that our ISM gas maps reproduce the $\gamma$-ray excess as well as,
or better than, simple Gaussian models. In Section~4, we discuss possible scenarios based on the
spectral and spatial distributions of $\gamma$ rays measured by \textit{Fermi}-LAT
in combination with the published multiwavelength data.
Finally, in Section~5, we present a summary of the study and prospects.

\section{Observations}

\subsection{Gamma-ray Observation, Data Selection, and Analysis Procedure}

\textit{Fermi}-LAT is a pair-tracking $\gamma$-ray telescope that detects photons from {$\sim$}20~MeV to more than 300~GeV.
Thanks to its wide field of view (${\sim}$2.4~sr) and good angular resolution ({$\sim 0\fdg1$} above 10~GeV),
\textit{Fermi}-LAT is an ideal telescope for searching for the counterpart of extended TeV $\gamma$-ray sources in the Milky Way.
Routine science operations with the LAT started on August 4, 2008.
We have accumulated events from August 4, 2008, to August 2, 2023 (i.e., 15 years) to study GeV properties of
the LHAASO/H.E.S.S. source.
We used the standard LAT analysis software, Fermitools\footnote{
\url{https://fermi.gsfc.nasa.gov/ssc/data/analysis/software/}
} v2.2.0.
We also employed an analysis pipeline, Fermipy\footnote{
\url{https://fermipy.readthedocs.io/en/latest/}
} v1.3.1, a Python package based on Fermitools that enables automated analyses \citep{Wood2017}.

We defined our region of interest (ROI) of $15\arcdeg \times 15\arcdeg$ in Galactic coordinates
centered at the position of LHAASO~J1912+1014u in the KM2A band (Galactic longitude $l$ and latitude $b$ to be
($l$, $b$)=($44\fdg78$, $-0\fdg06$)), and selected all PSF type events of the SOURCE class.
We required that the reconstructed zenith angles of the
arrival directions of the photons be less than $100^{\circ}$ to reduce contamination by photons from Earth's atmosphere.
We used the data from 0.4 to 409.6~GeV and binned them
with pixel sizes of $0\fdg25$, $0\fdg1$, and $0\fdg05$ in the 0.4--1.6~GeV, 1.6-12.8~GeV, and above 12.8~GeV ranges, respectively.
This procedure aims to fully utilize the angular resolution of the \textit{Fermi}-LAT while keeping the fitting computation cost reasonable. 

To analyze the source spectrum and morphology, we applied binned maximum likelihood to the ROI. We employed the latest response functions,
P8R3\_SOURCE\_V3 \citep{P8Ref1,P8Ref2}.
%%which match our dataset and event selection.
We used a standard diffuse background model (gll\_iem\_v07.fits
\footnote{\url{https://fermi.gsfc.nasa.gov/ssc/data/access/lat/BackgroundModels.html}}), an isotropic spectral template (iso\_P8R3\_SOURCE\_V3\_v1.txt), 
and point source models from the fourth \textit{Fermi}-LAT catalog (4FGL-DR4) as described by \cite{4FGLDR3} and \citet{4FGLDR4}.
Additionally, we developed a customized background model as detailed in Section~3.1.
We let the normalizations of 4FGL sources within $7\fdg5$ of the ROI's center free to vary, and the normalizations and spectral indices of the sources
within $2\fdg0$ free to vary.
We also allowed the Galactic diffuse background model (normalization and index) and isotropic spectral template (normalization) to vary freely.
In testing several templates for the GeV counterpart of 
the LHAASO/H.E.S.S. source, we
evaluated the test statistic (TS) defined as $\mathrm{TS}=2\left( \ln{L_{1}}-\ln{L_{0}} \right)$,
where $L_{1}$ and $L_{0}$ are the maximum likelihood values with and without the source template.
The value of TS follows the $\chi^{2}$ distribution of the degree of freedom 
to be the number of free parameters \citep{Mattox1996}.
When we compare models of the different number of free parameters, we also used the Akaike Information Criterion (AIC) defined with the same sign as TS.

\subsection{HI and CO Observations and Target Gas Templates}

Galactic proton PeVatrons will produce 
hard $\gamma$-ray emission in the GeV and TeV energy ranges through the interaction with nearby ISM clouds.
Dense clouds are primarily in the molecular phase (molecular clouds) and
 are conventionally traced indirectly via carbon monoxide (CO) line emission \cite[e.g.,][]{Dame2001}.
CO line observations enable us to decompose molecular clouds along the line of sight and estimate the distance to the source.

\cite{HESS2008} reported a possible association of HESS~J1912+101 with CO-line emission in the systemic velocity ($V_\mathrm{LSR}$)
ranging 50--70~$\mathrm{km~s^{-1}}$. Later, \cite{Su2017} claimed that molecular clouds with $V_\mathrm{LSR} \sim 60~\mathrm{km~s^{-1}}$,
together with shocked molecular gas and high-velocity atomic shells, are concentrated toward the shell-like TeV source. 
Subsequent studies of HESS~J1912+101 \citep[e.g.,][]{Sun2022,Li2023}
used CO maps of this velocity range. 
\citet{Sano2018}, on the other hand, pointed out that the high velocity of the clouds identified by \cite{Su2017} 
could be due to
an \HII\ region along the line of sight, and proposed an alternative velocity of $V_\mathrm{LSR} \sim 25~\mathrm{km~s^{-1}}$. 
The clouds of this velocity range also exhibit
a shell-like structure in the CO distribution, 
which coincides with HESS~J1912+101 (H. Sano et al. in preparation). 
To discuss the association of these ISM structures with the LHAASO/H.E.S.S. source, we used $^{12}$CO data from the high-resolution FUGIN CO survey \citep{Umemoto2017}.
The effective angular resolution is $\sim20\farcs2$.
Because an artificial striped pattern along the scanning direction was identified in the publicly available FUGIN data, we reprocessed the data
using the NOSTAR software package provided by the Nobeyama Radio Observatory \citep[see Section 5.1 in][]{Sawada2008}. 
Details of the reprocessing are described in \citet{Murase2026}.
We used $\WCO$ maps of $^{12}$CO in the velocity ranges 58.3--62.2~$\mathrm{km~s^{-1}}$
and 23.2--26.4~$\mathrm{km~s^{-1}}$ to model CO clouds proposed by \citet{Su2017} and
\citet{Sano2018}, respectively.
We adopted a nominal value for $\XCO$ \citep[$2.0 \times 10^{20}~\mathrm{cm^{-2} (K~km~s^{-1})^{-1}}$; e.g.,][]{Bertsch1993} to convert $\WCO$ to an $\Htwo$ column density, yielding the proton column density
of molecular gas as $\Np(\Htwo) = 2 \cdot \XCO \cdot \WCO$.

Through detailed correlation among ISM gas tracers in high-latitude regions, a considerable amount of gas, likely cold and optically thick {\HI}, was revealed around the molecular clouds traced by CO line emission \citep[e.g.,][]{Fukui2015,Mizuno2025}.
The proton column density of atomic gas, $\Np$({\HI}), is usually derived under the assumption that the 21~cm {\HI} line is optically thin; however, if optically thick {\HI} is present, the column density will be underestimated by up to a factor of about two \citep[][]{Fukui_2014,Fukui2015}. 
In this study, we derived the corrected $\Np$({\HI}) map using the empirical correlation between $\WHI$ (the integrated 21~cm {\HI} line intensity) and the dust optical depth as derived by \citet{Fukui2017}.
%%For details, see \citet{Sano2025}.
The velocity integration ranges of $\WHI$ are identical to those for the CO data. 
We used the {\HI} data from the VLA Galactic Plane Survey \citep[VGPS;][]{stil2006}, which has a spatial resolution 
(full width at half maximum) of $\sim60\arcsec$. 
%%The observation and data reduction procedures are described in \citet{stil2006}.

From the $\Np$({\HI}) map including the contribution of optically-thick {\HI}, and the $\Np(\Htwo)$ map as described above, we constructed proton column density ($\Np$) maps for the Su et al. (2017) %% \citet{Su2017} 
velocity range and the Sano et al. (2018) %% \citet{Sano2018} 
velocity range. 
We smoothed the $^{12}$CO data to match the {\HI} data resolution before adding
the $\Np$({\HI}) and $\Np(\Htwo)$ maps.
We will use the obtained $\Np$ maps in the $\gamma$-ray data analysis in Section~3.\\

\section{Data Analysis and Results}

\subsection{Baseline Model}

Our first step was to establish a baseline model to identify the GeV counterpart of 
the LHAASO/H.E.S.S. source.
We fit the data using a standard diffuse model, isotropic template, and 4FGL sources.
The results showed significant, extended residuals next to the LHAASO/H.E.S.S. source (toward positive Galactic longitude) 
in the 1.6--12.8~GeV range, as shown in Figure~1(a).
We also observed a striped residual above 12.8~GeV in areas connecting 
($l$, $b$) $\sim$ ($46\fdg0$, $0\fdg0$) to ($45\fdg5$, $-0\fdg5$),
as shown in Figure~1(b).
The substantial size of the residuals below 12.8~GeV indicates that
they are not due to unmodeled point sources, 
but rather to the inadequate modeling of the ISM gas. 
The standard diffuse model was constructed from atomic gas templates
(traced by the 21~cm {\HI} line), molecular gas templates (traced by the CO 2.6~mm line), and a dark neutral medium (DNM) template
(constructed from dust emission that is not well correlated with {\HI} and CO lines). 
The model uses a single DNM template because dust emission contains
no velocity information,
and residuals will appear unless the ratio of $\gamma$-ray intensity to the estimated DNM column density
in the ROI agrees well with that of the all-sky average.
We observed significant residuals where the DNM template intensity is high, as shown in Figure~1(c),
indicating an underestimation of the DNM in this area.
To address this, we included the DNM template in our model as an additional ISM gas template.
The spectrum for the additional DNM template is soft and follows a power law 
with a photon index of ${\sim}$2.7 above 2~GeV,
supporting it as unmodeled ISM gas.
The striped residual above 12.8~GeV remains and is likely associated with another sub-PeV source, LHAASO~J1914+1150u, rather than our target, LHAASO~J1912+1014u. We employed three point sources to model this residual phenomenologically (see Figure~1(b) for their positions). 
They showed hard spectra (best-fit indices of 1.6--2.2),
and we fixed their indices to 2.0 for simplicity.

These additions, which we refer to as the ``baseline model'',
significantly reduced the residuals 
%%\mycolor{(with $\mathrm{TS}>300$)}{red}
as shown in Figures~1(d) and (e). 
%%The spectrum of the additional DNM template is soft and follows a power law 
%%with a photon index of ${\sim}$2.7 above 1~GeV,
%%supporting it as unmodeled ISM gas.
However, largely extended and structured excess emission above 12.8~GeV remains in areas
toward the LHAASO/H.E.S.S. source,
as indicated in Figure~1(e).
We interpret this as the GeV counterpart 
of the LHAASO/H.E.S.S. source and
will examine its spatial and spectral properties
in the following sections.

We remind that there are two catalog sources marked with an identifier of ``c''
(coincident with interstellar clumps) within the LHAASO $r_{39}$ radii of WCDA and KM2A Gaussian, namely 
4FGL~J1911.7+1014c and 4FGL~J1914.7+1012c
(see Figure~1 for their locations).
They were not included to model the background 
in some of the past studies of \textit{Fermi}-LAT data on HESS~J1912+101 \citep[e.g.,][]{Zeng2021,Sun2022,Li2023}.
We found that their spectra are soft, similar to that of the additional DNM template. We interpret this as indicating that these sources are
not the GeV counterparts of the LHAASO/H.E.S.S. source, but rather arise from another unmodeled (clumpy) ISM gas component.
Accordingly, we continue to include them (and all other sources with a ``c'' identifier in our ROI) 
as point sources in the following analysis.

\begin{figure}[htbp]
\begin{tabular}{cc}
\begin{minipage}{0.49\textwidth}
\centering
\begin{overpic}[width=\textwidth]{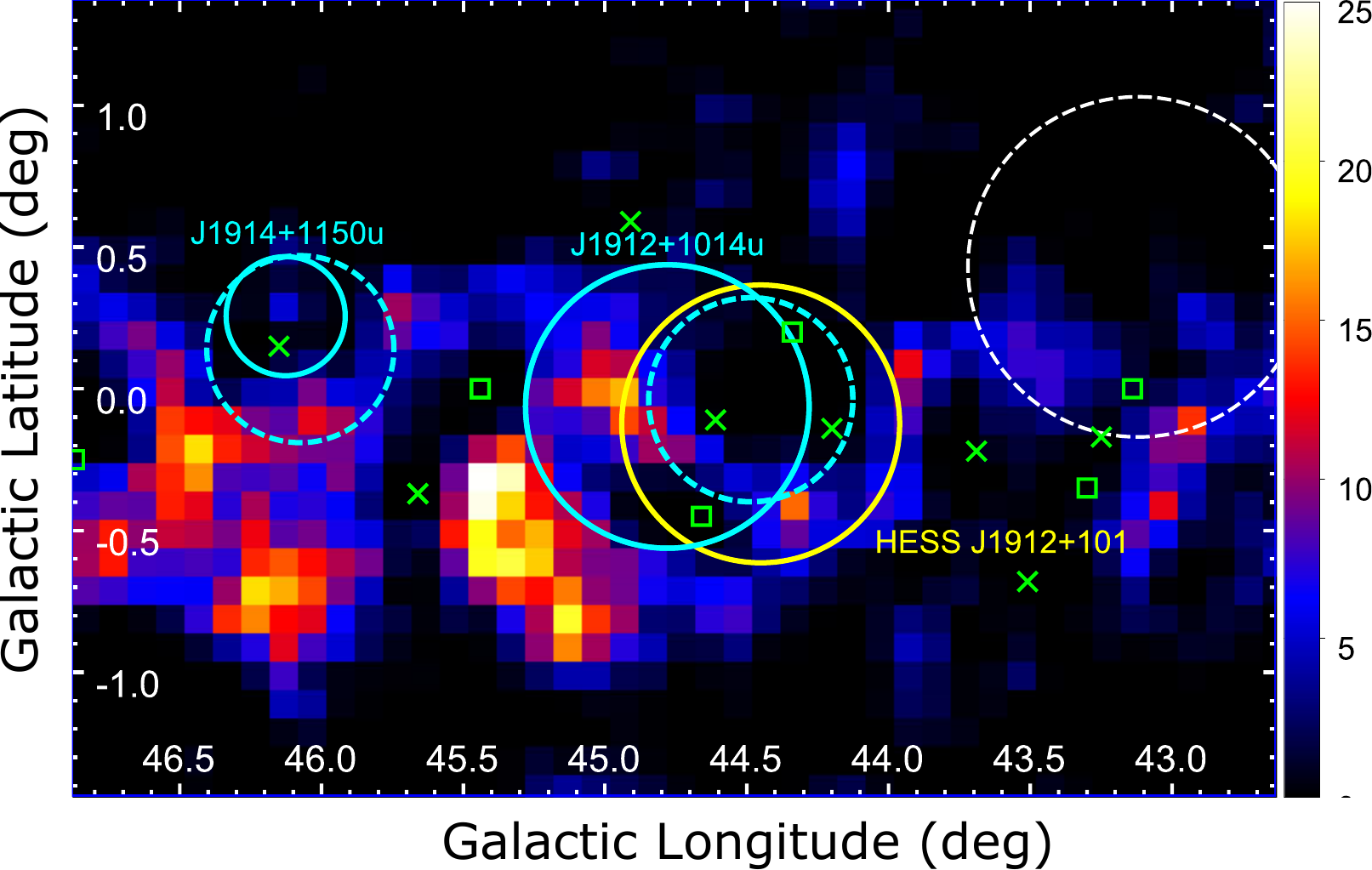}
\put(15,65){(a)}
\end{overpic}
\end{minipage}
\hspace{0.02\textwidth}
\begin{minipage}{0.49\textwidth}
\centering
\begin{overpic}[width=\textwidth]{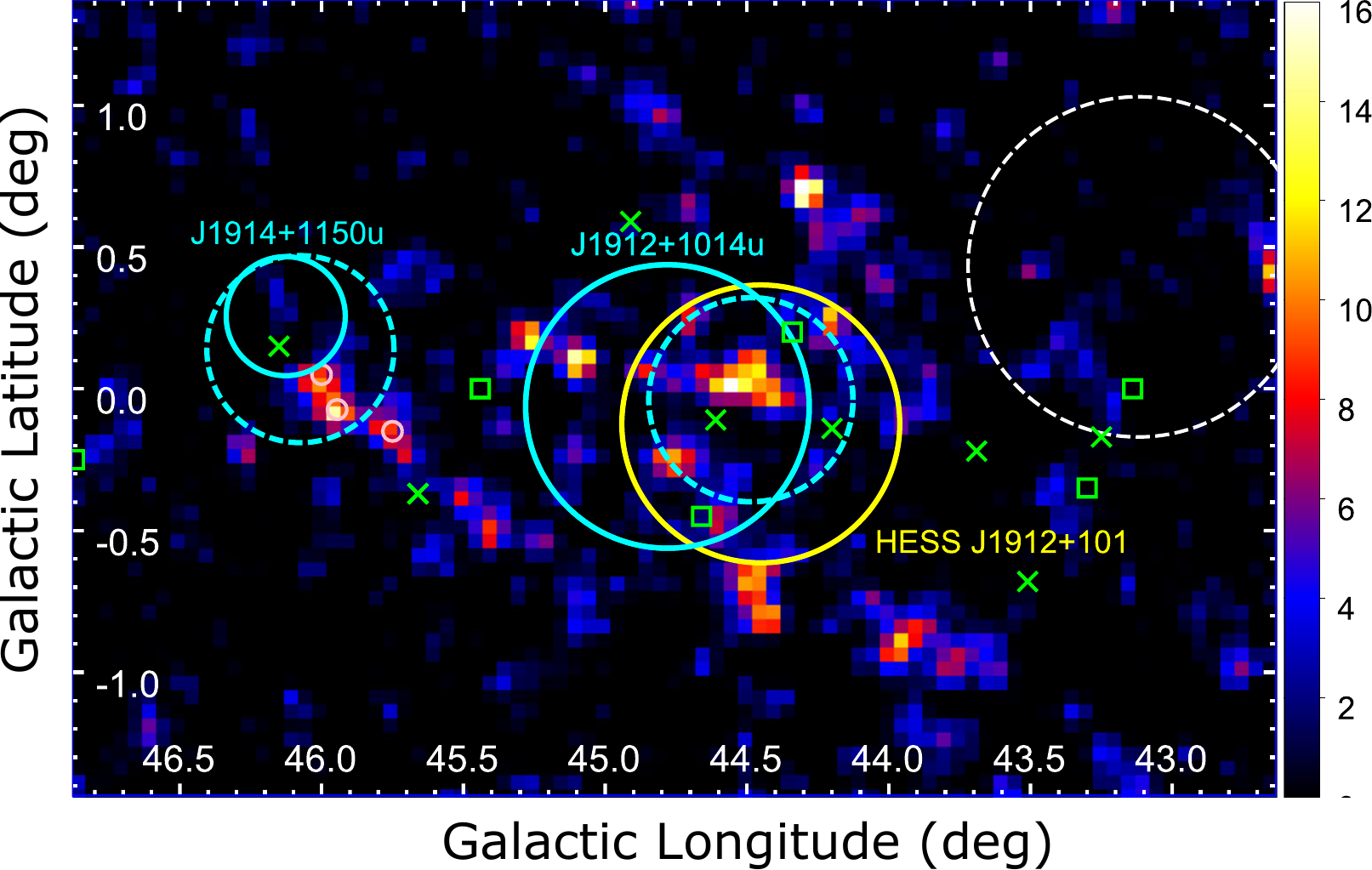}
\put(15,65){(b)}
\end{overpic}
\end{minipage} \\
\\
\\
\begin{minipage}{0.49\textwidth}
\centering
\begin{overpic}[width=\textwidth]{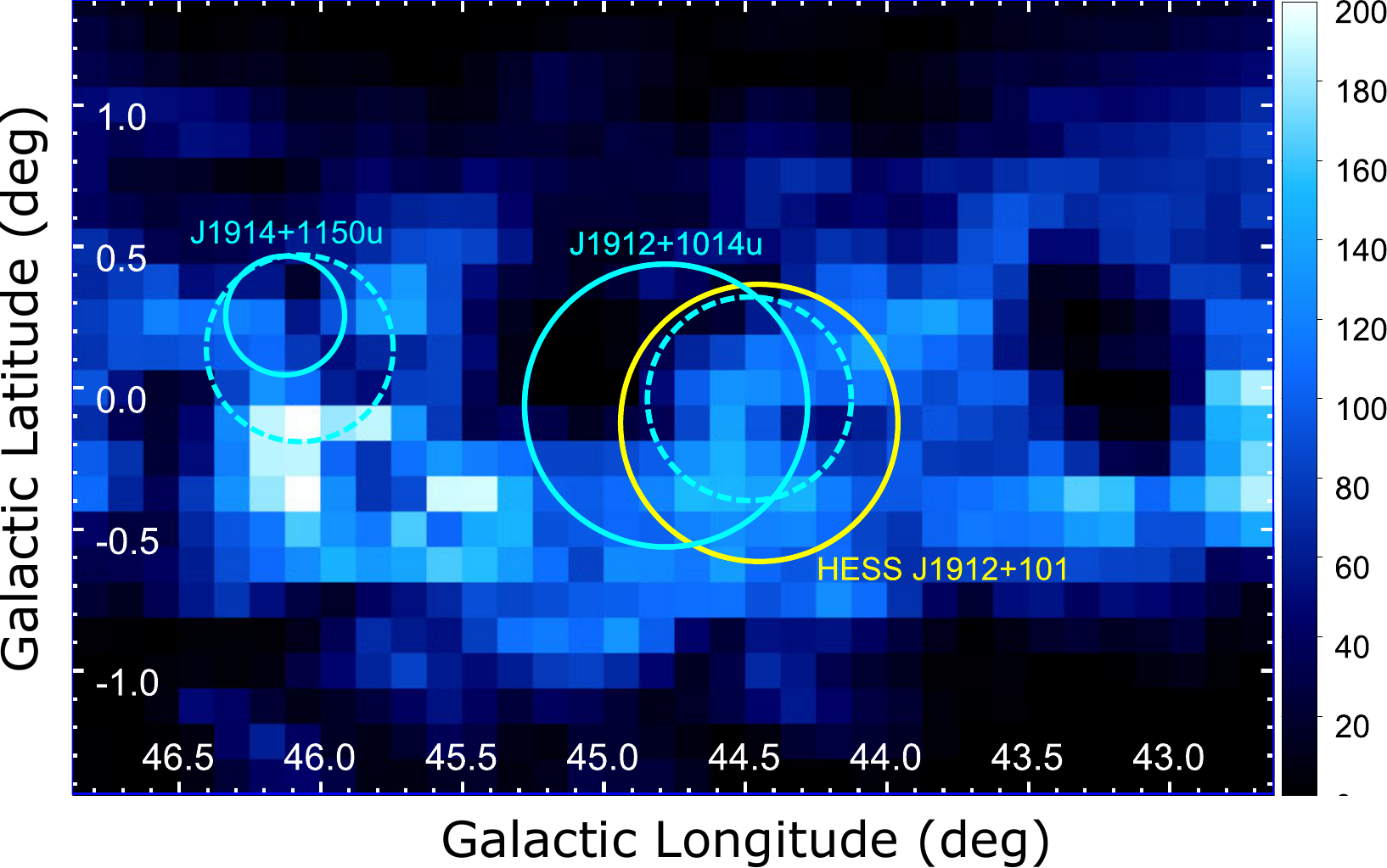}
\put(15,65){(c)}
\end{overpic}
\end{minipage} \\
%\begin{minipage}{0.5\textwidth}
%\centering
%\begin{overpic}[width=\textwidth]{Figs/Fig1d.png}
%\put(15,75){(d)}
%\end{overpic}
%\end{minipage}
\\
\\
\begin{minipage}{0.49\textwidth}
\centering
\begin{overpic}[width=\textwidth]{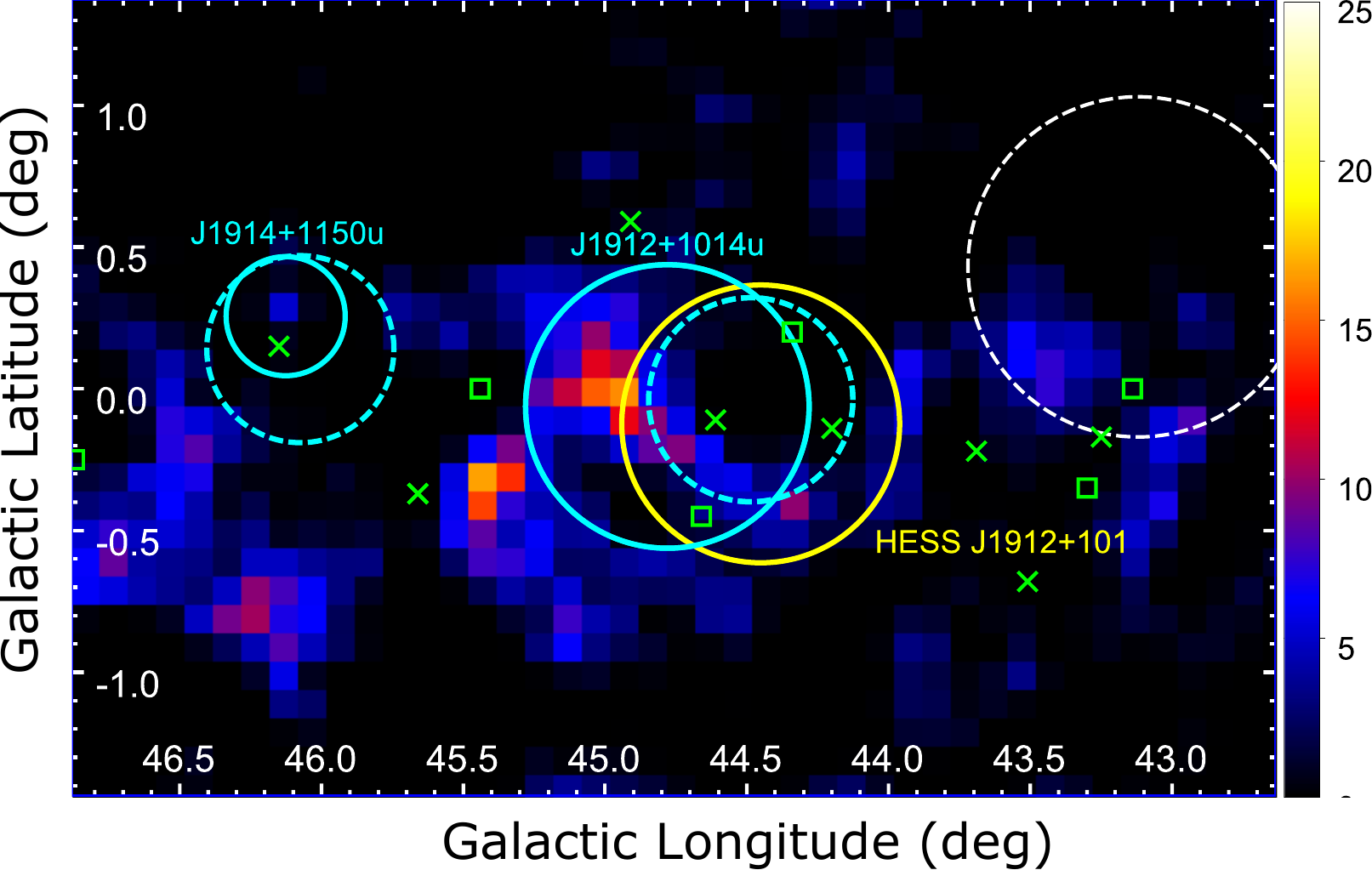}
\put(15,65){(d)}
\end{overpic}
\end{minipage}
\hspace{0.02\textwidth}
\begin{minipage}{0.5\textwidth}
\centering
\begin{overpic}[width=\textwidth]{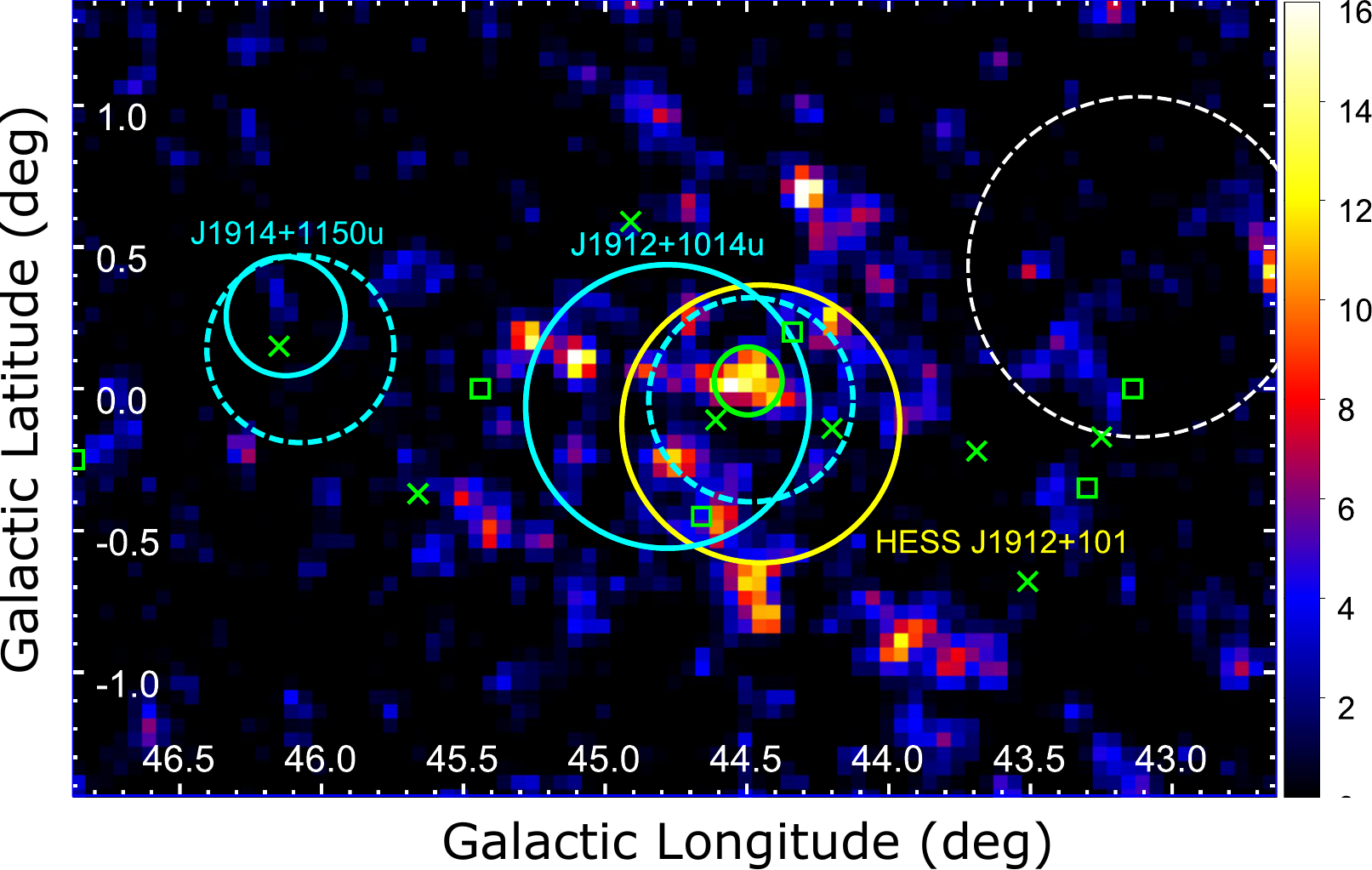}
\put(15,65){(e)}
\end{overpic}
\end{minipage}
\end{tabular}
\caption{
(a) The TS map in 1.6--12.8~GeV constructed using standard background models. 
The 4FGL sources are marked by squares (for those with a ``c'' identifier) and crosses (for others). 
The position and size of 4FGL J1908.6+0915e, modeled by a uniform disk, are shown by a white dashed circle. LHAASO sub-PeV sources, modeled by a 2D Gaussian, are indicated by cyan dashed (WCDA) and solid (KM2A) circles. The position and size of HESS~J1912+101, modeled by a spherical shell in \citet{HESS2018}, are shown by a yellow circle.
(b) The same as panel (a), but in $\ge 12.8~\mathrm{GeV}$ instead of 1.6--12.8~GeV. Three point sources to model the striped residual near LHAASO J1914+1150u are indicated by small pink circles.
(c) The DNM template map (in arbitrary units) used to construct the standard diffuse model, 
which is internally available to the \textit{Fermi}-LAT collaboration.
(d) The same as panel (a), but when modeled with the additional ISM gas template and three point sources.
(e) The same as panel (b), but when modeled with the additional ISM gas template and three point sources. A green circle at ($l$, $b$) $\sim$ ($44\fdg5$, $0\fdg0$) indicates
the small peak-like structure studied in Section~4.4.
\label{fig:f1}
}
\end{figure}

%%\clearpage

\subsection{Modeling the GeV Counterpart of the LHAASO/H.E.S.S. Source}

We observed an extended hard excess (diameter greater than $1\arcdeg$), likely the GeV counterpart of the LHAASO/H.E.S.S. source. 
As a preliminary analysis, we evaluated its size using a uniform-disk model.
The position and radius of it (determined using {\tt extension} method in Fermipy) are ($l$, $b$)=($44\fdg56 \pm 0\fdg08$, $-0\fdg05 \pm 0\fdg08$)) and 
$0\fdg86 \pm 0\fdg04$, respectively.
Modeling the excess with this uniform disk and a power-law spectrum yields 
a TS value of {$\sim$}100 and a hard spectrum with the photon index of {$\sim$}2.1,
as reported in Table~1. Accordingly, we detected a largely extended (radius ${\ge}0\fdg8$) 
GeV excess. The observed hard spectrum demonstrates that this excess originates from a CR accelerator rather than from unmodeled gas.
To investigate the physical origin of this excess,
we tested five templates prepared a priori,
each chosen for its potential to accurately represent the excess emission:
the H.E.S.S. intensity map, 
$\Np$ maps
of $V_\mathrm{LSR}$ in 58.3--62.2~$\mathrm{km~s^{-1}}$
\citep{Su2017} and 
$V_\mathrm{LSR}$ in 23.2--26.4~$\mathrm{km~s^{-1}}$
\citep{Sano2018}, 
and two Gaussian models of LHAASO~J1912+1014u in \citet{LHAASO_1stCat}.
The H.E.S.S. intensity map was constructed from the H.E.S.S. Galactic plane survey \citep{HESS2018_GPS} and is publicly available\footnote{
\url{https://www.mpi-hd.mpg.de/HESS/hgps/}
}.
This map will provide a template for the counterpart of HESS~J1912+101.
Two $\Np$ maps were constructed from {\HI} and CO data (Section~2.2). They will be
used to model the distribution of the target gas, either for protons (via hadronic interactions) or electrons (via electron Bremsstrahlung).
Two Gaussians are phenomenological models aimed to represent the GeV counterpart of LHAASO~J1912+1014u. 
The H.E.S.S. intensity map, the $\Np$ map of Su et al. (2017) velocity range,
and the $\Np$ map of Sano et al. (2018) velocity range are shown in Figures~2(a)--(c).

\begin{figure}[htbp]
\begin{tabular}{cc}
\begin{minipage}{0.49\textwidth}
\centering
\begin{overpic}[width=\textwidth]{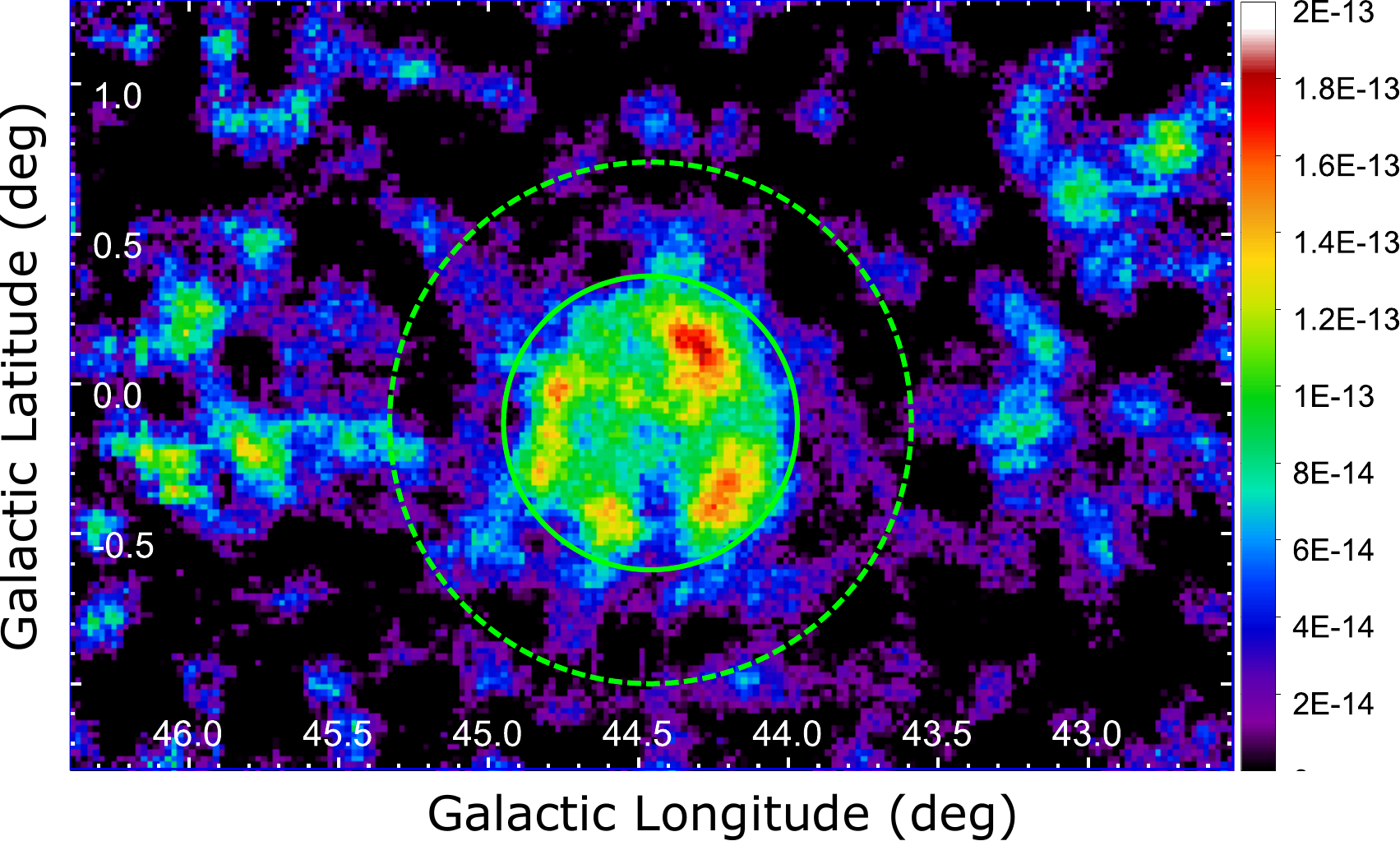}
\put(15,65){(a)}
\end{overpic}
\end{minipage}
\hspace{0.02\textwidth}
\begin{minipage}{0.49\textwidth}
\centering
\begin{overpic}[width=\textwidth]{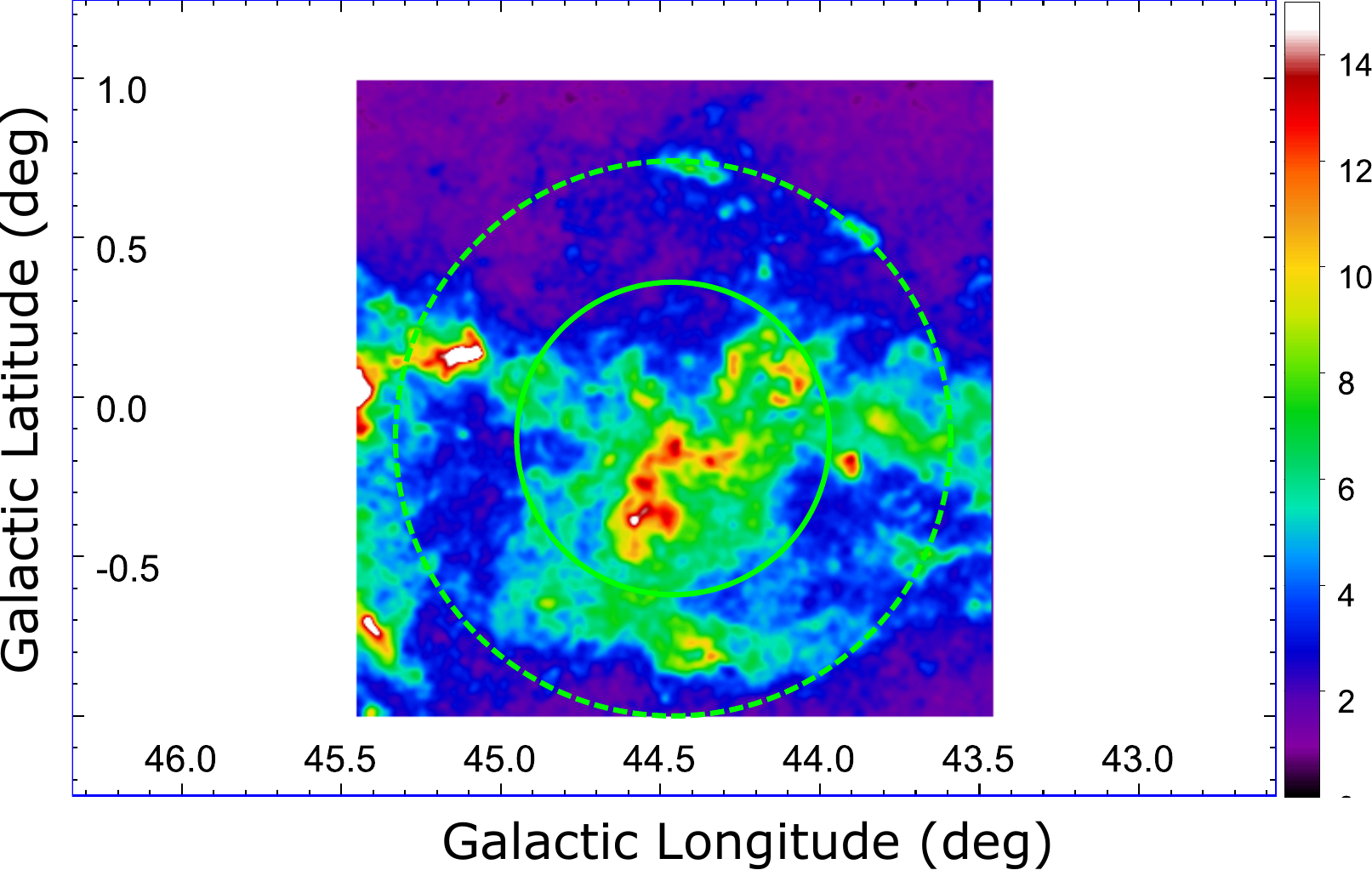}
\put(15,65){(b)}
\end{overpic}
\end{minipage} \\
\\
\\
\begin{minipage}{0.49\textwidth}
\centering
\begin{overpic}[width=\textwidth]{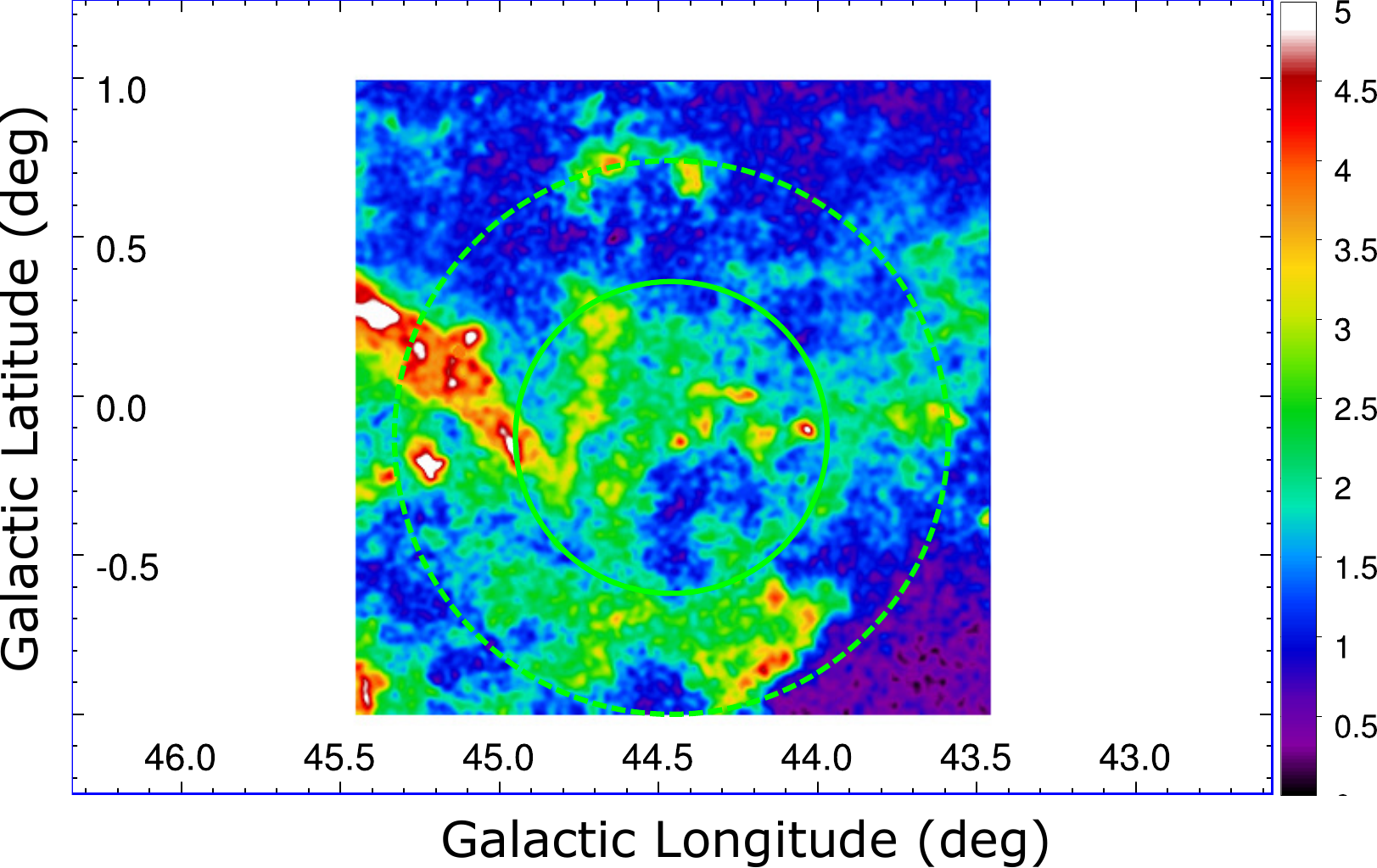}
\put(15,65){(c)}
\end{overpic}
\end{minipage}
\hspace{0.02\textwidth}
\begin{minipage}{0.49\textwidth}
\centering
\begin{overpic}[width=\textwidth]{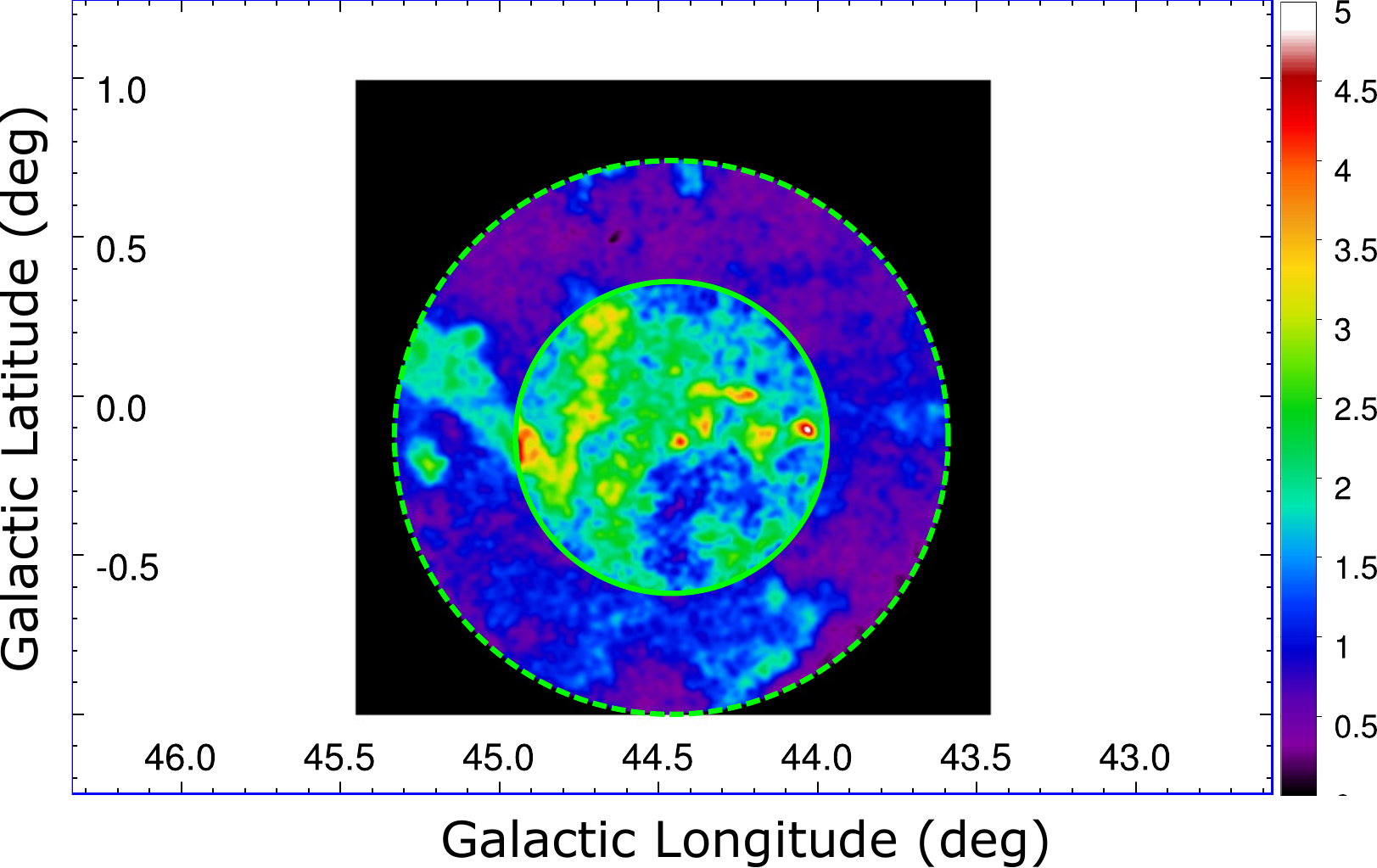}
\put(15,65){(d)}
\end{overpic}
\end{minipage}
\end{tabular}
\caption{
Template maps tested to reproduce the GeV excess emission toward the LHAASO/H.E.S.S. source.
(a) The H.E.S.S. intensity map. 
(b) The $\Np$ map of Su et al. (2017) velocity range ($58.4$--$62.2$ km $\mathrm{s^{-1}}$).
(c) The $\Np$ map of Sano et al. (2018) velocity range ($23.2$--$26.4$ km $\mathrm{s^{-1}}$).
(d) The same as (c), but $\Np$ in the annulus is scaled by 1/2.10 (see text for details).
There, the solid circle (radius of $0\fdg49$; the same as the yellow circle in Figure~1) shows the size of the spherical shell of HESS~J1912+101 used in the initial modeling, and the dashed circle (radius of $0\fdg87$) shows the area used in the final modeling. 
The H.E.S.S. intensity map is arbitrarily scaled to be compatible with Fermitools. 
The unit of $\Np$ maps is $\mathrm{10^{21}~cm^{-2}}$.
\label{fig:f2}
}
\end{figure}

We first examined the excess emission in the HESS~J1912+101 region using the H.E.S.S. intensity map and two $\Np$ maps. 
To compare the three templates 
in a self-consistent manner
and to compare our results with previous works \citep[e.g.,][]{Zhang2020,Sun2022,Li2023},
we selected a circle of $0\fdg49$ radius 
centered at ($l$, $b$) $ = $ ($44\fdg46$, $-0\fdg13$).
This area is determined to match the outer boundary of 
the spherical shell used to model HESS~J1912+101 in \citet{HESS2018}.
We modeled each template with a power-law spectrum. 
The TS values were 60.0, 60.8, and 68.6 for the H.E.S.S. intensity map, 
the Su et al. (2017) $\Np$ map, and the Sano et al. (2018) $\Np$ map, respectively, with
similar photon indices (${\sim}$2.05). 
%%While the Sano et al. (2018) $\Np$ map gives the best representation of the data, 
Accordingly, all three templates significantly improve the fit
(by more than 7$\sigma$).
We also found that without sources with a ``c'' identifier in the background model, the fit 
worsens (by about 20 in $\ln L$) and the spectrum 
softens below 10~GeV. We understand that this is why most pioneering
work on \textit{Fermi}-LAT data to search for the HESS~J1912+101 counterpart 
\citep[e.g.,][]{Zeng2021,Sun2022,Li2023}
reported soft spectra below 10~GeV.

Next, we examined the entire area of the observed excess emission (diameter greater than $1\arcdeg$). To do this, we employed the same three maps but with a radius of $0\fdg87$. 
This radius was chosen not to exceed the boundary of FUGIN CO maps and match the size of the uniform-disk model in the initial analysis.
We also used two LHAASO Gaussians
(one with $r_{39} = 0\fdg36$ and the other with $0\fdg50$). We found that the $\Np$ map of the Sano et al. (2018) velocity range prefers different normalizations inside and outside the
HESS~J1912+101 circle (of $0\fdg49$ radius)
\footnote{with TS=6.6 (giving 2.5$\sigma$ significance)
when using the $\Np$ maps inside the circle of $0\fdg49$ radius
and the annulus just outside instead of the single $\Np$ map
}
, and
we allowed the normalizations to vary independently 
(the number of degrees of freedom is larger by one than the others). 
The $\gamma$-ray emissivity ($\gamma$-ray intensity per unit $\Np$) inside the 
HESS~J1912+101 circle is 2.10 times larger.
The corrected $\Np$ map, in which $\Np$ in the annulus is scaled by 1/2.10 to make it compatible with
a uniform CR intensity, is shown in Figure~2(d).
Hereafter, we refer to this scaled map as the ``Sano et al. (2018) $\Np$ map''.
Allowing different normalizations does not improve the fit significantly 
for the Su et al. (2017) velocity range;
therefore, the original, unscaled map is used for their velocity range.
The best-fit photon index, TS, and AIC are summarized
in Table~1.
Therefore, all template maps significantly improve the fit 
(by more than 8$\sigma$) and yield similar values of the photon index. Among five models, 
the Su et al. (2017) $\Np$ map provides the best fit to \textit{Fermi}-LAT data, followed by 
the Sano et al. (2018) $\Np$ map and the LHAASO KM2A Gaussian.
These three maps imply different physical origins of the source,
with the Su et al. (2017) and Sano et al. (2018) $\Np$ maps indicating $\gamma$-ray emission from ISM gas at different distances and the
LHAASO KM2A Gaussian representing a smooth, symmetric distribution consistent with
$\gamma$-rays produced by IC scattering.
However, they exhibit only minor
differences in AIC. 
The spectral energy distributions (SEDs) in the LAT band of the three templates are also similar.
Accordingly, we will treat them in a fair way in Section~4.
We show the SED for the Sano et al. (2018) $\Np$ map together with the TS map as a representative case in Figure~3. 
SEDs and TS maps of the other two templates and numerical values of the fluxes are given in Appendix~1 (Table~4 and Figure~9).

With our baseline model, 
the extended excess emission toward the LHAASO/H.E.S.S. source
exhibits a non-uniform spatial distribution above 12.8~GeV,
with a prominent peak at
($l$, $b$) $\sim$ ($44\fdg5$, $0\fdg0$).
The coordinate is close to the centers 
of HESS~J1912+102 and LHAASO WCDA Gaussian, as well as the position of PSR~J1913+1011. 
This peak-like structure persists regardless of the models adopted for the LHAASO/H.E.S.S. source counterpart.
To examine this small structure, we modeled it with a uniform disk in our baseline model.
Using the {\tt extension} method
\footnote{
with the width\_max parameter set at 0.2 (in degrees)
}
in Fermipy, we obtained a position of
($l$, $b$)=($44\fdg50 \pm 0\fdg04$, $0\fdg03 \pm 0\fdg03$) and a radius of $0\fdg12 \pm 0\fdg03$ (Figure~1(e)).
The source spectrum is hard with a photon index of $1.76\pm0.05$.
However, its flux is only about 20\% of the total LHAASO/H.E.S.S. source counterpart in the 10--100~GeV band,
and its significance is marginal (the TS value is 10.0) even on top of our baseline model. Accordingly, we do not include this component in our spatial templates for the LHAASO/H.E.S.S. source counterpart.

Hereafter, we will first discuss the implications for CRs responsible for the overall GeV/TeV emission based on multiwavelength SED modeling
in Sections~4.1--4.3, and then revisit the small peak-like structure in Section~4.4.

\begin{deluxetable*}{cccc}[htbp]
\label{tab:fermi:TS}
%%\tablenum{1}
\tablecaption{Spatial models tested for the GeV excess toward the LHAASO/H.E.S.S. source. The values of TS and AIC are calculated with respect to the baseline model.}
\tablewidth{0pt}
\tablehead{
\colhead{Model} & \colhead{$\Delta$TS} & \colhead{$\Delta$AIC} & \colhead{index}
}
%%%\decimalcolnumbers
\startdata
Radial Disk & 98.6 & 88.6 & $2.08 \pm 0.01$ \\
\hline
H.E.S.S. int. map & 78.6 & 74.6 & $2.11 \pm 0.01$ \\
Su et al. (2017) $\Np$ map & 99.6 & 95.6 & $2.08 \pm 0.01$ \\
Sano et al. (2018) $\Np$ map & 100.8 & 94.8 & $2.06 \pm 0.01$ \\
WCDA Gaussian & 90.4 & 86.4 & $2.09 \pm 0.01$ \\
KM2A Gaussian & 98.4 & 94.4 & $2.08 \pm 0.01$ \\
\enddata
%%\tablecomments{This table ``hides'' the third column in the \latex\ when compiled.
%%The Distance is also centered on the decimals.  Note that when using decimal
%%alignment you need to include the {\tt\string\decimals} command before
%%{\tt\string\startdata} and all of the values in that column have to have a
%%space before the next ampersand.}
\end{deluxetable*}

\begin{figure}[htbp]
\begin{tabular}{cc}
\begin{minipage}{0.49\textwidth}
\centering
\begin{overpic}[width=\textwidth]{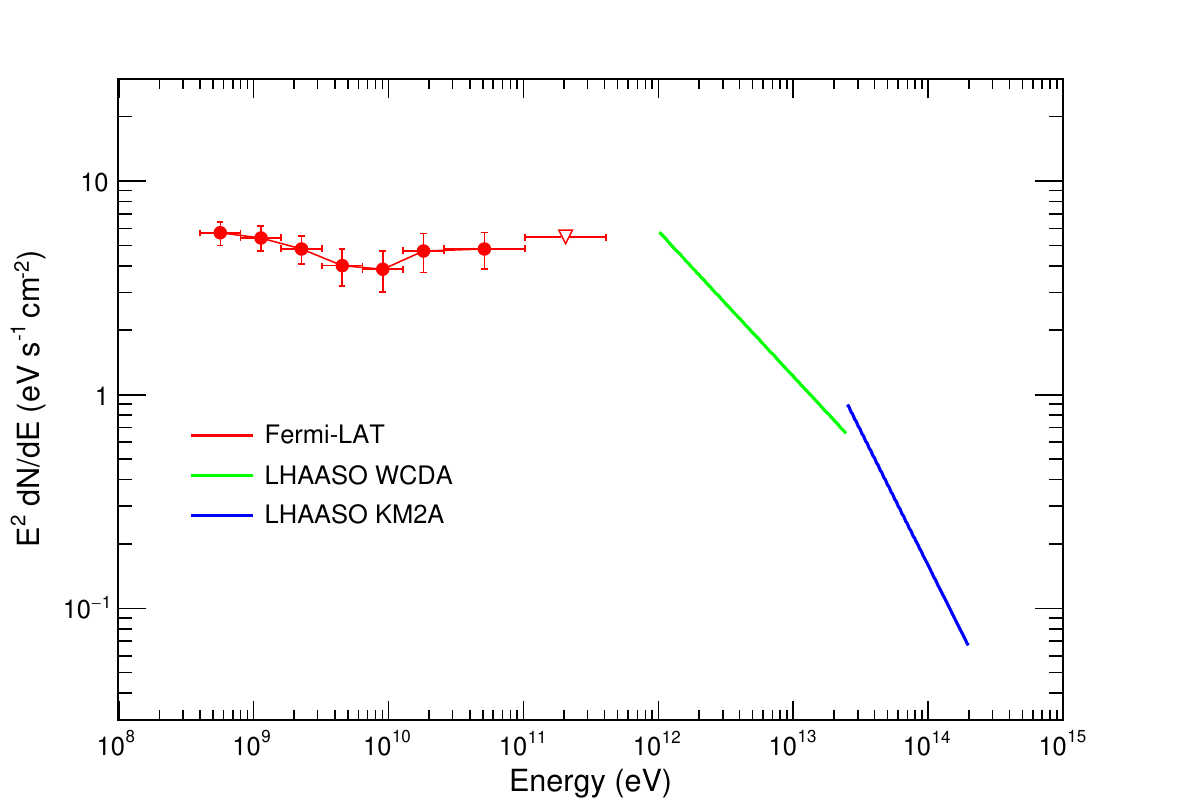}
\put(15,65){(a)}
\end{overpic}
\end{minipage}
%%\vspace{0.02\textwidth}
\begin{minipage}{0.49\textwidth}
\centering
\begin{overpic}[width=\textwidth]{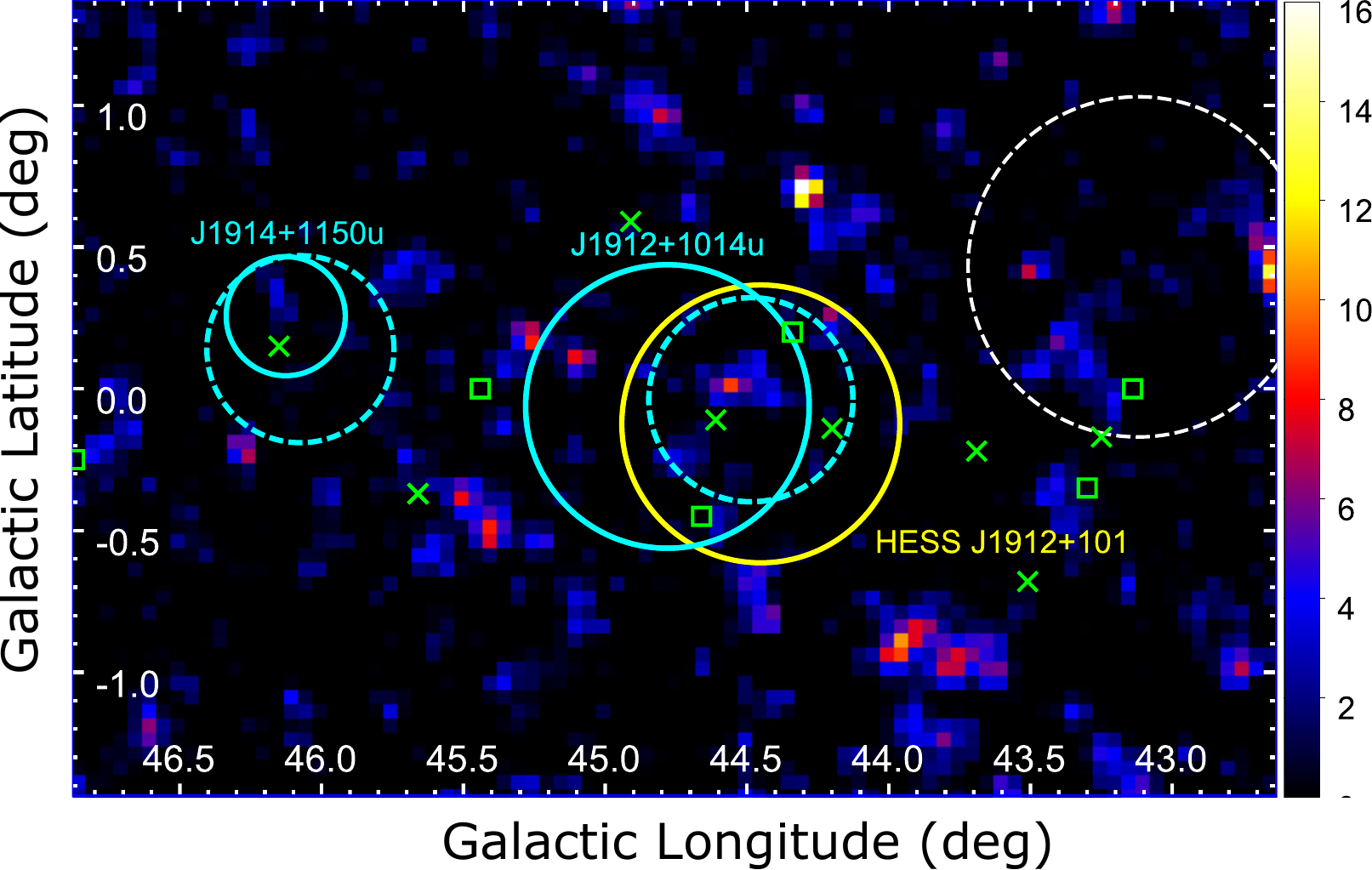}
\put(15,65){(b)}
\end{overpic}
\end{minipage}
\end{tabular}
\caption{
(a) \textit{Fermi}-LAT GeV spectrum of the LHAASO/H.E.S.S. source counterpart (red points) obtained using the Sano et al. (2018) $\Np$ map (radius of $0\fdg87$) as a template. 
The best-fit spectral models for LHAASO~J1912+1014u, as reported by
\citet{LHAASO_1stCat}, are also plotted.
(b) The TS map in {$\ge$}12.8~GeV obtained using the Sano et al. (2018) $\Np$ map.
\label{fig:f3}
}
\end{figure}

%%\clearpage

\section{Discussion}

Using the GeV spectrum together with data at other wavelengths, we will model the non-thermal spectrum of the LHAASO/H.E.S.S. source to investigate its $\gamma$-ray origin.
We refrain from assuming species for the parent CRs during model fitting but employ a phenomenological approach based on observational data.
Possible mechanisms for $\gamma$-ray radiation are the IC scattering or electron Bremsstrahlung (eB) for electrons, and neutral pion decay produced in pp collisions (pp) for protons.
Which of these radiation components becomes dominant depends primarily on the electron-to-proton flux ratio ($K_\mathrm{ep}$) and the target gas density ($n_\mathrm{gas}$), as well as the energy density of the seed photon for IC.
In particular, since $K_\mathrm{ep}$ and $n_\mathrm{gas}$ may vary by more than an order of magnitude depending on the assumed environment, we scan these two parameters over wide ranges, $K_\mathrm{ep}$ $\in [10^{-5}, 10^{2}]$ and $n_\mathrm{gas} \in [10^{-3}, 10^{3}]$~cm$^{-3}$, and attempt to identify viable scenarios for plausible accelerators and environments, such as an SNR 
(in a high gas density environment) and a PWN (in a low gas density environment).

We assume that CR spectra can be characterized by a power-law function with an exponential cutoff: 
% $f\mathrm{(}E\mathrm{)}=A\mathrm{(}E/E_{0}\mathrm{)}^{-\alpha} \mathrm{exp(}-E/E_\mathrm{cut}\mathrm{)}$.
$f\mathrm{(}E\mathrm{)} \propto \mathrm{(}E/E_{0}\mathrm{)}^{-\alpha} \mathrm{exp}(-E/E_\mathrm{cut}\mathrm{)}$.
We obtain the spectral index $\alpha$, the cutoff energy $E_\mathrm{cut}$, and the normalizations
(as the total energy of protons ($W_\mathrm{p}$) and electrons ($W_\mathrm{e}$) above 1~GeV) through SED model fitting.
The spectral indices are common between CR protons and electrons, while the electron spectrum is scaled by the factor $K_\mathrm{ep}$ relative to the proton spectrum.
Regarding the cutoff energies of protons ($E_\mathrm{cut, p}$) and electrons ($E_\mathrm{cut, e}$), we first consider a no-cooling case where 
$E_\mathrm{cut, p} = E_\mathrm{cut, e}$ in Section~\ref{sec:model:no_cooling_case}, 
and then fix $E_\mathrm{cut, e}$ to a representative value in Section~\ref{sec:model:cooling_case}, because electrons undergo radiative cooling more efficiently than protons, resulting in a smaller maximum energy in SNRs \citep[e.g.,][]{Ohira2012}.

\subsection{SED Modeling Setup}

We calculate photon spectra using the Naima framework \citep{NAIMA}.
We also employ the Markov Chain Monte Carlo %%(MCMC) 
algorithm \citep{Foreman2013PASP} implemented in Naima, to optimize the CR spectral parameters.
% In the MCMC fitting, only the normalization flux, spectral index, and cutoff energy are allowed to vary freely, while $K_\mathrm{ep}$ and $n_\mathrm{gas}$ are fixed to given values within the aforementioned ranges.
This fitting takes into account the \textit{Fermi}-LAT spectrum derived in Section~3 and the TeV measurements of HESS \citep{HESS2018_GPS}, 
HAWC \citep{HAWC2017_2HWC}, and LHAASO \citep{LHAASO_1stCat}.
For the \textit{Fermi}-LAT GeV data points, we use the results from the %% \citet{Sano2018} 
Sano et al. (2018) $N_\mathrm{p}$ map, which agree with those from the other viable spatial templates (i.e., Su et al. (2017) $\Np$ map and the KM2A Gaussian) %%\citet{Su2017} 
within the statistical $1\sigma$ errors. 
Results of the HAWC \citep{HAWC2017_2HWC} and LHAASO \citep{LHAASO_1stCat} measurements are reported not as data points but as fitted power-law functions, and we consider the flux and its statistical uncertainty at the normalization energy in the SED fitting.

\citet{Sano2018} and \citet{Su2017} respectively proposed an association of HESS~J1912+101 with the ISM gas in the $23.2$--$26.4$ km $\mathrm{s^{-1}}$ and $58.3$--$62.2$~km $\mathrm{s^{-1}}$ velocity ranges (SNR scenario).
Based on the Galactic rotation curve model of \citet{Brand1993}, the distance to the Earth is estimated to be $1.7\pm0.1$~kpc (near side) and $10.5\pm0.1$~kpc (far side) for the %%\citet{Sano2018} case,
Sano et al. (2018) velocity range,
whereas it is estimated to be $4.6\pm0.2$~kpc (near side) and $7.5\pm0.2$~kpc (far side) for the %%\citet{Su2017} case.
Su et al. (2017) velocity range.
Alternatively, an energetic pulsar PSR~J1913+1011 could accelerate electrons and positrons, producing GeV/TeV $\gamma$-rays;
its distance is estimated at 4.5~kpc from the dispersion measure \citep{Manchester05}.
Accordingly, we adopt $D=1.7$~kpc in the following calculations, but also discuss the case of larger distances.

As for seed photon fields in the IC process, we consider the cosmic microwave background (CMB) and Galactic far-infrared (FIR) radiation.
We used the interstellar radiation field model by \citet{Porter2017}, 
and adopted the FIR energy density and temperature (0.23~eV cm$^{-3}$ and 42.9~K, respectively) at $D=1.7~\mathrm{kpc}$.
Assuming $D=4.6~\mathrm{kpc}$ gives similar values, 
0.36~eV cm$^{-3}$ and 43.6~K. Since the difference is small and the
CMB photons constitute the dominant target photon field for interactions with TeV electrons, the impact on our conclusions is negligible.

For the target gas (in eB or pp processes), 
we estimate the number density in the vicinity of the $\gamma$-ray source for the %% \citet{Sano2018} $N_\mathrm{p}$ map data.
Sano et al. (2018) $\Np$ map.
Assuming a spherical volume with a radius of $0\fdg87$ from the center of the H.E.S.S. source ($l=44\fdg46$, $b=-0\fdg13$), the proton number density is found to be $n_\mathrm{gas} = 11.3~\mathrm{cm}^{-3}$.
The number density is $16.2~\mathrm{cm^{-3}}$ for the Su et al. (2017) $\Np$ map. Accordingly, we adopt
$n_\mathrm{gas} = 10~\mathrm{cm}^{-3}$ as a nominal case.

\subsection{No Cooling Case} \label{sec:model:no_cooling_case}

\begin{figure*}[htbp]
\plotone{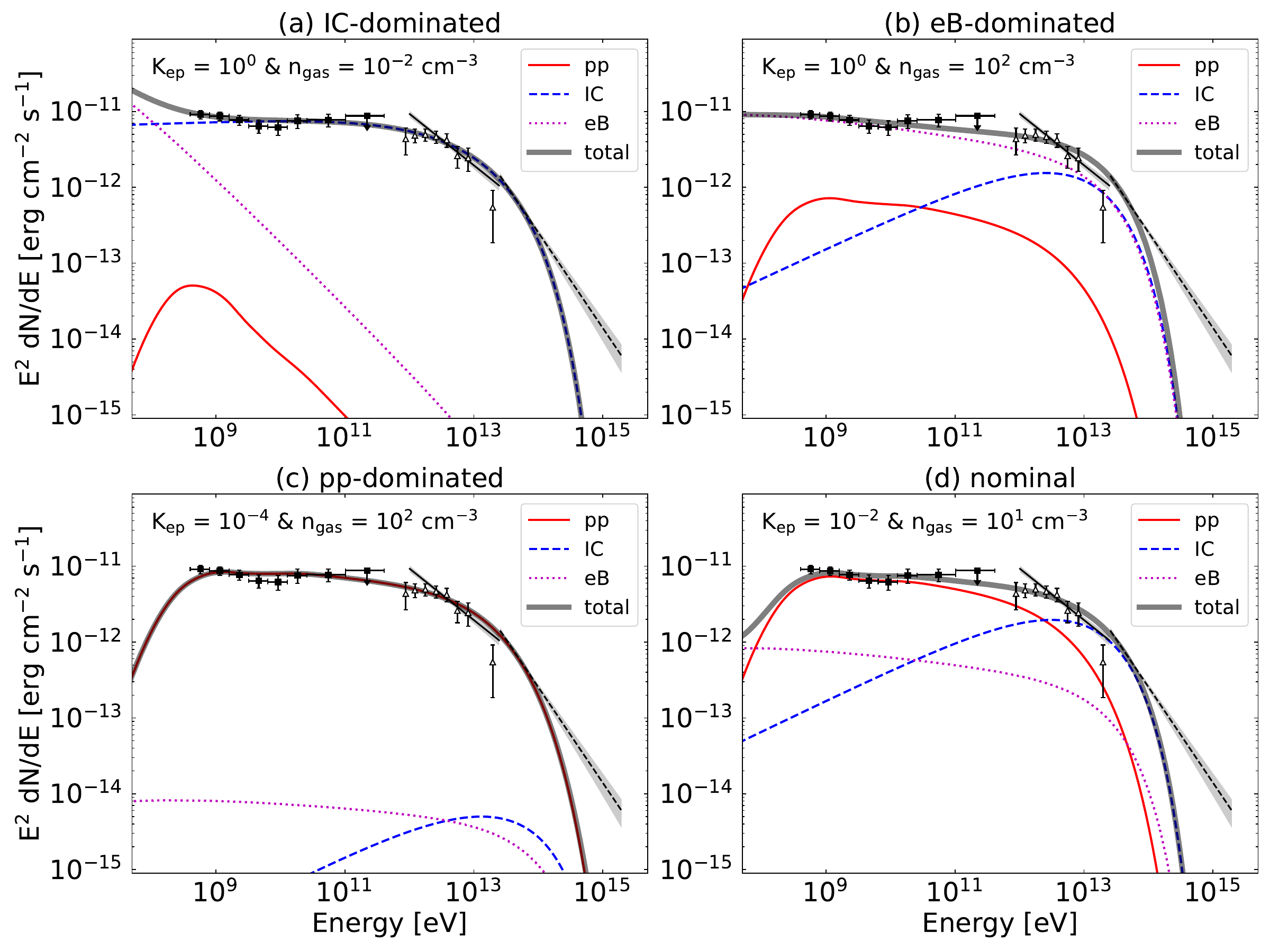}
\caption{
SED fit results for four representative $K_{\mathrm{ep}}$-$n_{\mathrm{gas}}$ parameter sets assuming no electron cooling.
(a) IC-dominated case with ($K_{\mathrm{ep}}$, $n_{\mathrm{gas}}$) = ($10^{0}$, $10^{-2}$~$\mathrm{cm^{-3}}$),
(b) eB-dominated case with ($K_{\mathrm{ep}}$, $n_{\mathrm{gas}}$) = ($10^{0}$, $10^{2}$~$\mathrm{cm^{-3}}$),
(c) pp-dominated case with ($K_{\mathrm{ep}}$, $n_{\mathrm{gas}}$) = ($10^{-4}$, $10^{2}$~$\mathrm{cm^{-3}}$), and
(d) nominal case with ($K_{\mathrm{ep}}$, $n_{\mathrm{gas}}$) = ($10^{-2}$, $10^{1}$~$\mathrm{cm^{-3}}$).
In each panel, the filled squares and open triangles represent the spectrum in the \textit{Fermi} band (this work) and the H.E.S.S. band \citep{HESS2018}, respectively.
The line segments show the reported spectrum by LHAASO (WCDA:1--25~TeV; KM2A:25--200~TeV). 
Although we account for the HAWC spectrum in the SED fitting, we do not show it (very close to the WCDA spectrum) for clarity.
The red solid, blue dashed, magenta dotted, and thick black solid lines represent the spectra of pp, IC, eB, and total $\gamma$-ray emission, respectively.
}
\label{fig:no_cooling:spec}
\end{figure*}

We first considered the case where $E_\mathrm{cut, p} = E_\mathrm{cut, e}$,
and evaluated the CR parameters by fitting the $\gamma$-ray data.
A summary of the parameter scan is provided in Appendix~\ref{sec:appendix:naima}.
We identify the sets of $n_\mathrm{gas}$ and $K_\mathrm{ep}$ that lead to IC-, eB-, and pp-dominated regimes;
the dominant emission process changes from IC to eB and finally to pp as
$n_\mathrm{gas}/K_\mathrm{ep}$ increases.
In Figure~\ref{fig:no_cooling:spec}, we present four representative cases (IC-, eB-, pp-dominated, and nominal cases) and list the model parameters in Table~\ref{tab:naima}.
Here, the nominal case is $K_\mathrm{ep}=0.01$ as measured at Earth \citep[e.g.,][]{Cummings2016ApJ} and $n_\mathrm{gas} = 10~\mathrm{cm^{-2}}$ as estimated by the radio data.
This model reproduces the data in the \textit{Fermi} and LHAASO KM2A bands by the pp and IC components, respectively.

All of these cases can reproduce the $\gamma$-ray data, but the IC-dominated case raises concerns because it requires a substantial CR energy budget, $W_\mathrm{e} = 7.9\times 10^{49}~($D/1.7$~\mathrm{kpc})^{2} \sim 60 \times 10^{49}~(D/4.5~\mathrm{kpc})^{2}$ in erg. The most promising source of CR electrons in this region is a PWN
driven by PSR~J1913+1011 located at $D \sim 4.5$~kpc.
%%Although such a PWN has not been detected even the radio to X-ray bands, 
Indeed, this pulsar has a high spin-down luminosity of $L_\mathrm{sd} = 2.9 \times 10^{36}~\mathrm{erg~s^{-1}}$ \citep{Morris2002MNRAS}, 
and a faint diffuse emission 
around the pulsar was detected at 6~GHz by \citet{Duvidovich2023}, suggesting that the PWN (or TeV halo) scenario is worth considering. However, a
conservative upper limit on the CR energy with the characteristic age of $t_\mathrm{sd} = 1.7 \times 10^{5}$~yr is $L_\mathrm{sd} \times t_\mathrm{sd} \sim 1.5 \times 10^{49}$~erg, which is much lower than the model requirements.
Even if we adopt a distance of 1~kpc, a more energetic (or older) pulsar than PSR~J1913+1011 is required.

As an alternative scenario, an SNR can be a source of CRs. Although no SNR has been identified in this region\footnote{
\citet{Kassim1988ApJ} and \citet{Gorham1990ApJ} reported an SNR candidate, G44.6+0.1, based on the Clark Lake survey. However, this source has not been confirmed in recent observations by \citet{Reich2019}.}, a typical SNR, if present, could supply up to $\lesssim 10^{50}$~erg ($E_\mathrm{SN}/10^{51}$~erg) ($\eta/0.1$), where $E_\mathrm{SN}$ is the explosion energy of a supernova, and $\eta$ is the acceleration efficiency.
The CR energy for the eB-dominated case scales as
$W_\mathrm{e} = 1.0\times 10^{47}~(n_\mathrm{gas}/100~\mathrm{cm^{-3}})^{-1}($D/1.7$~\mathrm{kpc})^{2} \sim 1 \times 10^{48}~(n_\mathrm{gas}/10~\mathrm{cm^{-3}})^{-1}(D/1.7~\mathrm{kpc})^{2}$ in erg.
For the pp-dominated case, the required energy scales as
$W_\mathrm{p} = 1.3\times 10^{48}~(n_\mathrm{gas}/100~\mathrm{cm^{-3}})^{-1}($D/1.7$~\mathrm{kpc})^{2} \sim 1 \times 10^{49}~(n_\mathrm{gas}/10~\mathrm{cm^{-3}})^{-1}(D/1.7~\mathrm{kpc})^{2}$ in erg.
An SNR either at $D=1.7$~kpc or 4.6~kpc, as proposed by \citet{Sano2018} and \citet{Su2017}, respectively, can account for the observed $\gamma$-ray spectrum in terms of the energetics.

A caveat of the leptonic scenario (IC- or eB-dominated cases) is that radiative cooling of high-energy electrons poses a challenge for the required cutoff energy.
Suppose that a PWN powered by the PSR~J1913+1011 produce $\gamma$ rays via IC scattering (set the energetics aside for now).
If we adopt the characteristic age ($1.7 \times 10^{5}$~yr) and a magnetic field strength of 6~$\mathrm{\mu G}$ comparable to the interstellar field, the maximum electron energy limited by synchrotron losses is estimated to be $E_\mathrm{max, e} \sim 2.1
~\mathrm{TeV} \mathrm{(}B/6~\mathrm{\mu G)^{-2}(}t_\mathrm{age}/170~\mathrm{kyr)}^{-1}$, much smaller than the cutoff energy required by the model fitting ($E_\mathrm{cut} \sim 100$~TeV; see Table~\ref{tab:naima}).
Alternatively, $\gamma$ rays may be generated by an SNR via electron bremsstrahlung,
and the SNR age can be estimated from the size of the remnant \citep[e.g.,][]{Suzuki2022, Ranasinghe2023ApJS}.
Adopting the angular diameter of the shell-like structure observed in the TeV band by the H.E.S.S. telescope ($R_{\mathrm{out}}$ = 0.$^{\circ}$49; taken from Table~3 of \citet{HESS2018}) and
$D=1.7~\mathrm{kpc}$, 
the diameter of the putative SNR is ${\sim}$29~pc (D/1.7~kpc).
Based on the empirical relation \citep[Eq.~2 in][]{Ranasinghe2023ApJS}, 
the SNR age will be ${\sim}$13~kyr (D/1.7~kpc)$^{2.3}$.
Assuming this age, the maximum electron energy limited by synchrotron losses is 27~TeV, again smaller than the model requirements ($E_\mathrm{cut} \sim 50~\mathrm{TeV}$). Adopting the larger distance $D=4.6$~kpc,
as proposed by \citet{Su2017}, worsens the situation. \footnote{
One may argue that the uncertainty of the age-size relation
(see Figure~5 in \citet{Ranasinghe2023ApJS})
may allow the age as young as 7~kyr for $D=1.7~\mathrm{kpc}$, which is compatible with the required
cutoff energy (50~TeV). Even so, the leptonic scenario is challenged by X-ray and radio upper limits, as discussed in Section~4.4.2.
}

To assess the plausibility of the proton cutoff energy, the diffusion timescale is more important than cooling effects.
According to \citet{Gabici2007}, the diffusion timescale can be written as $\tau_\mathrm{diff} = 11~\mathrm{yr}~\chi^{-1}~(R/14.5~\mathrm{pc})^{2}(E/200~\mathrm{TeV})^{-0.5}(B/6~\mathrm{\mu G})^{0.5}$, where $\chi$ is a parameter that represents the suppression of CR diffusion.
In vicinities of CR sources such as SNRs and PWNe, $\chi$ is theoretically expected to decrease down to $10^{-4}$ \citep{DAngelo2018MNRAS}, and observationally found to be $\lesssim 10^{-2}$ \citep[e.g.,][]{Aharonian2008A&A_W28, Oka2025ApJ, HAWC2017Sci}.
If we assume $\chi = 10^{-3}$, $\tau_\mathrm{diff} \sim 11$~kyr, which is compatible with the SNR age estimated above.

In short, while we can reproduce $\gamma$-ray data with various spectral models,
the IC-dominated case is disfavored in terms of the energetics, and the eB-dominated case is disfavored
in terms of the electron cooling. 

\begin{deluxetable*}{llccccc}
\tablecaption{
Parameters obtained from SED fitting for four representative cases for $D=1.7~\mathrm{kpc}$.
The upper rows show the results obtained with no electron cooling ($E_\mathrm{cut, e} = E_\mathrm{cut, p}$), while the lower rows show those obtained by considering electron cooling with $E_\mathrm{cut, e}$ fixed at 30~TeV.
The errors represent the statistical $1\sigma$ uncertainties.
\label{tab:naima}}
\tablewidth{0pt}
\tablehead{
\colhead{} & \colhead{Case} & \colhead{$K_\mathrm{ep}$} & \colhead{$n_\mathrm{gas}$~[cm$^{-3}$]} & \colhead{$W_\mathrm{p,>1\,GeV}$~[erg]} & \colhead{$\alpha$} & \colhead{$E_\mathrm{cut}$~[TeV]}
}
\startdata
\multirow{2}{*}[-1.5em]{\shortstack{no electron cooling \\ $E_\mathrm{cut, e}$ = $E_\mathrm{cut, p}$}}
& IC      & $10^{0}$  & $10^{-2}$ & ($7.9^{+0.7}_{-1.1}$) $\times 10^{49}$ & $2.94^{+0.02}_{-0.02}$ & $105^{+14}_{-14}$ \\
& eB      & $10^{0}$  & $10^{2}$  & ($1.0^{+0.1}_{-0.1}$) $\times 10^{47}$ & $2.20^{+0.02}_{-0.02}$ & $51^{+3}_{-3}$ \\
& pp      & $10^{-4}$ & $10^{2}$  & ($1.3^{+0.1}_{-0.1}$) $\times 10^{48}$ & $2.14^{+0.02}_{-0.02}$ & $205^{+30}_{-18}$ \\
& nominal & $10^{-2}$ & $10^{1}$  & ($1.1^{+0.1}_{-0.1}$) $\times 10^{49}$ & $2.18^{+0.02}_{-0.02}$ & $58^{+4}_{-4}$  \\
\hline
\multirow{2}{*}{\shortstack{electron cooling \\ $E_\mathrm{cut, e}$ = 30~TeV}}
& pp      & $10^{-4}$ & $10^{2}$  & ($1.3^{+0.1}_{-0.1}$) $\times 10^{48}$ & $2.14^{+0.02}_{-0.02}$ & $200^{+17}_{-17}$ \\
& nominal & $10^{-2}$ & $10^{1}$  & ($1.1^{+0.1}_{-0.1}$) $\times 10^{49}$ & $2.20^{+0.02}_{-0.02}$ & $320^{+50}_{-50}$ \\
\enddata
\tablecomments{
For the leptonic scenario (IC- and eB-dominated cases),
$W_\mathrm{e,>1\,GeV}$ is equal to $W_\mathrm{p,>1\,GeV}$.
}
\end{deluxetable*}

%%\clearpage

\subsection{Considering Cooling of CR Electrons} \label{sec:model:cooling_case}

\begin{figure*}[htbp]
\plotone{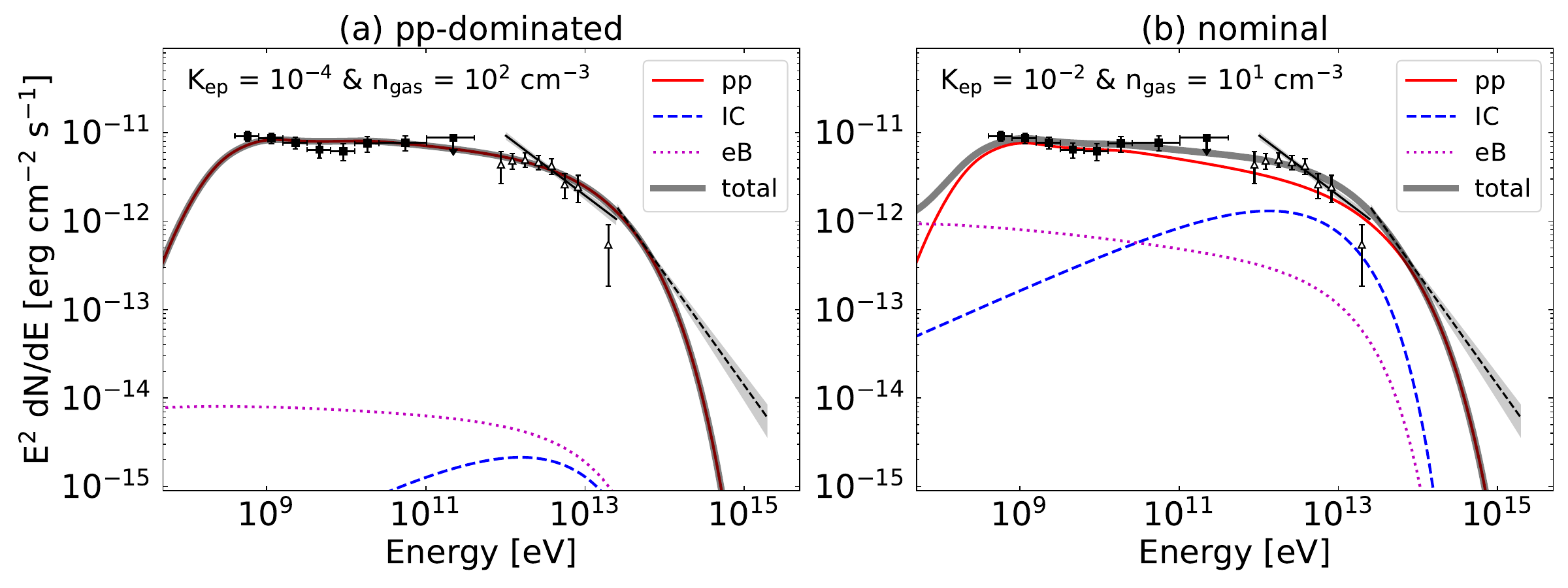}
\caption{
SED fit results for two representative $K_{\mathrm{ep}}$-$n_{\mathrm{gas}}$ parameter sets assuming electron cooling with $E_\mathrm{cut, e}$ fixed to 30~TeV.
(a) pp-dominated case with ($K_{\mathrm{ep}}$, $n_{\mathrm{gas}}$) = ($10^{-4}$, $10^{2}$~$\mathrm{cm^{-3}}$), and
(b) nominal case with ($K_{\mathrm{ep}}$, $n_{\mathrm{gas}}$) = ($10^{-2}$, $10^{1}$~$\mathrm{cm^{-3}}$).
See the caption of Figure~\ref{fig:no_cooling:spec} for the meanings of lines and markers.
}
\label{fig:cooling:spec}
\end{figure*}

To resolve the inconsistency in the electron cutoff energy, %discussed in Section~\ref{sec:model:no_cooling_case}, 
we consider cases with electron cooling.
We set $E_\mathrm{cut,e} = 30~\mathrm{TeV}$, a nominal cutoff energy for a putative SNR at $D=1.7~\mathrm{kpc}$,
and reassessed the CR parameters through the SED fitting.
We found that the leptonic scenario (IC- and eB-dominated cases) inevitably fail to fit the LHAASO data of 50~TeV, whereas the pp-dominated and nominal cases can reproduce the data across the entire energy range, as shown in Figure~\ref{fig:cooling:spec}.
The parameters are also tabulated in Table~\ref{tab:naima}.
The result for the pp-dominated case is nearly the same as that obtained without electron cooling.
In the nominal case, the proton cutoff energy increases by approximately a factor of five and the neutral pion decay $\gamma$-ray emission becomes dominant throughout the GeV and TeV band, making the source a CR proton PeVatron. Inferred spectral index is {$\sim$}2.2 and integrated energy above 1~GeV scales as
$W_\mathrm{p} = 1.1\times 10^{49}~(n_\mathrm{gas}/10~\mathrm{cm^{-3}})^{-1}($D/1.7$~\mathrm{kpc})^{2}$ in erg.
This gives $W_\mathrm{p}$ of ${\sim} 1 \times 10^{49}~\mathrm{erg}$ and 
${\sim} 5 \times 10^{49}~\mathrm{erg}$ for the Sano et al. (2018) $\Np$ map and Su et al. (2017) $\Np$ map, respectively. 

\begin{figure}[htbp]
\plotone{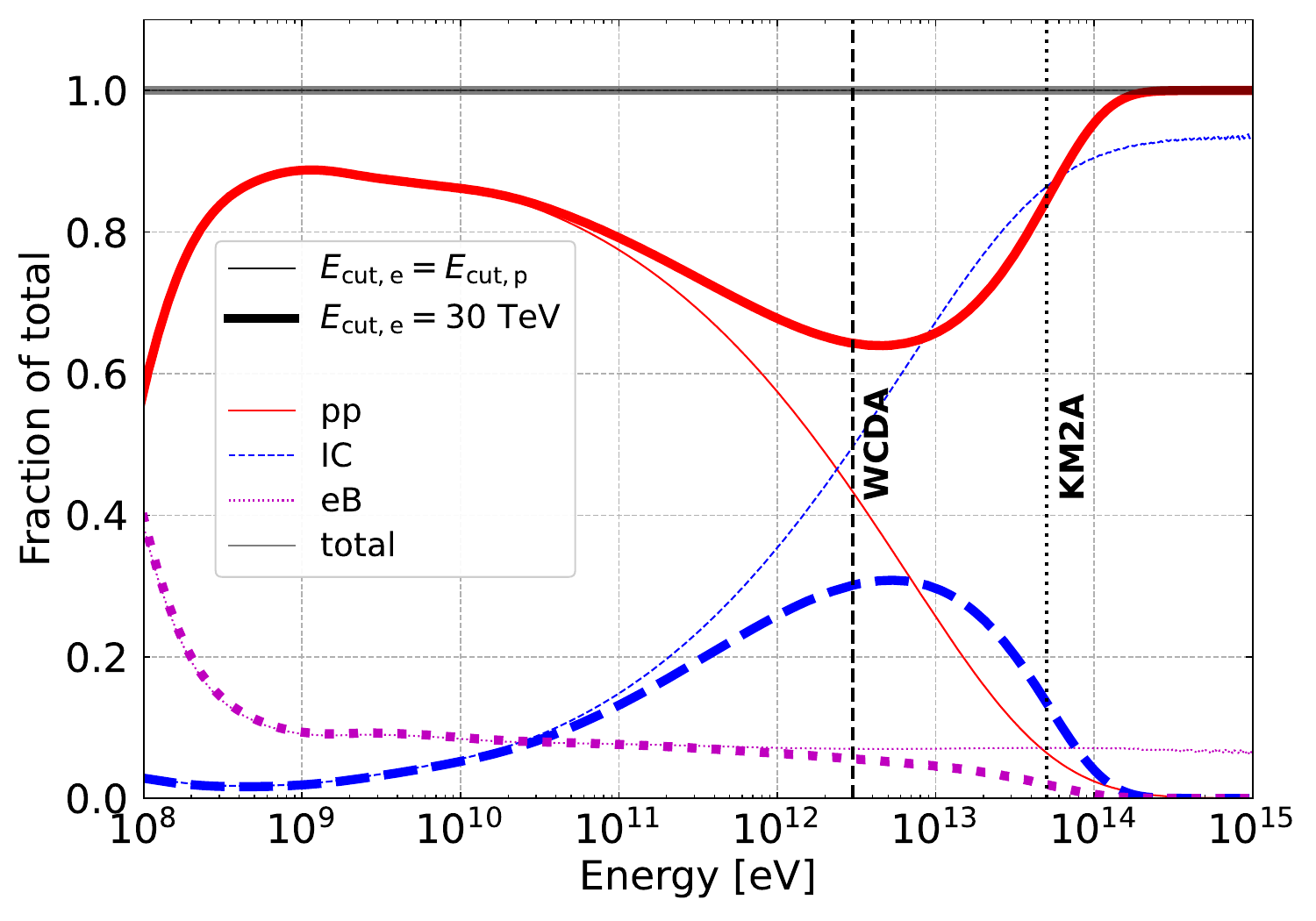}
\caption{
Relative contribution by each emission component (IC, eB, and pp) in the nominal case.
Thin lines and thick lines show the results with $E_\mathrm{cut, e} = E_\mathrm{cut, p}$ and $E_\mathrm{cut, e} = 30$~TeV, respectively.
The dashed and dotted vertical lines indicate the normalization energies of the fitted functions for WCDA and KM2A, as reported in \citet{LHAASO_1stCat}.
}
\label{fig:naima:fraction}
\end{figure}

Taking into account the electron cooling also affects the energy dependence of the relative contribution of each emission component (IC, eB, and pp) in the nominal case (Figure~\ref{fig:naima:fraction}).
With $E_\mathrm{cut, e}=30$~TeV, the IC-to-pp flux ratio is largest between 1 and 10~TeV, which coincides with the energy range observed by H.E.S.S. and LHAASO WCDA.
Figure~\ref{fig:naima:fraction} implies that while the $\gamma$-ray morphology traces the dense gas distribution in GeV and $>$50~TeV energies, it may be affected by the increasing contribution from IC emission in 1--10~TeV.
This energy dependence could account for the lower TS values with the H.E.S.S. intensity map and the WCDA Gaussian template (Table~\ref{tab:fermi:TS}).

%%\clearpage

\subsection{Constraints from X-ray and Radio Data}
Although the hadronic scenario is preferred for energetics and cooling, the leptonic scenario remains viable
if a putative pulsar is more energetic (or older), or a putative SNR is younger.
If part of the $\gamma$-ray emission originates from the accelerated electrons, synchrotron emission is expected from the radio to X-ray band. A stringent radio upper limit for the counterpart of HESS~J1912+101 was reported from radio polarization observation by \citet{Reich2019}. In the following,
we first search for the X-ray counterpart (Section~4.4.1) and then discuss constraints 
on the leptonic scenario from X-ray and radio data (Section~4.4.2).

\subsubsection{Search for X-ray Counterpart}

%With our baseline model, a significant excess in {$\ge$}12.8~GeV
%appeared at ($l$, $b$) $\sim$ ($44\fdg5$, $0\fdg0$), close to the centers 
%of HESS~J1912+102 and LHAASO WCDA Gaussian and the position of the PSR~J1913+1011. The excess %remains with whatever models we used for the LHAASO/H.E.S.S. counterpart.
%To examine this excess, we modeled it with a uniform disk in our baseline model;
%we used Fermipy ({\tt extension} method) and obtained a position of
%($l$, $b$)=($44\fdg50$, $0\fdg03$) and a radius of $0\fdg12$ (Figure~1(e)).
%The source spectrum is hard with a photon index of $1.76\pm0.05$,
%and the flux is about 20\% of the total LHAASO/H.E.S.S. counterpart in 10--100~GeV.

\begin{figure*}[htbp]
\plotone{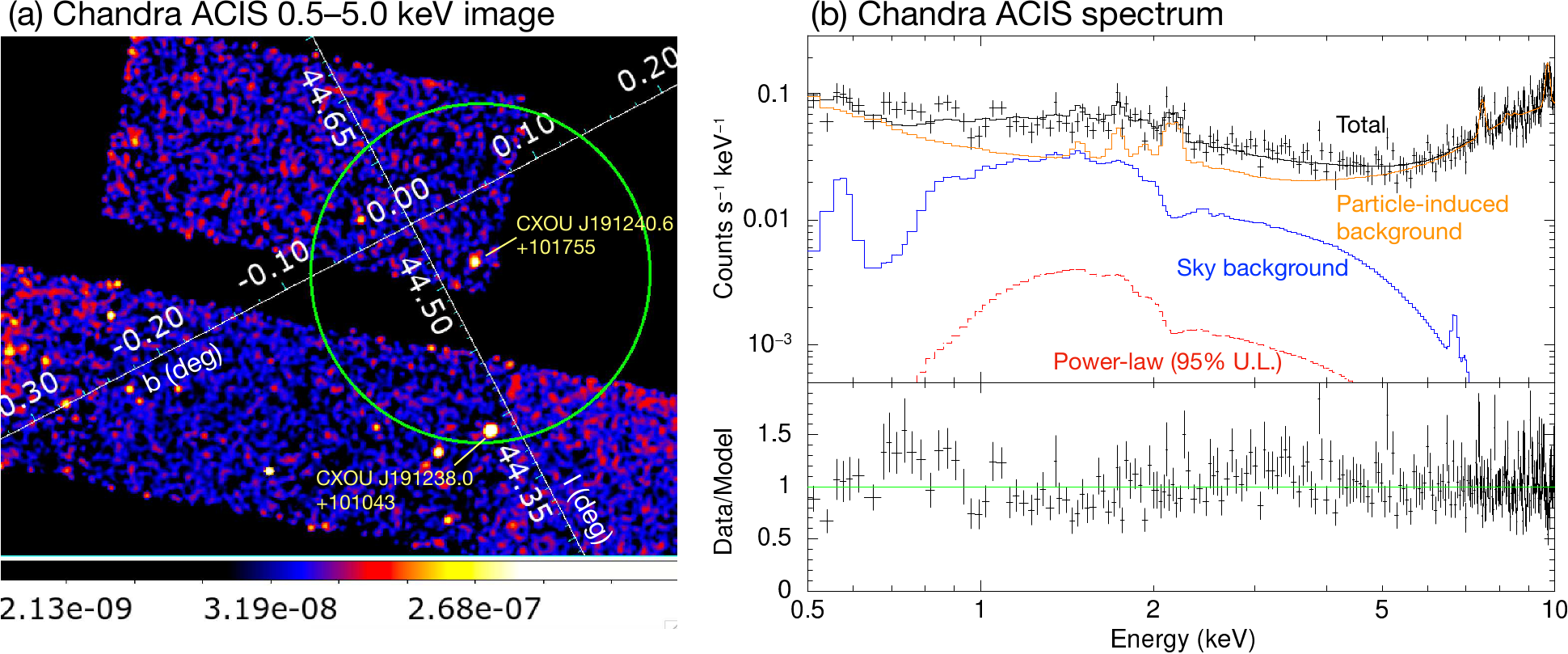}
\caption{(a) A 0.5--5.0~keV Chandra image and (b) energy spectrum with the best-fit spectral model. The spectrum was extracted from the green circle in panel (a). The blue, orange, and black lines indicate the sky background, particle-induced background, and total model, respectively. The upper limit for the power-law component, used as the counterpart to the $\gamma$-ray excess, is shown by the red dashed line.
%%likes below the range of the vertical axis.
\label{fig:xray}}
\end{figure*}

As discussed in Section~3.2, we identified a small peak-like structure at ($l$, $b$){$\sim$}($44\fdg50$, $0\fdg03$) with a radius of ${\sim}0\fdg12$ (Figure~1(e)),
located near the center of the overall excess.
Assuming that this feature originates from an unknown PWN via IC scattering,
%or an enhanced population of CR electrons accelerated by an unknown SNR (via IC scattering or electron bremsstrahlung), 
we searched for an X-ray counterpart using archival Chandra data.
We used the Chandra data ObsID 3854, which partially covers the vicinity of the position ($l$, $b$)=($44\fdg50$, $0\fdg03$). Here, we consider a $0\fdg12$-radius circle that is partially covered by the ACIS-I2, S1, and S2 chips.
The effective exposure, after reprocessing with {\tt chandra\_repro} in the CIAO package v4.16 \citep{fruscione06}, is {$\sim$}20~ks. We confirmed that there is no visible diffuse emission within the circular region relative to the background. 
(Figure~\ref{fig:xray} (a) )
We note that PSR~J1913+1011 was not detected in the Chandra data, as only an upper limit on the flux is reported by \cite{chang08}.
On the other hand, two sources CXOU~J191240.6+101755 at ($l$, $b$)=($44\fdg50566, 0\fdg02790$) and CXOU~J191238.0+101043 at ($l$, $b$)=($44\fdg39439, -00\fdg01816$) are located in this region \citep{chang08}.
We found that they are both significantly extended relative to the Chandra point spread function 
with the source extents of ${\sim}5\arcsec$
when modeled with a single Gaussian (Table~\ref{tab:xray_ps}), suggesting they may be unknown PWNe. 
We note that the radial distribution of CXOU~J191238.0+101043 is flatter than a single Gaussian function in radii larger than 5\farcs0.
We then extracted the Chandra spectra from a 
19\farcs0 radius circle centered on these sources.
We performed a simple phenomenological fit with a power-law model in the 1--5~keV band, where the background is nearly negligible. We used XSPEC version 12.14.1 \citep{arnaud96} for the spectroscopy. The derived flux in the 2--10~keV band and power-law index are ${\sim}1.5\times10^{-14}$~erg~s$^{-1}$~cm$^{-2}$ and ${\sim}2.6$ for CXOU~J191240.6+101755, 
and ${\sim}60\times10^{-14}$~erg~s$^{-1}$~cm$^{-2}$ and ${\sim}0.5$ for CXOU~J191238.0+101043, respectively (Table~\ref{tab:xray_ps}).
The observed low flux ({$\sim$}1/100 of the flux of hard excess in 10--100~GeV)
and soft X-ray spectrum of CXOU~J191240.6+101755 could be consistently explained by synchrotron (radio to X-ray) and IC ($\gamma$-ray) scenarios with small magnetic-field values and a synchrotron emission cutoff at {$\le$}1~keV.
Even so, the source's contribution to the entire $\gamma$-ray excess emission in the 10--100~GeV range is at the 20\% level.
Another source, CXOU~J191238.0+101043 is unrelated to the $\gamma$-ray excess, with a very hard X-ray spectrum that contradicts the spectral index ({$\sim$}2) of the GeV excess.
Accordingly, we will not further consider these sources as potential counterparts of the entire $\gamma$-ray excess.

\begin{table}[htb!]
    \centering
    \caption{Parameters of CXOU~J191240.6+101755 and CXOU~J191238.0+101043}
    \begin{tabular}{l l l l}
    \hline\hline
      Source & Power-law index & 2--10 keV flux & Gaussian sigma \\ %($''$)
      & & ($10^{-14}$~erg~s$^{-1}$~cm$^{-2}$) & (\arcsec)\\
      \hline
      CXOU~J191240.6+101755 & $2.6 \pm 0.4$ & $1.5 \pm 0.3$ & $6.0 \pm 1.2$ \\
      CXOU~J191238.0+101043 & $0.45 \pm 0.14$ & $60 \pm 8$ & $3.8 \pm 0.4$ \\
      \hline
    \end{tabular}
    \label{tab:xray_ps}
\end{table}

We then evaluated the maximum allowed non-thermal diffuse X-ray emission in this region by modeling the sky and particle-induced backgrounds, following the methodology used in recent work on faint diffuse sources (e.g., \citealt{suzuki20, suzuki25, kuboike25}). 
We extracted the X-ray spectrum from a $0\fdg12$-radius circle centered on ($l$, $b$)=($44\fdg50$, $0\fdg03$).
We used the {\tt mkacispback} tool v2023-09-23 \citep{suzuki21}, to generate the expected particle-induced background spectrum for this region. As for the sky background, we considered the Local hot bubble (foreground emission), the Milky Way halo, the Cosmic X-ray background, and the Galactic ridge X-ray emission. The first two were modeled using the collisional ionization equilibrium plasma model ({\tt apec} in XSPEC) with electron temperatures of 0.1~keV and 0.7~keV, respectively. The Cosmic X-ray background component was modeled as a power law with a slope of 1.42 and a surface brightness of $5.4\times10^{-15}$~erg~cm$^{-2}$~arcmin$^{-2}$. The Galactic ridge emission is modeled using an {\tt apec} model with an electron temperature of 6.6~keV and a fixed flux, calculated based on \cite{uchiyama13}. Following \cite{uchiyama13}, we assumed a $\pm20\%$ uncertainty for the flux.
We then modeled the presumable non-thermal X-ray emission associated with the $\gamma$-ray excess using a power-law model (slope fixed at 2.0). We applied the free Galactic absorption ({\tt tbabs} model in XSPEC) to the sky components other than the Local hot bubble.
We found that this spectral model reasonably explains the data in the 0.5--10.0~keV band, as shown in Figure~\ref{fig:xray} (b). The 95\% upper limit flux of the power-law model in the 2--10~keV band was derived to be $\sim 1\times10^{-13}$~erg~s$^{-1}$~cm$^{-2}$ for the entire analysis region (area = 84.5~arcmin$^2$), corresponding to a surface brightness of ${\sim}1\times10^{-15}$~erg~s$^{-1}$~cm$^{-2}$ arcmin$^{-2}$. The Galactic absorption column density was determined to be $\sim 1\times10^{22}$~cm$^{-2}$, which is reasonable considering the source location \citep{hi4pi16, kalberla05, dickey90}.
% We confirmed that the other free parameters of the sky background components are also reasonable.
Since this upper limit is mainly determined by the uncertainty in the
background (Galactic ridge X-ray emission), we can scale it to the entire area of the LHAASO/H.E.S.S. source.
If we assume the size of HESS~J1912+101 ($0\fdg49$ radius), the upper limit will be
${\sim}2 \times 10^{-12}~\mathrm{erg~s^{-1}~cm^{-2}}$ in the SED.

%%\clearpage
\subsubsection{Limits on the Leptonic Scenario}

Although we were unable to detect enhanced diffuse X-ray emission, we obtained a tight upper limit.
Together with the previously reported radio upper limit, this further constrains the leptonic scenario. 

\begin{figure}[htbp]
\plotone{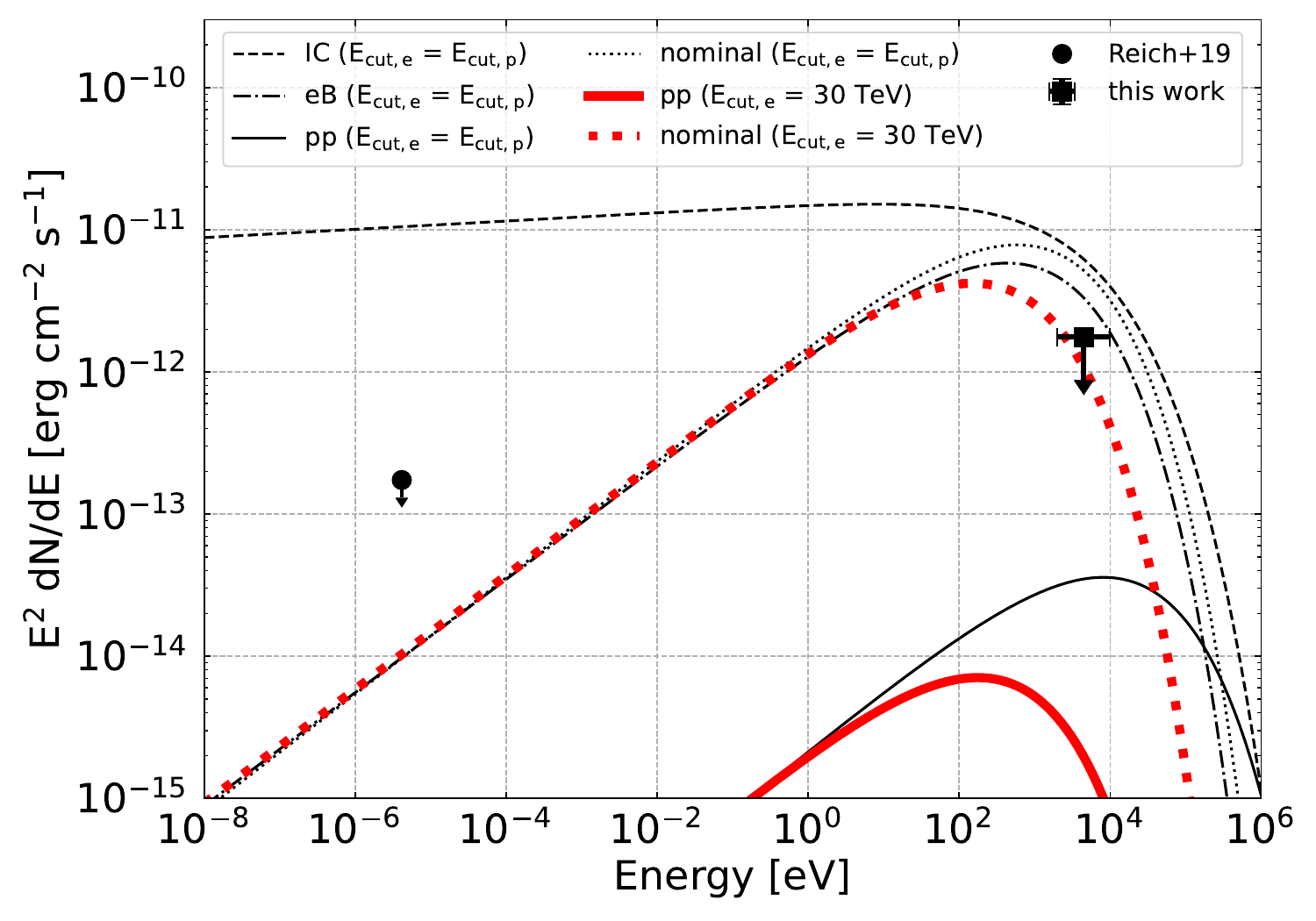}
\caption{
Synchrotron spectra of each representative case. 
The dashed, dot-dashed, solid, and dotted lines represent the IC-dominated, eB-dominated, pp-dominated, and nominal cases, respectively.
The thick red and thin black lines indicate the cases with $E_{\mathrm{cut,e}} = E_{\mathrm{cut,p}}$ and $E_{\mathrm{cut,e}} = 30$~TeV, respectively.
The circle shows the upper limit taken from \citet{Reich2019}, while the square shows the upper limit derived in this work.
\label{fig:naima:synchrotron}}
\end{figure}

We calculated synchrotron spectra for each representative case.
Figure~\ref{fig:naima:synchrotron} shows a comparison between the synchrotron spectra and the upper limits in the X-ray and radio bands, the latter of which is taken from \citet{Reich2019}.
If we assume a magnetic-field strength of 6~$\mathrm{\mu G}$, typical of that in the cold atomic or molecular clouds with
$n_\mathrm{gas} \le 300~\mathrm{cm^{-3}}$ \citep{Heiles2005ApJ, Crutcher2012ARA&A}, the IC-, eB-dominated cases and the nominal case with $E_{\mathrm{cut,e}} = E_{\mathrm{cut,p}}$ are incompatible with the X-ray upper limit.
The IC-dominated case is also inconsistent with the radio upper limit, requiring an unrealistically low magnetic field below 1~$\mu$G.
These constraints effectively exclude the leptonic scenario in a typical magnetic field environment, reinforcing the proton PeVatron scenario for the LHAASO/H.E.S.S. source.

It should be noted that the magnetic field strength can be lower than the typical interstellar value, particularly in low-density environments \citep[e.g.,][]{Seta2025}. A PWN powered by PSR~J1849-0001 is an example \citep{LHAASOJ1848-0001}. If the magnetic field strength is as low as 3~$\mathrm{\mu G}$, all representative cases are consistent with the X-ray upper limit, making the eB-dominated case plausible. 
However, the high gas density implicitly assumed in the eB-dominated case 
is incompatible with the required low magnetic-field value.
In the IC-dominated case, the compatibility with radio and X-ray upper limits is achieved
if the magnetic field is extremely low ({$\le$}1~$\mathrm{\mu G}$). Alternatively, 
a moderately low magnetic field ({$\sim$}3~$\mathrm{\mu G}$) combined with
a low-energy break in synchrotron emission at ${\sim}10^{-3}~\mathrm{eV}$, as observed in middle-aged PWN systems
\citep[e.g.,][]{Temim2013,Temim2015,Gong2025}, also satisfies these constraints.
We remind that although the IC-dominated case aligns with radio and X-ray data under these configurations,
it remains severely challenged by energetic considerations (see Section~4.2).

In summary, while the leptonic scenario cannot be formally excluded,
the hadronic scenario offers a more natural explanation for the observed data in the 
radio and X-ray bands as well as GeV and TeV $\gamma$-ray bands.

%%\clearpage

\section{Summary and Prospects}
LHAASO has reported more than 40 sub-PeV sources, which are promising candidates for either proton or electron PeVatron.
Multi-wavelength observations are crucial for identifying the dominant particle species and evaluating the CR energy content of these objects.
In this study, we investigated the LHAASO~J1912+1014u (and HESS~J1912+101) region using \textit{Fermi}-LAT GeV $\gamma$-ray data and FUGIN CO data to assess the proton PeVatron scenario.

We analyzed 15 years of \textit{Fermi}-LAT data in the 0.4--409.6~GeV range. By improving the
standard \textit{Fermi}-LAT diffuse emission model, we substantially reduced the large residuals that appear when using the standard diffuse model.
This allowed us to reveal a significant excess 
above the diffuse background and nearby source models,
likely representing {$\ge$}10~GeV emission associated with the LHAASO/H.E.S.S. source.
The GeV emission exhibits a hard spectrum (index of {$\sim$}2.1) and is well modeled using ISM gas templates
with velocities of about 25~$\mathrm{km~s^{-1}}$ or 60~$\mathrm{km~s^{-1}}$.

Using the GeV spectrum together with the reported TeV measurements, we have phenomenologically evaluated the CR spectra responsible for the LHAASO/H.E.S.S. source.
By exploring combinations of $K_{\mathrm{ep}}$ and $n_{\mathrm{gas}}$, we have identified parameter sets in which the IC, eB, and pp emission components become dominant, respectively.
In the observationally motivated nominal case ($K_{\mathrm{ep}} = 1\%$, $n_{\mathrm{gas}} = 10~\mathrm{cm^{-3}}$), the pp and IC components contribute at comparable levels.

Considerations of the CR energy budget disfavor the IC-dominated case.
%%From the perspective of the CR energy budget, the IC-dominated case is unlikely.
Although the eB-dominated case is energetically viable,
%%acceptable in terms of energetics,
the electron cutoff energy required to reproduce the GeV--TeV $\gamma$-ray spectrum
exceeds the maximum energy expected from radiative cooling for the age of a putative SNR.
When adopting a plausible electron cutoff energy of $E_{\mathrm{cut,e}} = 30$~TeV, 
the leptonic scenarios (IC- and eB-dominated cases) become inconsistent with the data.
In the nominal case, the IC contribution is substantial only in the 1--10~TeV energy range, indicating that the source is predominantly powered by hadronic processes.
The inferred CR proton spectral index is {$\sim$}2.2, and the total CR proton energy above 1~GeV is (1--5)~$\times 10^{49}~\mathrm{erg}$, depending on the assumed distance of the source.

We further found that, under a typical interstellar magnetic field of $6~\mathrm{\mu G}$,
the synchrotron emission predicted in the IC-dominated, eB-dominated, and nominal no-cooling cases exceeds the upper limit on diffuse X-ray emission.
This provides additional support for the proton PeVatron scenario.

In conclusion, the nominal case including electron cooling offers the most self-consistent explanation of both the multi-wavelength spectral data and the GeV--TeV spatial morphology, strongly favoring the proton PeVatron scenario. As demonstrated in this work,
detailed comparisons between ISM gas distributions and \textit{Fermi}-LAT {$\gamma$}-ray data are crucial for identifying the
particle species and constraining the CR properties of sub-PeV sources. 
Radio and X-ray observations, whether detections or upper limits, also provide valuable constraints.
We note that our estimate of the electron cutoff energy is limitated by the currently available data,
and that the LHAASO KM2A spectrum, reported as a simple power-law model,
was incorporated approximately in our SED modeling. Deeper observations from radio to TeV $\gamma$-ray will be essential for fully characterizing this object and advancing our understanding of the origin of Galactic CRs up to the knee energy.

%% Please use the acknowledgment and contribution environments. This will 
%% be anonomyized when the "anonymous" style option is used. 
\begin{acknowledgments}
The \textit{Fermi} LAT Collaboration acknowledges generous ongoing support
from a number of agencies and institutes that have supported both the
development and the operation of the LAT as well as scientific data analysis.
These include the National Aeronautics and Space Administration and the
Department of Energy in the United States, the Commissariat \`a l'Energie Atomique
and the Centre National de la Recherche Scientifique / Institut National de Physique
Nucl\'eaire et de Physique des Particules in France, the Agenzia Spaziale Italiana
and the Istituto Nazionale di Fisica Nucleare in Italy, the Ministry of Education,
Culture, Sports, Science and Technology (MEXT), High Energy Accelerator Research
Organization (KEK) and Japan Aerospace Exploration Agency (JAXA) in Japan, and
the K.~A.~Wallenberg Foundation, the Swedish Research Council and the
Swedish National Space Board in Sweden.

Additional support for science analysis during the operations phase is gratefully 
acknowledged from the Istituto Nazionale di Astrofisica in Italy and the Centre 
National d'\'Etudes Spatiales in France. This work performed in part under DOE 
Contract DE-AC02-76SF00515.

This work was also supported in part by a University Research Support Grant from the National Astronomical Observatory of Japan (NAOJ).

Part of this work was supported by JSPS KAKENHI Grant Numbers 23K25882 and 23H04895 (T. Mizuno), 22H00152 and 24H00246 (H. Sano), 
and 24K17093 (H. Suzuki).

This paper employs a list of Chandra datasets, obtained by the Chandra X-ray Observatory, contained in~\dataset[DOI: 10.25574/cdc.578]{https://doi.org/10.25574/cdc.578}.

\end{acknowledgments}

\begin{contribution}
%%This section gives authors the space to recognize author contributions. The text inside this environment is NOT counted towards the total word quanta. At a minimum, manuscripts are expected to include this text:

T. Mizuno and N. Nakahara analyzed \textit{Fermi}-LAT $\gamma$-ray data and investigated the spectrum and morphology of the source. 
H. Sano and T. Murase reprocessed FUGIN CO data and prepared target-gas maps, including contributions from optically thick {\HI}.
H. Suzuki analyzed Chandra data and constrained the diffuse X-ray emission from the source by accounting for astrophysical and instrumental backgrounds.
T. Oka carried out SED modeling by performing a comprehensive fit to the multiwavelength data.
All coauthors participated in interpreting the results and providing feedback to the manuscript.
%%All authors contributed equally to the Terra Mater collaboration.

%% But authors are expected to provide more specific details, e.g. 
%%
%%SC was responsible for writing and submitting the manuscript.
%%WWM came up with the initial research concept and edited the manuscript.
%%OTS obtained the funding and edited the manuscript.
%%EBF provided the formal analysis and validation. He also edited the manuscript.
%%GEH Supervised the undergraduates, wrote the software and administers the project github and Zenodo repositories.
%%
%% Authors can use the Contributor Role Taxonomy (CRediT) at
%% for ideas on how write a good statement tailored to their needs.

\end{contribution}


\facilities{\textit{Fermi}-LAT, CXO, Nobeyama 45-m radio telescope (NRO45), The Very Large Array (VLA)}

\software{%astropy \citep{2013A&A...558A..33A,2018AJ....156..123A,2022ApJ...935..167A},  
          %Cloudy \citep{2013RMxAA..49..137F}, 
          %Source Extractor \citep{1996A&AS..117..393B},
          Fermitools \citep{Fermitools},
          Fermipy \citep{Wood2017},
          Naima \citep{NAIMA},
          astropy \citep{astropy_2013,astropy_2018,astropy_2022},
          numpy \citep{harris_2020},
          Scipy \citep{Virtanen2020SciPy-NMeth}.
          XSPEC \citep{arnaud96},
          CIAO \citep{fruscione06},
          mkacispback \citep{suzuki21}
          }


%%\clearpage

\appendix

\section{Summary of \textit{Fermi}-LAT Spectrum}

We give the numerical values of the SEDs in the LAT band (Table~4).
We also show the SED plot and TS map for the Su et al. (2017) $\Np$ map and 
LHAASO KM2A Gaussian (Figure~9).

\begin{deluxetable*}{cccc}[htbp]
%%\tablenum{1}
\tablecaption{Spectrum of the GeV excess toward the LHAASO/H.E.S.S. source}
\tablewidth{0pt}
\tablehead{
\colhead{Energy} & \colhead{KM2A Gaussian} & \colhead{Su et al. (2017) $\Np$ map} & \colhead{Sano et al. (2018) $\Np$ map} \\
\cline{2-4}
(GeV) & & ($\mathrm{10^{-9}~GeV~s^{-1}~cm^{-2}}$) & }
%%%\decimalcolnumbers
\startdata
0.4--0.8 & $6.13\pm0.77$ & $5.85 \pm 0.74$ & $5.72 \pm 0.73$ \\
0.8--1.6 & $6.13\pm0.78$ & $5.49 \pm 0.74$ & $5.42 \pm 0.72$ \\
1.6--3.2 & $5.54\pm0.81$ & $4.65 \pm 0.76$ & $4.81 \pm 0.73$ \\
3.2--6.4 & $4.09\pm0.88$ & $4.12 \pm 0.81$ & $4.01 \pm 0.78$ \\
6.4--12.8 & $4.77\pm0.97$& $3.67 \pm 0.87$ & $3.86 \pm 0.84$ \\
12.8--25.6 & $5.89\pm1.16$ & $4.66 \pm 1.01$ & $4.71 \pm 0.98$ \\
25.6--102.4 & $4.70\pm1.06$ & $4.99 \pm 0.94$ & $4.82 \pm 0.92$ \\
102.4--409.6 & $\le 5.73$ & $\le 5.62$ & $\le 5.47$ \\
\enddata
%%\tablecomments{This table ``hides'' the third column in the \latex\ when compiled.
%%The Distance is also centered on the decimals.  Note that when using decimal
%%alignment you need to include the {\tt\string\decimals} command before
%%{\tt\string\startdata} and all of the values in that column have to have a
%%space before the next ampersand.}
\end{deluxetable*}

\begin{figure}[htbp]
\begin{tabular}{cc}
\begin{minipage}{0.49\textwidth}
\centering
\begin{overpic}[width=\textwidth]{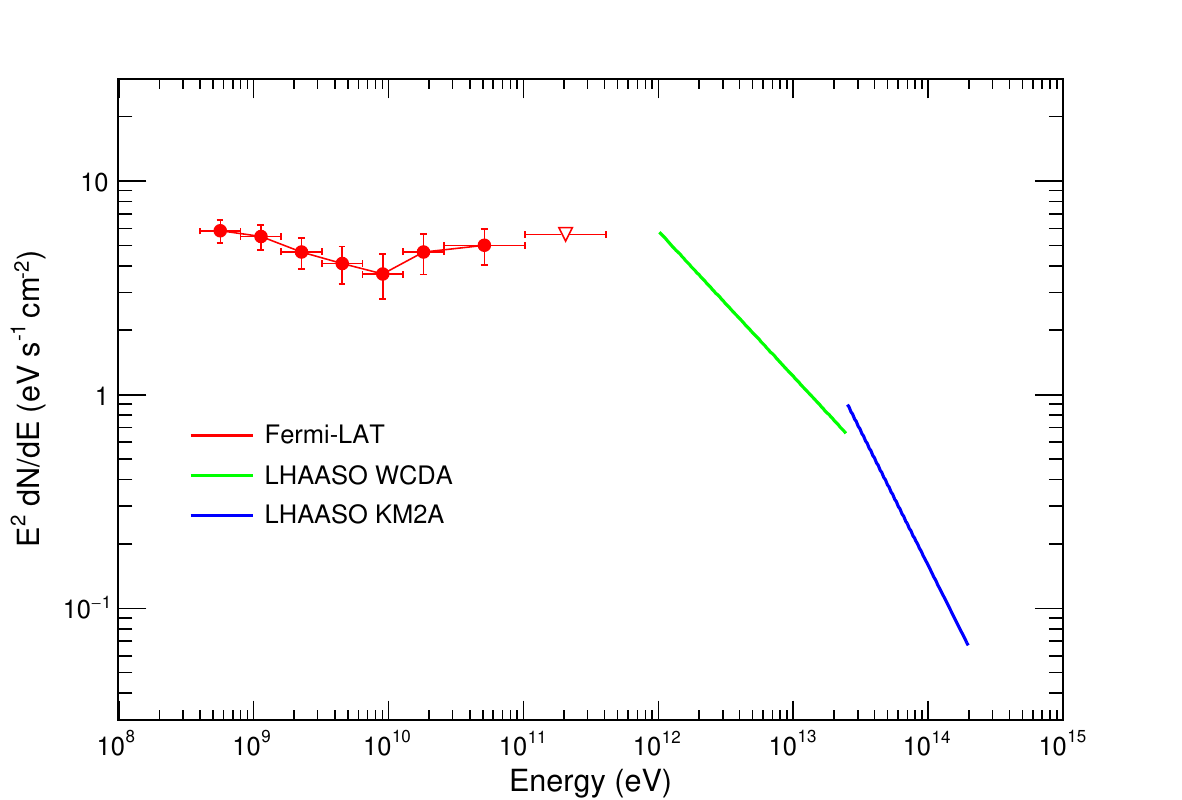}
\put(15,65){(a)}
\end{overpic}
\end{minipage}
%%\hspace{0.02\textwidth}
\begin{minipage}{0.49\textwidth}
\centering
\begin{overpic}[width=\textwidth]{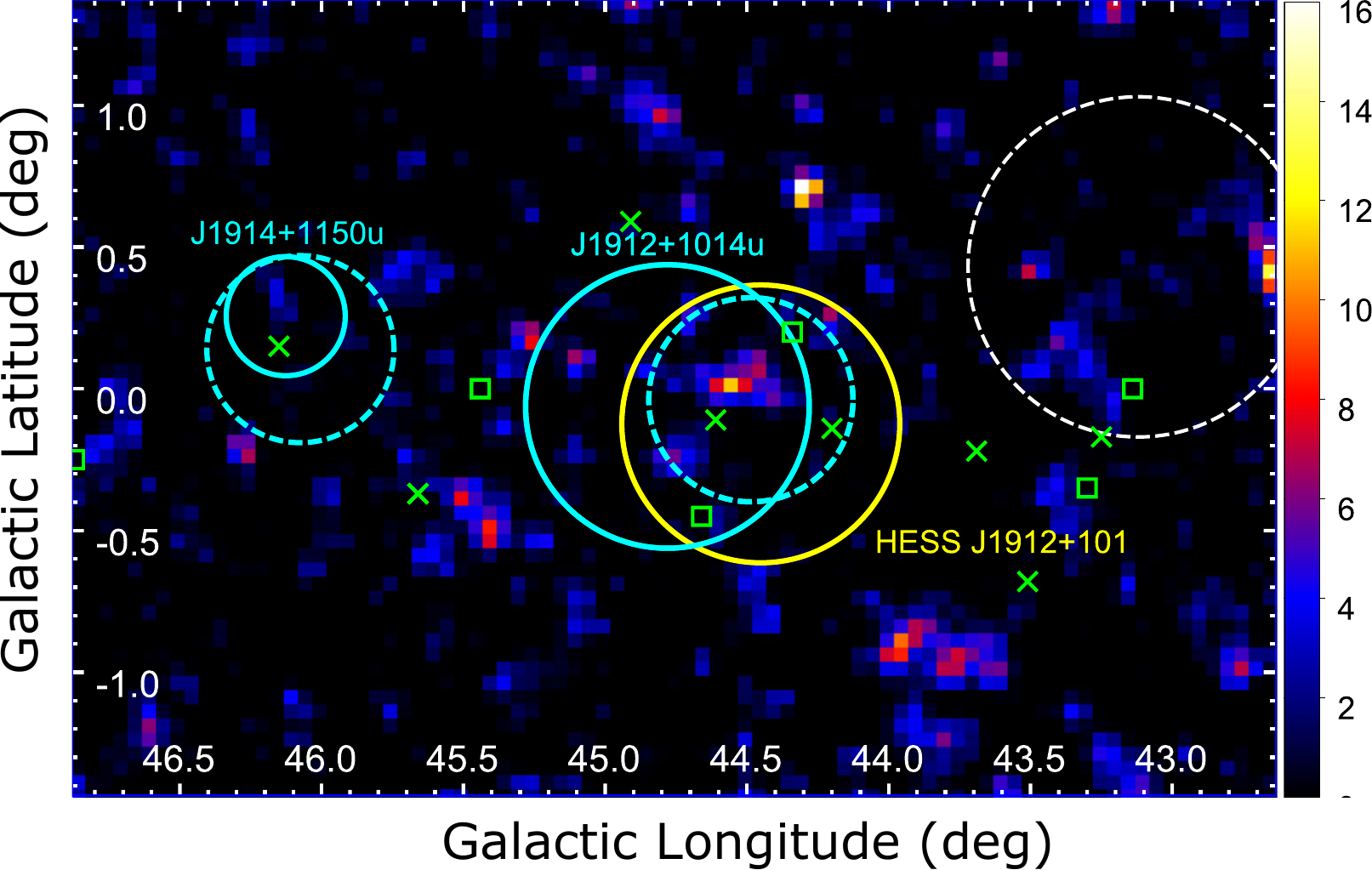}
\put(15,65){(b)}
\end{overpic}
\end{minipage} \\
%%%\\
%%%\\
%%
\begin{minipage}{0.49\textwidth}
\centering
\begin{overpic}[width=\textwidth]{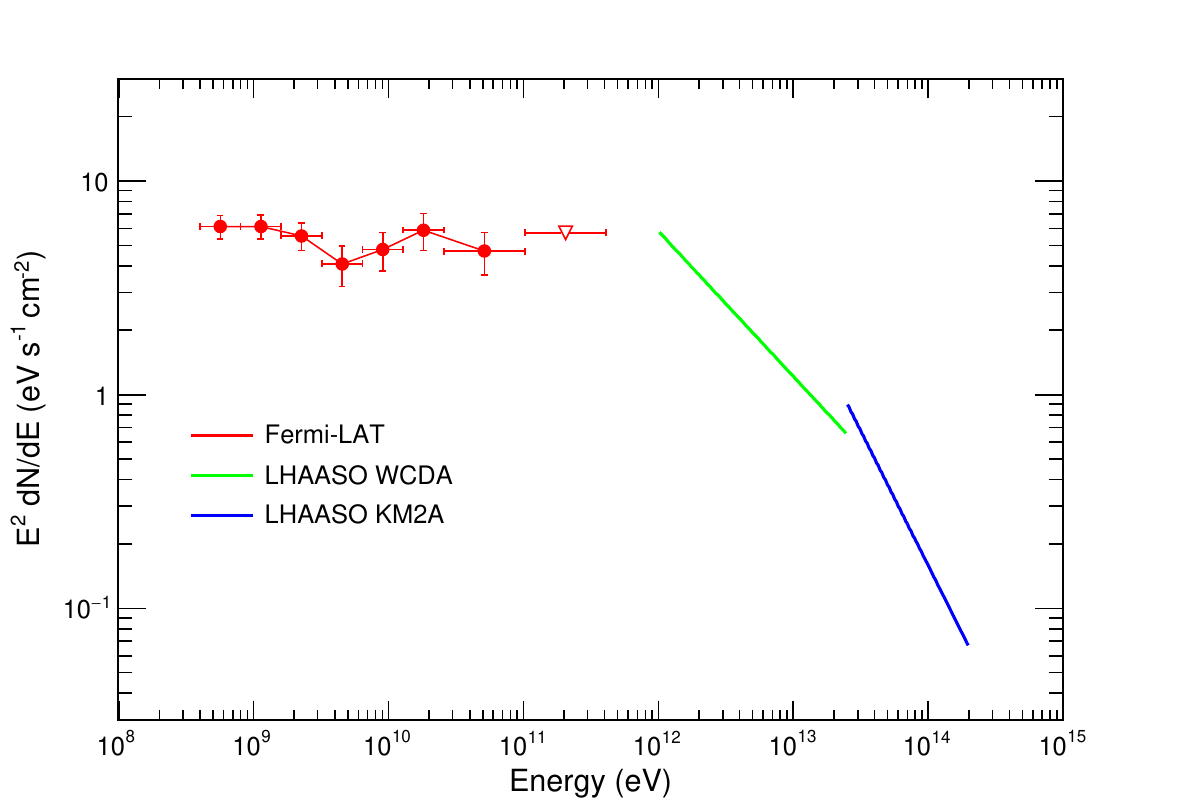}
\put(15,65){(c)}
\end{overpic}
\end{minipage}
%%\hspace{0.02\textwidth}
\begin{minipage}{0.49\textwidth}
\centering
\begin{overpic}[width=\textwidth]{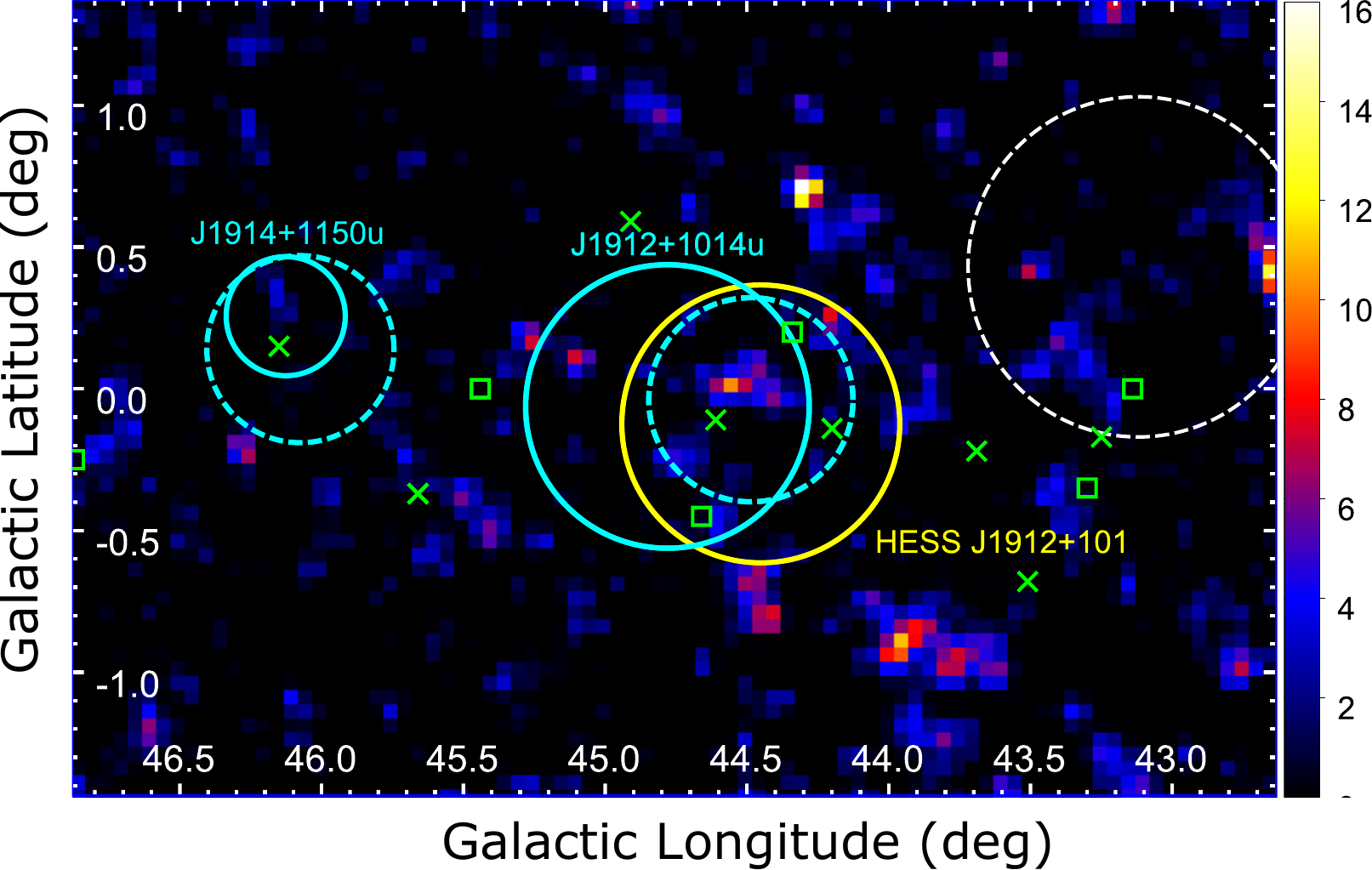}
\put(15,65){(d)}
\end{overpic}
\end{minipage}
\end{tabular}
\caption{
(a) \textit{Fermi}-LAT GeV spectrum of the LHAASO/H.E.S.S. source counterpart (red points) obtained using the Su et al. (2017) $\Np$ map (radius of $0\fdg87$) as a template. 
The best-fit spectral models for LHAASO~J1912+1014u in
\citet{LHAASO_1stCat} are also plotted.
(b) The TS map in {$\ge$}12.8~GeV obtained using the Su et al. (2017) $\Np$ map.
(c) and (d) are the same as panels (a) and (b), respectively, but obtained using LHAASO KM2A Gaussian.
\label{fig:fA}
}
\end{figure}

%%\clearpage

\section{Summary of CR Parameter Scan} \label{sec:appendix:naima}

\begin{figure*}[htbp]
\plotone{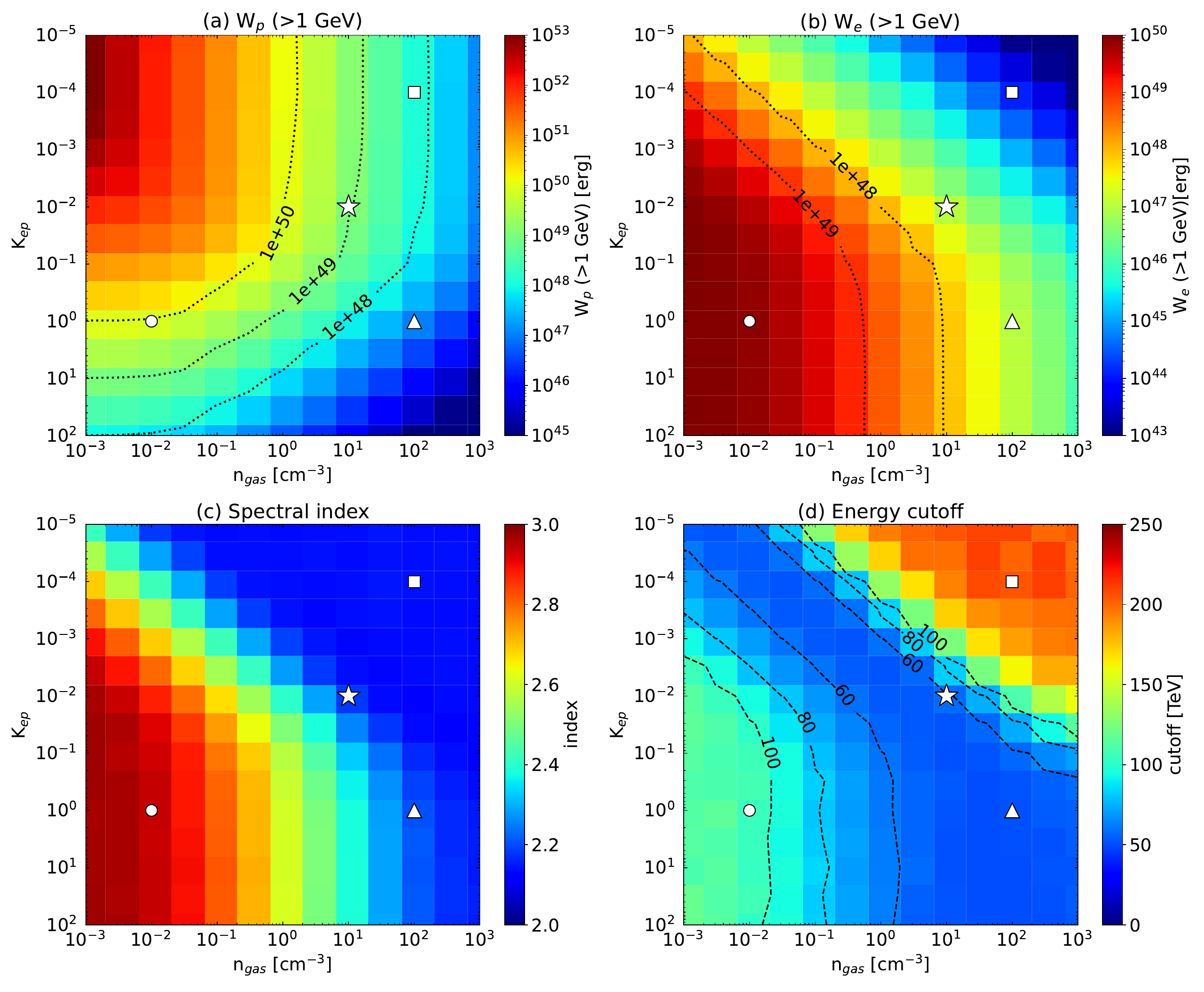}
\caption{Summary of the parameter scan with no electron cooling: (a) $W_\mathrm{p}$, (b) $W_\mathrm{e}$, (c) spectral index, and (d) $E_\mathrm{cut}$.
The white circle, triangle, square, and star represent $K_\mathrm{ep}$ and $n_\mathrm{gas}$ for IC-, eB-, pp-dominated, and nominal cases.
\label{fig:naima:parameter_scan}}
\end{figure*}

Figure~\ref{fig:naima:parameter_scan} summarizes the parameter scan for the no-cooling case, showing the best-fit values of $W_\mathrm{p}$, $W_\mathrm{e}$, index, and $E_\mathrm{cut}$ as a function of $K_\mathrm{ep}$ and $n_\mathrm{gas}$.
Interestingly, the panel of $E_\mathrm{cut}$ is divided into three distinct layers, with a boundary at ${\sim} $80~TeV.
As $n_\mathrm{gas}/K_\mathrm{ep}$ increases, the dominant emission process changes from IC to eB, and finally to pp.
The nominal case lies near the boundary between the eB-dominated and pp-dominated regimes.

%%\clearpage

%%\section{Author publication charges} \label{sec:pubcharge}
%%Keep this part for now for the author's convenience

%%In April 2011 the traditional way of calculating author charges based on 
%%the number of printed pages was changed.  The reason for the change
%%was due to a recognition of the growing number of article items that could not 
%%be represented in print. Now author charges are determined by a number of
%%digital ``quanta''.  A single quantum is defined as 350 words, one figure, one table,
%%and one digital asset.  For the latter this includes machine readable
%%tables, data behind a figure, figure sets, animations, and interactive figures.  The current cost
%%for the different quanta types is available at 
%%\url{https://journals.aas.org/article-charges-and-copyright/#author_publication_charges}. 
%%Authors may use the ApJL length calculator to get a {\tt rough} estimate of 
%%the number of word and float quanta in their manuscript. The calculator 
%%is located at \url{https://authortools.aas.org/ApJL/betacountwords.html}.

%% For this sample we use BibTeX plus aasjournalv7.bst to generate the

%%\clearpage

\bibliography{references}{}

@ARTICLE{kalberla05,
       author = {{Kalberla}, P.~M.~W. and {Burton}, W.~B. and {Hartmann}, Dap and {Arnal}, E.~M. and {Bajaja}, E. and {Morras}, R. and {P{\"o}ppel}, W.~G.~L.},
        title = "{The Leiden/Argentine/Bonn (LAB) Survey of Galactic HI. Final data release of the combined LDS and IAR surveys with improved stray-radiation corrections}",
      journal = {\aap},
         year = 2005,
        month = sep,
       volume = {440},
       number = {2},
        pages = {775-782},
          doi = {10.1051/0004-6361:20041864},
archivePrefix = {arXiv},
       eprint = {astro-ph/0504140},
 primaryClass = {astro-ph},
       adsurl = {https://ui.adsabs.harvard.edu/abs/2005A&A...440..775K}
}

@article{dickey90,
	adsurl = {https://ui.adsabs.harvard.edu/abs/1990ARA&A..28..215D},
	author = {{Dickey}, John M. and {Lockman}, Felix J.},
	doi = {10.1146/annurev.aa.28.090190.001243},
	journal = {\araa},
	month = jan,
	pages = {215-261},
	title = {{H I in the galaxy.}},
	volume = {28},
	year = 1990
}

@article{hi4pi16,
	adsurl = {https://ui.adsabs.harvard.edu/abs/2016A&A...594A.116H},
	archiveprefix = {arXiv},
	author = {{HI4PI Collaboration} and {Ben Bekhti}, N. and {Fl{¥"o}er}, L. and {Keller}, R. and {Kerp}, J. and {Lenz}, D. and {Winkel}, B. and {Bailin}, J. and {Calabretta}, M.~R. and {Dedes}, L. and {Ford}, H.~A. and {Gibson}, B.~K. and {Haud}, U. and {Janowiecki}, S. and {Kalberla}, P.~M.~W. and {Lockman}, F.~J. and {McClure-Griffiths}, N.~M. and {Murphy}, T. and {Nakanishi}, H. and {Pisano}, D.~J. and {Staveley-Smith}, L.},
	doi = {10.1051/0004-6361/201629178},
	eid = {A116},
	eprint = {1610.06175},
	journal = {\aap},
	month = oct,
	pages = {A116},
	primaryclass = {astro-ph.GA},
	title = {{HI4PI: A full-sky H I survey based on EBHIS and GASS}},
	volume = {594},
	year = 2016
}

@ARTICLE{Suzuki2022,
       author = {{Suzuki}, Hiromasa and {Bamba}, Aya and {Yamazaki}, Ryo and {Ohira}, Yutaka},
        title = "{Observational Constraints on the Maximum Energies of Accelerated Particles in Supernova Remnants: Low Maximum Energies and a Large Variety}",
      journal = {\apj},
         year = 2022,
        month = jan,
       volume = {924},
       number = {2},
          eid = {45},
        pages = {45},
          doi = {10.3847/1538-4357/ac33b5},
archivePrefix = {arXiv},
       eprint = {2110.13304},
 primaryClass = {astro-ph.HE},
       adsurl = {https://ui.adsabs.harvard.edu/abs/2022ApJ...924...45S}
}

@ARTICLE{Manchester05,
       author = {{Manchester}, R.~N. and {Hobbs}, G.~B. and {Teoh}, A. and {Hobbs}, M.},
        title = "{The Australia Telescope National Facility Pulsar Catalogue}",
      journal = {\aj},
         year = 2005,
        month = apr,
       volume = {129},
       number = {4},
        pages = {1993-2006},
          doi = {10.1086/428488},
archivePrefix = {arXiv},
       eprint = {astro-ph/0412641},
 primaryClass = {astro-ph},
       adsurl = {https://ui.adsabs.harvard.edu/abs/2005AJ....129.1993M}
}

@INPROCEEDINGS{fruscione06,
       author = {{Fruscione}, Antonella and {McDowell}, Jonathan C. and {Allen}, Glenn E. and {Brickhouse}, Nancy S. and {Burke}, Douglas J. and {Davis}, John E. and {Durham}, Nick and {Elvis}, Martin and {Galle}, Elizabeth C. and {Harris}, Daniel E. and {Huenemoerder}, David P. and {Houck}, John C. and {Ishibashi}, Bish and {Karovska}, Margarita and {Nicastro}, Fabrizio and {Noble}, Michael S. and {Nowak}, Michael A. and {Primini}, Frank A. and {Siemiginowska}, Aneta and {Smith}, Randall K. and {Wise}, Michael},
        title = "{CIAO: Chandra's data analysis system}",
    booktitle = {Observatory Operations: Strategies, Processes, and Systems},
         year = 2006,
       editor = {{Silva}, David R. and {Doxsey}, Rodger E.},
       series = {Society of Photo-Optical Instrumentation Engineers (SPIE) Conference Series},
       volume = {6270},
        month = jun,
          eid = {62701V},
        pages = {62701V},
          doi = {10.1117/12.671760},
       adsurl = {https://ui.adsabs.harvard.edu/abs/2006SPIE.6270E..1VF}
}

@INPROCEEDINGS{arnaud96,
       author = {{Arnaud}, K.~A.},
        title = "{XSPEC: The First Ten Years}",
    booktitle = {Astronomical Data Analysis Software and Systems V},
         year = 1996,
       editor = {{Jacoby}, George H. and {Barnes}, Jeannette},
       series = {Astronomical Society of the Pacific Conference Series},
       volume = {101},
        month = jan,
        pages = {17},
       adsurl = {https://ui.adsabs.harvard.edu/abs/1996ASPC..101...17A}
}

@ARTICLE{kuboike25,
       author = {{Kuboike}, Yui and {Sato}, Toshiki and {Suzuki}, Hiromasa and {Matsunaga}, Kai and {Uchida}, Hiroyuki and {Hughes}, John P. and {Plucinsky}, Paul P.},
        title = "{The Origin of the Mg-rich Supernova Remnant J0550-6823 and the Frequency of Similar Events in the Large Magellanic Cloud}",
      journal = {arXiv e-prints},
         year = 2025,
        month = nov,
          eid = {arXiv:2511.14126},
        pages = {arXiv:2511.14126},
          doi = {10.48550/arXiv.2511.14126},
archivePrefix = {arXiv},
       eprint = {2511.14126},
 primaryClass = {astro-ph.HE},
       adsurl = {https://ui.adsabs.harvard.edu/abs/2025arXiv251114126K}
}

@ARTICLE{suzuki25,
       author = {{Suzuki}, Hiromasa and {Tsuji}, Naomi and {Kanemaru}, Yoshiaki and {Shidatsu}, Megumi and {Olivera-Nieto}, Laura and {Safi-Harb}, Samar and {Kimura}, Shigeo S. and {de la Fuente}, Eduardo and {Casanova}, Sabrina and {Mori}, Kaya and {Wang}, Xiaojie and {Kato}, Sei and {Tateishi}, Dai and {Uchiyama}, Hideki and {Tanaka}, Takaaki and {Uchida}, Hiroyuki and {Inoue}, Shun and {Huang}, Dezhi and {Lemoine-Goumard}, Marianne and {Miura}, Daiki and {Ogawa}, Shoji and {Kobayashi}, Shogo B. and {Done}, Chris and {Parra}, Maxime and {D{\'\i}az Trigo}, Maria and {Mu{\~n}oz-Darias}, Teo and {Armas Padilla}, Montserrat and {Tomaru}, Ryota and {Ueda}, Yoshihiro},
        title = "{Detection of Extended X-Ray Emission around the PeVatron Microquasar V4641 Sgr with XRISM}",
      journal = {\apjl},
         year = 2025,
        month = jan,
       volume = {978},
       number = {2},
          eid = {L20},
        pages = {L20},
          doi = {10.3847/2041-8213/ad9d11},
archivePrefix = {arXiv},
       eprint = {2412.08089},
 primaryClass = {astro-ph.HE},
       adsurl = {https://ui.adsabs.harvard.edu/abs/2025ApJ...978L..20S}
}

@ARTICLE{uchiyama13,
       author = {{Uchiyama}, Hideki and {Nobukawa}, Masayoshi and {Tsuru}, Takeshi Go and {Koyama}, Katsuji},
        title = "{K-Shell Line Distribution of Heavy Elements along the Galactic Plane Observed with Suzaku}",
      journal = {\pasj},
         year = 2013,
        month = feb,
       volume = {65},
          eid = {19},
        pages = {19},
          doi = {10.1093/pasj/65.1.19},
archivePrefix = {arXiv},
       eprint = {1209.0067},
 primaryClass = {astro-ph.HE},
       adsurl = {https://ui.adsabs.harvard.edu/abs/2013PASJ...65...19U}
}

@ARTICLE{suzuki20,
       author = {{Suzuki}, Hiromasa and {Bamba}, Aya and {Enokiya}, Rei and {Yamaguchi}, Hiroya and {Plucinsky}, Paul P. and {Odaka}, Hirokazu},
        title = "{Uniform Distribution of the Extremely Overionized Plasma Associated with the Supernova Remnant G359.1-0.5}",
      journal = {\apj},
         year = 2020,
        month = apr,
       volume = {893},
       number = {2},
          eid = {147},
        pages = {147},
          doi = {10.3847/1538-4357/ab80ba},
archivePrefix = {arXiv},
       eprint = {2003.07576},
 primaryClass = {astro-ph.HE},
       adsurl = {https://ui.adsabs.harvard.edu/abs/2020ApJ...893..147S}
}

@ARTICLE{suzuki21,
       author = {{Suzuki}, H. and {Plucinsky}, P.~P. and {Gaetz}, T.~J. and {Bamba}, A.},
        title = "{Spatial and temporal variations of the Chandra ACIS particle-induced background and development of a spectral-model generation tool}",
      journal = {\aap},
         year = 2021,
        month = nov,
       volume = {655},
          eid = {A116},
        pages = {A116},
          doi = {10.1051/0004-6361/202141458},
archivePrefix = {arXiv},
       eprint = {2108.11234},
 primaryClass = {astro-ph.HE},
       adsurl = {https://ui.adsabs.harvard.edu/abs/2021A&A...655A.116S}
}

@ARTICLE{chang08,
       author = {{Chang}, Chulhoon and {Konopelko}, Alexander and {Cui}, Wei},
        title = "{Search for Pulsar Wind Nebula Associations with Unidentified TeV {\ensuremath{\gamma}}-Ray Sources}",
      journal = {\apj},
         year = 2008,
        month = aug,
       volume = {682},
       number = {2},
        pages = {1177-1184},
          doi = {10.1086/589225},
archivePrefix = {arXiv},
       eprint = {0709.3614},
 primaryClass = {astro-ph},
       adsurl = {https://ui.adsabs.harvard.edu/abs/2008ApJ...682.1177C}
}

@ARTICLE{stil2006,
       author = {{Stil}, J.~M. and {Taylor}, A.~R. and {Dickey}, J.~M. and {Kavars}, D.~W. and {Martin}, P.~G. and {Rothwell}, T.~A. and {Boothroyd}, A.~I. and {Lockman}, Felix J. and {McClure-Griffiths}, N.~M.},
        title = "{The VLA Galactic Plane Survey}",
      journal = {\aj},
         year = 2006,
        month = sep,
       volume = {132},
       number = {3},
        pages = {1158-1176},
          doi = {10.1086/505940},
archivePrefix = {arXiv},
       eprint = {astro-ph/0605422},
 primaryClass = {astro-ph},
       adsurl = {https://ui.adsabs.harvard.edu/abs/2006AJ....132.1158S}
}

@article{Fukui_2014,
doi = {10.1088/0004-637X/796/1/59},
url = {https://doi.org/10.1088/0004-637X/796/1/59},
year = {2014},
month = {nov},
publisher = {The American Astronomical Society},
volume = {796},
number = {1},
pages = {59},
author = {Fukui, Yasuo and Okamoto, Ryuji and Kaji, Ryohei and Yamamoto, Hiroaki and Torii, Kazufumi and Hayakawa, Takahiro and Tachihara, Kengo and Dickey, John M. and Okuda, Takeshi and Ohama, Akio and Kuroda, Yutaka and Kuwahara, Toshihisa},
title = {{\rm H\,\scriptsize{I}}, CO, AND PLANCK/IRAS DUST PROPERTIES IN THE HIGH LATITUDE CLOUD COMPLEX, MBM 53, 54, 55 AND HLCG 92 − 35. POSSIBLE EVIDENCE FOR AN OPTICALLY THICK {\rm H\,\scriptsize{I}} ENVELOPE AROUND THE CO CLOUDS},
journal = {The Astrophysical Journal}
}

@ARTICLE{Sawada2008,
       author = {{Sawada}, Tsuyoshi and {Ikeda}, Norio and {Sunada}, Kazuyoshi and {Kuno}, Nario and {Kamazaki}, Takeshi and {Morita}, Koh-Ichiro and {Kurono}, Yasutaka and {Koura}, Norikazu and {Abe}, Katsumi and {Kawase}, Sachiko and {Maekawa}, Jun and {Horigome}, Osamu and {Yanagisawa}, Kiyohiko},
        title = "{On-The-Fly Observing System of the Nobeyama 45-m and ASTE 10-m Telescopes}",
      journal = {\pasj},
         year = 2008,
        month = jun,
       volume = {60},
        pages = {445},
          doi = {10.1093/pasj/60.3.445},
archivePrefix = {arXiv},
       eprint = {0712.1283},
 primaryClass = {astro-ph},
       adsurl = {https://ui.adsabs.harvard.edu/abs/2008PASJ...60..445S}
}

@ARTICLE{P8Ref1,
       author = {{Atwood}, W. and {Albert}, A. and {Baldini}, L. and {Tinivella}, M. and {Bregeon}, J. and {Pesce-Rollins}, M. and {Sgr{\`o}}, C. and {Bruel}, P. and {Charles}, E. and {Drlica-Wagner}, A. and {Franckowiak}, A. and {Jogler}, T. and {Rochester}, L. and {Usher}, T. and {Wood}, M. and {Cohen-Tanugi}, J. and {Zimmer}, S.},
        title = "{Pass 8: Toward the Full Realization of the Fermi-LAT Scientific Potential}",
      journal = {arXiv e-prints},
         year = 2013,
        month = mar,
          eid = {arXiv:1303.3514},
        pages = {arXiv:1303.3514},
          doi = {10.48550/arXiv.1303.3514},
archivePrefix = {arXiv},
       eprint = {1303.3514},
 primaryClass = {astro-ph.IM},
       adsurl = {https://ui.adsabs.harvard.edu/abs/2013arXiv1303.3514A}
}

@ARTICLE{P8Ref2,
       author = {{Bruel}, P. and {Burnett}, T.~H. and {Digel}, S.~W. and {Johannesson}, G. and {Omodei}, N. and {Wood}, M.},
        title = "{Fermi-LAT improved Pass\raisebox{-0.5ex}\textasciitilde8 event selection}",
      journal = {arXiv e-prints},
         year = 2018,
        month = oct,
          eid = {arXiv:1810.11394},
        pages = {arXiv:1810.11394},
          doi = {10.48550/arXiv.1810.11394},
archivePrefix = {arXiv},
       eprint = {1810.11394},
 primaryClass = {astro-ph.IM},
       adsurl = {https://ui.adsabs.harvard.edu/abs/2018arXiv181011394B}
}

@ARTICLE{murase2026,
       author = {{Murase}, Takeru and {Sano}, Hidetoshi and {Matsubara}, Kohei and {Fukui}, Yasuo and {Nishi}, Junya and {Einecke}, Sabrina and {Filipovi{\'c}}, Miroslav and {Kasai}, Rina and {Matsusaka}, Ren and {Rowell}, Gavin and {Sodoh}, Hiroshi and {Suzuki}, Hiromasa and {Shibata}, Yosuke and {Tsuge}, Kisetsu and {Takaba}, Hiroshi and {Handa}, Toshihiro},
        title = "{Gas Kinematics and Cosmic-Ray Acceleration in the Gamma-ray SNRs W41 and G22.7-0.2}",
      journal = {arXiv e-prints},
         year = 2026,
        month = jan,
          eid = {arXiv:2601.18040},
        pages = {arXiv:2601.18040},
          doi = {10.48550/arXiv.2601.18040},
archivePrefix = {arXiv},
       eprint = {2601.18040},
 primaryClass = {astro-ph.HE},
       adsurl = {https://ui.adsabs.harvard.edu/abs/2026arXiv260118040M},
      note= {ApJ accepted},
}

@ARTICLE{astropy_2013,
       author = {{Astropy Collaboration} and {Robitaille}, Thomas P. and {Tollerud}, Erik J. and {Greenfield}, Perry and {Droettboom}, Michael and {Bray}, Erik and {Aldcroft}, Tom and {Davis}, Matt and {Ginsburg}, Adam and {Price-Whelan}, Adrian M. and {Kerzendorf}, Wolfgang E. and {Conley}, Alexander and {Crighton}, Neil and {Barbary}, Kyle and {Muna}, Demitri and {Ferguson}, Henry and {Grollier}, Fr{\'e}d{\'e}ric and {Parikh}, Madhura M. and {Nair}, Prasanth H. and {Unther}, Hans M. and {Deil}, Christoph and {Woillez}, Julien and {Conseil}, Simon and {Kramer}, Roban and {Turner}, James E.~H. and {Singer}, Leo and {Fox}, Ryan and {Weaver}, Benjamin A. and {Zabalza}, Victor and {Edwards}, Zachary I. and {Azalee Bostroem}, K. and {Burke}, D.~J. and {Casey}, Andrew R. and {Crawford}, Steven M. and {Dencheva}, Nadia and {Ely}, Justin and {Jenness}, Tim and {Labrie}, Kathleen and {Lim}, Pey Lian and {Pierfederici}, Francesco and {Pontzen}, Andrew and {Ptak}, Andy and {Refsdal}, Brian and {Servillat}, Mathieu and {Streicher}, Ole},
        title = "{Astropy: A community Python package for astronomy}",
      journal = {\aap},
         year = 2013,
        month = oct,
       volume = {558},
          eid = {A33},
        pages = {A33},
          doi = {10.1051/0004-6361/201322068},
archivePrefix = {arXiv},
       eprint = {1307.6212},
 primaryClass = {astro-ph.IM},
       adsurl = {https://ui.adsabs.harvard.edu/abs/2013A&A...558A..33A}
}

@ARTICLE{astropy_2018,
       author = {{Astropy Collaboration} and {Price-Whelan}, A.~M. and {Sip{\H{o}}cz}, B.~M. and {G{\"u}nther}, H.~M. and {Lim}, P.~L. and {Crawford}, S.~M. and {Conseil}, S. and {Shupe}, D.~L. and {Craig}, M.~W. and {Dencheva}, N. and {Ginsburg}, A. and {VanderPlas}, J.~T. and {Bradley}, L.~D. and {P{\'e}rez-Su{\'a}rez}, D. and {de Val-Borro}, M. and {Aldcroft}, T.~L. and {Cruz}, K.~L. and {Robitaille}, T.~P. and {Tollerud}, E.~J. and {Ardelean}, C. and {Babej}, T. and {Bach}, Y.~P. and {Bachetti}, M. and {Bakanov}, A.~V. and {Bamford}, S.~P. and {Barentsen}, G. and {Barmby}, P. and {Baumbach}, A. and {Berry}, K.~L. and {Biscani}, F. and {Boquien}, M. and {Bostroem}, K.~A. and {Bouma}, L.~G. and {Brammer}, G.~B. and {Bray}, E.~M. and {Breytenbach}, H. and {Buddelmeijer}, H. and {Burke}, D.~J. and {Calderone}, G. and {Cano Rodr{\'\i}guez}, J.~L. and {Cara}, M. and {Cardoso}, J.~V.~M. and {Cheedella}, S. and {Copin}, Y. and {Corrales}, L. and {Crichton}, D. and {D'Avella}, D. and {Deil}, C. and {Depagne}, {\'E}. and {Dietrich}, J.~P. and {Donath}, A. and {Droettboom}, M. and {Earl}, N. and {Erben}, T. and {Fabbro}, S. and {Ferreira}, L.~A. and {Finethy}, T. and {Fox}, R.~T. and {Garrison}, L.~H. and {Gibbons}, S.~L.~J. and {Goldstein}, D.~A. and {Gommers}, R. and {Greco}, J.~P. and {Greenfield}, P. and {Groener}, A.~M. and {Grollier}, F. and {Hagen}, A. and {Hirst}, P. and {Homeier}, D. and {Horton}, A.~J. and {Hosseinzadeh}, G. and {Hu}, L. and {Hunkeler}, J.~S. and {Ivezi{\'c}}, {\v{Z}}. and {Jain}, A. and {Jenness}, T. and {Kanarek}, G. and {Kendrew}, S. and {Kern}, N.~S. and {Kerzendorf}, W.~E. and {Khvalko}, A. and {King}, J. and {Kirkby}, D. and {Kulkarni}, A.~M. and {Kumar}, A. and {Lee}, A. and {Lenz}, D. and {Littlefair}, S.~P. and {Ma}, Z. and {Macleod}, D.~M. and {Mastropietro}, M. and {McCully}, C. and {Montagnac}, S. and {Morris}, B.~M. and {Mueller}, M. and {Mumford}, S.~J. and {Muna}, D. and {Murphy}, N.~A. and {Nelson}, S. and {Nguyen}, G.~H. and {Ninan}, J.~P. and {N{\"o}the}, M. and {Ogaz}, S. and {Oh}, S. and {Parejko}, J.~K. and {Parley}, N. and {Pascual}, S. and {Patil}, R. and {Patil}, A.~A. and {Plunkett}, A.~L. and {Prochaska}, J.~X. and {Rastogi}, T. and {Reddy Janga}, V. and {Sabater}, J. and {Sakurikar}, P. and {Seifert}, M. and {Sherbert}, L.~E. and {Sherwood-Taylor}, H. and {Shih}, A.~Y. and {Sick}, J. and {Silbiger}, M.~T. and {Singanamalla}, S. and {Singer}, L.~P. and {Sladen}, P.~H. and {Sooley}, K.~A. and {Sornarajah}, S. and {Streicher}, O. and {Teuben}, P. and {Thomas}, S.~W. and {Tremblay}, G.~R. and {Turner}, J.~E.~H. and {Terr{\'o}n}, V. and {van Kerkwijk}, M.~H. and {de la Vega}, A. and {Watkins}, L.~L. and {Weaver}, B.~A. and {Whitmore}, J.~B. and {Woillez}, J. and {Zabalza}, V. and {Astropy Contributors}},
        title = "{The Astropy Project: Building an Open-science Project and Status of the v2.0 Core Package}",
      journal = {\aj},
         year = 2018,
        month = sep,
       volume = {156},
       number = {3},
          eid = {123},
        pages = {123},
          doi = {10.3847/1538-3881/aabc4f},
archivePrefix = {arXiv},
       eprint = {1801.02634},
 primaryClass = {astro-ph.IM},
       adsurl = {https://ui.adsabs.harvard.edu/abs/2018AJ....156..123A}
}

@ARTICLE{astropy_2022,
       author = {{Astropy Collaboration} and {Price-Whelan}, Adrian M. and {Lim}, Pey Lian and {Earl}, Nicholas and {Starkman}, Nathaniel and {Bradley}, Larry and {Shupe}, David L. and {Patil}, Aarya A. and {Corrales}, Lia and {Brasseur}, C.~E. and {N{\"o}the}, Maximilian and {Donath}, Axel and {Tollerud}, Erik and {Morris}, Brett M. and {Ginsburg}, Adam and {Vaher}, Eero and {Weaver}, Benjamin A. and {Tocknell}, James and {Jamieson}, William and {van Kerkwijk}, Marten H. and {Robitaille}, Thomas P. and {Merry}, Bruce and {Bachetti}, Matteo and {G{\"u}nther}, H. Moritz and {Aldcroft}, Thomas L. and {Alvarado-Montes}, Jaime A. and {Archibald}, Anne M. and {B{\'o}di}, Attila and {Bapat}, Shreyas and {Barentsen}, Geert and {Baz{\'a}n}, Juanjo and {Biswas}, Manish and {Boquien}, M{\'e}d{\'e}ric and {Burke}, D.~J. and {Cara}, Daria and {Cara}, Mihai and {Conroy}, Kyle E. and {Conseil}, Simon and {Craig}, Matthew W. and {Cross}, Robert M. and {Cruz}, Kelle L. and {D'Eugenio}, Francesco and {Dencheva}, Nadia and {Devillepoix}, Hadrien A.~R. and {Dietrich}, J{\"o}rg P. and {Eigenbrot}, Arthur Davis and {Erben}, Thomas and {Ferreira}, Leonardo and {Foreman-Mackey}, Daniel and {Fox}, Ryan and {Freij}, Nabil and {Garg}, Suyog and {Geda}, Robel and {Glattly}, Lauren and {Gondhalekar}, Yash and {Gordon}, Karl D. and {Grant}, David and {Greenfield}, Perry and {Groener}, Austen M. and {Guest}, Steve and {Gurovich}, Sebastian and {Handberg}, Rasmus and {Hart}, Akeem and {Hatfield-Dodds}, Zac and {Homeier}, Derek and {Hosseinzadeh}, Griffin and {Jenness}, Tim and {Jones}, Craig K. and {Joseph}, Prajwel and {Kalmbach}, J. Bryce and {Karamehmetoglu}, Emir and {Ka{\l}uszy{\'n}ski}, Miko{\l}aj and {Kelley}, Michael S.~P. and {Kern}, Nicholas and {Kerzendorf}, Wolfgang E. and {Koch}, Eric W. and {Kulumani}, Shankar and {Lee}, Antony and {Ly}, Chun and {Ma}, Zhiyuan and {MacBride}, Conor and {Maljaars}, Jakob M. and {Muna}, Demitri and {Murphy}, N.~A. and {Norman}, Henrik and {O'Steen}, Richard and {Oman}, Kyle A. and {Pacifici}, Camilla and {Pascual}, Sergio and {Pascual-Granado}, J. and {Patil}, Rohit R. and {Perren}, Gabriel I. and {Pickering}, Timothy E. and {Rastogi}, Tanuj and {Roulston}, Benjamin R. and {Ryan}, Daniel F. and {Rykoff}, Eli S. and {Sabater}, Jose and {Sakurikar}, Parikshit and {Salgado}, Jes{\'u}s and {Sanghi}, Aniket and {Saunders}, Nicholas and {Savchenko}, Volodymyr and {Schwardt}, Ludwig and {Seifert-Eckert}, Michael and {Shih}, Albert Y. and {Jain}, Anany Shrey and {Shukla}, Gyanendra and {Sick}, Jonathan and {Simpson}, Chris and {Singanamalla}, Sudheesh and {Singer}, Leo P. and {Singhal}, Jaladh and {Sinha}, Manodeep and {Sip{\H{o}}cz}, Brigitta M. and {Spitler}, Lee R. and {Stansby}, David and {Streicher}, Ole and {{\v{S}}umak}, Jani and {Swinbank}, John D. and {Taranu}, Dan S. and {Tewary}, Nikita and {Tremblay}, Grant R. and {de Val-Borro}, Miguel and {Van Kooten}, Samuel J. and {Vasovi{\'c}}, Zlatan and {Verma}, Shresth and {de Miranda Cardoso}, Jos{\'e} Vin{\'\i}cius and {Williams}, Peter K.~G. and {Wilson}, Tom J. and {Winkel}, Benjamin and {Wood-Vasey}, W.~M. and {Xue}, Rui and {Yoachim}, Peter and {Zhang}, Chen and {Zonca}, Andrea and {Astropy Project Contributors}},
        title = "{The Astropy Project: Sustaining and Growing a Community-oriented Open-source Project and the Latest Major Release (v5.0) of the Core Package}",
      journal = {\apj},
         year = 2022,
        month = aug,
       volume = {935},
       number = {2},
          eid = {167},
        pages = {167},
          doi = {10.3847/1538-4357/ac7c74},
archivePrefix = {arXiv},
       eprint = {2206.14220},
 primaryClass = {astro-ph.IM},
       adsurl = {https://ui.adsabs.harvard.edu/abs/2022ApJ...935..167A}
}

@Article{harris_2020,
 title         = {Array programming with {NumPy}},
 author        = {Charles R. Harris and K. Jarrod Millman and St{\'{e}}fan J.
                 van der Walt and Ralf Gommers and Pauli Virtanen and David
                 Cournapeau and Eric Wieser and Julian Taylor and Sebastian
                 Berg and Nathaniel J. Smith and Robert Kern and Matti Picus
                 and Stephan Hoyer and Marten H. van Kerkwijk and Matthew
                 Brett and Allan Haldane and Jaime Fern{\'{a}}ndez del
                 R{\'{i}}o and Mark Wiebe and Pearu Peterson and Pierre
                 G{\'{e}}rard-Marchant and Kevin Sheppard and Tyler Reddy and
                 Warren Weckesser and Hameer Abbasi and Christoph Gohlke and
                 Travis E. Oliphant},
 year          = {2020},
 month         = sep,
 journal       = {Nature},
 volume        = {585},
 number        = {7825},
 pages         = {357--362},
 doi           = {10.1038/s41586-020-2649-2},
 publisher     = {Springer Science and Business Media {LLC}},
 url           = {https://doi.org/10.1038/s41586-020-2649-2}
}

@ARTICLE{Virtanen2020SciPy-NMeth,
  author  = {Virtanen, Pauli and Gommers, Ralf and Oliphant, Travis E. and
            Haberland, Matt and Reddy, Tyler and Cournapeau, David and
            Burovski, Evgeni and Peterson, Pearu and Weckesser, Warren and
            Bright, Jonathan and {van der Walt}, St{\'e}fan J. and
            Brett, Matthew and Wilson, Joshua and Millman, K. Jarrod and
            Mayorov, Nikolay and Nelson, Andrew R. J. and Jones, Eric and
            Kern, Robert and Larson, Eric and Carey, C J and
            Polat, {\.I}lhan and Feng, Yu and Moore, Eric W. and
            {VanderPlas}, Jake and Laxalde, Denis and Perktold, Josef and
            Cimrman, Robert and Henriksen, Ian and Quintero, E. A. and
            Harris, Charles R. and Archibald, Anne M. and
            Ribeiro, Ant{\^o}nio H. and Pedregosa, Fabian and
            {van Mulbregt}, Paul and {SciPy 1.0 Contributors}},
  title   = {{{SciPy} 1.0: Fundamental Algorithms for Scientific
            Computing in Python}},
  journal = {Nature Methods},
  year    = {2020},
  volume  = {17},
  pages   = {261--272},
  adsurl  = {https://rdcu.be/b08Wh},
  doi     = {10.1038/s41592-019-0686-2},
}

@ARTICLE{Duvidovich2023,
       author = {{Duvidovich}, L. and {Petriella}, A.},
        title = "{Radio and infrared study of the supernova remnant candidate HESS J1912+101}",
      journal = {\aap},
         year = 2023,
        month = apr,
       volume = {672},
          eid = {A195},
        pages = {A195},
          doi = {10.1051/0004-6361/202245819},
archivePrefix = {arXiv},
       eprint = {2303.11115},
 primaryClass = {astro-ph.HE},
       adsurl = {https://ui.adsabs.harvard.edu/abs/2023A&A...672A.195D}
}

@inproceedings{Sano2018,
author = {{Sano}, H. and {Yoshiike}, S. and {Yamane}, Y. and {Nagaya}, T. and {Nishimura}, J. and {Yamamoto}, H. and {Tamura}, Y. and {Tachihara}, K. and {Fukui}, Y.},
title = "{talk title}",
booktitle = {Astronomical Society of Japan Annual Meeting (2018 Spring) Abstract book},
year = {2018},
pages = {Q15a},
}

@ARTICLE{Fukui2017,
       author = {{Fukui}, Y. and {Sano}, H. and {Sato}, J. and {Okamoto}, R. and {Fukuda}, T. and {Yoshiike}, S. and {Hayashi}, K. and {Torii}, K. and {Hayakawa}, T. and {Rowell}, G. and {Filipovi{\'c}}, M.~D. and {Maxted}, N. and {McClure-Griffiths}, N.~M. and {Kawamura}, A. and {Yamamoto}, H. and {Okuda}, T. and {Mizuno}, N. and {Tachihara}, K. and {Onishi}, T. and {Mizuno}, A. and {Ogawa}, H.},
        title = "{A Detailed Study of the Interstellar Protons toward the TeV {\ensuremath{\gamma}}-Ray SNR RX J0852.0-4622 (G266.2-1.2, Vela Jr.): The Third Case of the {\ensuremath{\gamma}}-Ray and ISM Spatial Correspondence}",
      journal = {\apj},
         year = 2017,
        month = nov,
       volume = {850},
       number = {1},
          eid = {71},
        pages = {71},
          doi = {10.3847/1538-4357/aa9219},
archivePrefix = {arXiv},
       eprint = {1708.07911},
 primaryClass = {astro-ph.HE},
       adsurl = {https://ui.adsabs.harvard.edu/abs/2017ApJ...850...71F}
}

@ARTICLE{Fukui2015,
       author = {{Fukui}, Y. and {Torii}, K. and {Onishi}, T. and {Yamamoto}, H. and {Okamoto}, R. and {Hayakawa}, T. and {Tachihara}, K. and {Sano}, H.},
        title = "{Optically Thick H I Dominant in the Local Interstellar Medium: An Alternative Interpretation to ``Dark Gas''}",
      journal = {\apj},
         year = 2015,
        month = jan,
       volume = {798},
       number = {1},
          eid = {6},
        pages = {6},
          doi = {10.1088/0004-637X/798/1/6},
archivePrefix = {arXiv},
       eprint = {1403.0999},
 primaryClass = {astro-ph.GA},
       adsurl = {https://ui.adsabs.harvard.edu/abs/2015ApJ...798....6F}
}

@ARTICLE{Mizuno2025,
       author = {{Mizuno}, Tsunefumi and {Hayashi}, Katsuhiro and {Ochi}, Hinako and {Moskalenko}, Igor V. and {Orlando}, Elena and {Strong}, Andrew W.},
        title = "{Cosmic ray and interstellar gas properties in the solar neighborhood revealed by diffuse gamma-rays}",
      journal = {\pasj},
         year = 2025,
        month = oct,
       volume = {77},
       number = {5},
        pages = {965-984},
          doi = {10.1093/pasj/psaf069},
archivePrefix = {arXiv},
       eprint = {2506.16252},
 primaryClass = {astro-ph.HE},
       adsurl = {https://ui.adsabs.harvard.edu/abs/2025PASJ...77..965M}
}

@ARTICLE{Bertsch1993,
       author = {{Bertsch}, D.~L. and {Dame}, T.~M. and {Fichtel}, C.~E. and {Hunter}, S.~D. and {Sreekumar}, P. and {Stacy}, J.~G. and {Thaddeus}, P.},
        title = "{Diffuse Gamma-Ray Emission in the Galactic Plane from Cosmic-Ray, Matter, and Photon Interactions}",
      journal = {\apj},
         year = 1993,
        month = oct,
       volume = {416},
        pages = {587},
          doi = {10.1086/173261},
       adsurl = {https://ui.adsabs.harvard.edu/abs/1993ApJ...416..587B}
}

@article{Reich2019,
doi = {10.1088/1674-4527/19/3/45},
url = {https://doi.org/10.1088/1674-4527/19/3/45},
year = {2019},
month = {mar},
publisher = {National Astronomical Observatories, CAS and IOP Publishing Ltd.},
volume = {19},
number = {3},
pages = {045},
author = {Reich, Wolfgang and Sun, Xiao-Hui},
title = {Polarised radio emission associated with HESS J1912+101},
journal = {Research in Astronomy and Astrophysics}
}

@ARTICLE{HAWC2017_2HWC,
       author = {{Abeysekara}, A.~U. and {Albert}, A. and {Alfaro}, R. and {Alvarez}, C. and {{\'A}lvarez}, J.~D. and {Arceo}, R. and {Arteaga-Vel{\'a}zquez}, J.~C. and {Ayala Solares}, H.~A. and {Barber}, A.~S. and {Baughman}, B. and {Bautista-Elivar}, N. and {Becerra Gonzalez}, J. and {Becerril}, A. and {Belmont-Moreno}, E. and {BenZvi}, S.~Y. and {Berley}, D. and {Bernal}, A. and {Braun}, J. and {Brisbois}, C. and {Caballero-Mora}, K.~S. and {Capistr{\'a}n}, T. and {Carrami{\~n}ana}, A. and {Casanova}, S. and {Castillo}, M. and {Cotti}, U. and {Cotzomi}, J. and {Couti{\~n}o de Le{\'o}n}, S. and {de la Fuente}, E. and {De Le{\'o}n}, C. and {Diaz Hernandez}, R. and {Dingus}, B.~L. and {DuVernois}, M.~A. and {D{\'\i}az-V{\'e}lez}, J.~C. and {Ellsworth}, R.~W. and {Engel}, K. and {Fiorino}, D.~W. and {Fraija}, N. and {Garc{\'\i}a-Gonz{\'a}lez}, J.~A. and {Garfias}, F. and {Gerhardt}, M. and {Gonz{\'a}lez Mu{\~n}oz}, A. and {Gonz{\'a}lez}, M.~M. and {Goodman}, J.~A. and {Hampel-Arias}, Z. and {Harding}, J.~P. and {Hernandez}, S. and {Hernandez-Almada}, A. and {Hinton}, J. and {Hui}, C.~M. and {H{\"u}ntemeyer}, P. and {Iriarte}, A. and {Jardin-Blicq}, A. and {Joshi}, V. and {Kaufmann}, S. and {Kieda}, D. and {Lara}, A. and {Lauer}, R.~J. and {Lee}, W.~H. and {Lennarz}, D. and {Le{\'o}n Vargas}, H. and {Linnemann}, J.~T. and {Longinotti}, A.~L. and {Raya}, G. Luis and {Luna-Garc{\'\i}a}, R. and {L{\'o}pez-Coto}, R. and {Malone}, K. and {Marinelli}, S.~S. and {Martinez}, O. and {Martinez-Castellanos}, I. and {Mart{\'\i}nez-Castro}, J. and {Mart{\'\i}nez-Huerta}, H. and {Matthews}, J.~A. and {Miranda-Romagnoli}, P. and {Moreno}, E. and {Mostaf{\'a}}, M. and {Nellen}, L. and {Newbold}, M. and {Nisa}, M.~U. and {Noriega-Papaqui}, R. and {Pelayo}, R. and {Pretz}, J. and {P{\'e}rez-P{\'e}rez}, E.~G. and {Ren}, Z. and {Rho}, C.~D. and {Rivi{\`e}re}, C. and {Rosa-Gonz{\'a}lez}, D. and {Rosenberg}, M. and {Ruiz-Velasco}, E. and {Salazar}, H. and {Salesa Greus}, F. and {Sandoval}, A. and {Schneider}, M. and {Schoorlemmer}, H. and {Sinnis}, G. and {Smith}, A.~J. and {Springer}, R.~W. and {Surajbali}, P. and {Taboada}, I. and {Tibolla}, O. and {Tollefson}, K. and {Torres}, I. and {Ukwatta}, T.~N. and {Vianello}, G. and {Villase{\~n}or}, L. and {Weisgarber}, T. and {Westerhoff}, S. and {Wisher}, I.~G. and {Wood}, J. and {Yapici}, T. and {Younk}, P.~W. and {Zepeda}, A. and {Zhou}, H.},
        title = "{The 2HWC HAWC Observatory Gamma-Ray Catalog}",
      journal = {\apj},
         year = 2017,
        month = jul,
       volume = {843},
       number = {1},
          eid = {40},
        pages = {40},
          doi = {10.3847/1538-4357/aa7556},
archivePrefix = {arXiv},
       eprint = {1702.02992},
 primaryClass = {astro-ph.HE},
       adsurl = {https://ui.adsabs.harvard.edu/abs/2017ApJ...843...40A}
}

@INPROCEEDINGS{NAIMA,
       author = {{Zabalza}, V.},
        title = "{Naima: a Python package for inference of particle distribution properties from nonthermal spectra}",
    booktitle = {34th International Cosmic Ray Conference (ICRC2015)},
         year = 2015,
       series = {International Cosmic Ray Conference},
       volume = {34},
        month = jul,
          eid = {922},
        pages = {922},
          doi = {10.22323/1.236.0922},
archivePrefix = {arXiv},
       eprint = {1509.03319},
 primaryClass = {astro-ph.HE},
       adsurl = {https://ui.adsabs.harvard.edu/abs/2015ICRC...34..922Z}
}

@ARTICLE{Porter2017,
       author = {{Porter}, T.~A. and {J{\'o}hannesson}, G. and {Moskalenko}, I.~V.},
        title = "{High-energy Gamma Rays from the Milky Way: Three-dimensional Spatial Models for the Cosmic-Ray and Radiation Field Densities in the Interstellar Medium}",
      journal = {\apj},
         year = 2017,
        month = sep,
       volume = {846},
       number = {1},
          eid = {67},
        pages = {67},
          doi = {10.3847/1538-4357/aa844d},
archivePrefix = {arXiv},
       eprint = {1708.00816},
 primaryClass = {astro-ph.HE},
       adsurl = {https://ui.adsabs.harvard.edu/abs/2017ApJ...846...67P}
}

@ARTICLE{Brand1993,
       author = {{Brand}, J. and {Blitz}, L.},
        title = "{The velocity field of the outer galaxy.}",
      journal = {\aap},
         year = 1993,
        month = aug,
       volume = {275},
        pages = {67-90},
       adsurl = {https://ui.adsabs.harvard.edu/abs/1993A&A...275...67B}
}

@article{Ohira2012,
    author = {Ohira, Yutaka and Yamazaki, Ryo and Kawanaka, Norita and Ioka, Kunihito},
    title = {Escape of cosmic-ray electrons from supernova remnants},
    journal = {Monthly Notices of the Royal Astronomical Society},
    volume = {427},
    number = {1},
    pages = {91-102},
    year = {2012},
    month = {11},
    issn = {0035-8711},
    doi = {10.1111/j.1365-2966.2012.21908.x},
    url = {https://doi.org/10.1111/j.1365-2966.2012.21908.x},
    eprint = {https://academic.oup.com/mnras/article-pdf/427/1/91/18242166/427-1-91.pdf},
}

@ARTICLE{Su2017,
       author = {{Su}, Yang and {Zhou}, Xin and {Yang}, Ji and {Chen}, Yang and {Chen}, Xuepeng and {Gong}, Yan and {Zhang}, Shaobo},
        title = "{Is HESS J1912+101 Associated with an Old Supernova Remnant?}",
      journal = {\apj},
         year = 2017,
        month = aug,
       volume = {845},
       number = {1},
          eid = {48},
        pages = {48},
          doi = {10.3847/1538-4357/aa7f2a},
archivePrefix = {arXiv},
       eprint = {1707.09807},
 primaryClass = {astro-ph.GA},
       adsurl = {https://ui.adsabs.harvard.edu/abs/2017ApJ...845...48S}
}

@ARTICLE{Dame2001,
       author = {{Dame}, T.~M. and {Hartmann}, Dap and {Thaddeus}, P.},
        title = "{The Milky Way in Molecular Clouds: A New Complete CO Survey}",
      journal = {\apj},
         year = 2001,
        month = feb,
       volume = {547},
       number = {2},
        pages = {792-813},
          doi = {10.1086/318388},
archivePrefix = {arXiv},
       eprint = {astro-ph/0009217},
 primaryClass = {astro-ph},
       adsurl = {https://ui.adsabs.harvard.edu/abs/2001ApJ...547..792D}
}

@ARTICLE{Mattox1996,
       author = {{Mattox}, J.~R. and {Bertsch}, D.~L. and {Chiang}, J. and {Dingus}, B.~L. and {Digel}, S.~W. and {Esposito}, J.~A. and {Fierro}, J.~M. and {Hartman}, R.~C. and {Hunter}, S.~D. and {Kanbach}, G. and {Kniffen}, D.~A. and {Lin}, Y.~C. and {Macomb}, D.~J. and {Mayer-Hasselwander}, H.~A. and {Michelson}, P.~F. and {von Montigny}, C. and {Mukherjee}, R. and {Nolan}, P.~L. and {Ramanamurthy}, P.~V. and {Schneid}, E. and {Sreekumar}, P. and {Thompson}, D.~J. and {Willis}, T.~D.},
        title = "{The Likelihood Analysis of EGRET Data}",
      journal = {\apj},
         year = 1996,
        month = apr,
       volume = {461},
        pages = {396},
          doi = {10.1086/177068},
       adsurl = {https://ui.adsabs.harvard.edu/abs/1996ApJ...461..396M}
}

@ARTICLE{4FGLDR4,
       author = {{Ballet}, J. and {Bruel}, P. and {Burnett}, T.~H. and {Lott}, B. and {The Fermi-LAT collaboration}},
        title = "{Fermi Large Area Telescope Fourth Source Catalog Data Release 4 (4FGL-DR4)}",
      journal = {arXiv e-prints},
         year = 2023,
        month = jul,
          eid = {arXiv:2307.12546},
        pages = {arXiv:2307.12546},
          doi = {10.48550/arXiv.2307.12546},
archivePrefix = {arXiv},
       eprint = {2307.12546},
 primaryClass = {astro-ph.HE},
       adsurl = {https://ui.adsabs.harvard.edu/abs/2023arXiv230712546B}
}

@ARTICLE{4FGLDR3,
       author = {{Abdollahi}, S. and {Acero}, F. and {Baldini}, L. and {Ballet}, J. and {Bastieri}, D. and {Bellazzini}, R. and {Berenji}, B. and {Berretta}, A. and {Bissaldi}, E. and {Blandford}, R.~D. and {Bloom}, E. and {Bonino}, R. and {Brill}, A. and {Britto}, R.~J. and {Bruel}, P. and {Burnett}, T.~H. and {Buson}, S. and {Zaharijas}, G.},
        title = "{Incremental Fermi Large Area Telescope Fourth Source Catalog}",
      journal = {\apjs},
         year = 2022,
        month = jun,
       volume = {260},
       number = {2},
          eid = {53},
        pages = {53},
          doi = {10.3847/1538-4365/ac6751},
archivePrefix = {arXiv},
       eprint = {2201.11184},
 primaryClass = {astro-ph.HE},
       adsurl = {https://ui.adsabs.harvard.edu/abs/2022ApJS..260...53A}
}

@software{Fermitools,
       author = {{Fermi Science Support Development Team}},
        title = "{Fermitools: Fermi Science Tools}",
 howpublished = {Astrophysics Source Code Library, record ascl:1905.011},
         year = 2019,
        month = may,
          eid = {ascl:1905.011},
archivePrefix = {ascl},
       eprint = {1905.011},
       adsurl = {https://ui.adsabs.harvard.edu/abs/2019ascl.soft05011F}
}

@INPROCEEDINGS{Wood2017,
       author = {{Wood}, M. and {Caputo}, R. and {Charles}, E. and {Di Mauro}, M. and {Magill}, J. and {Perkins}, J.~S. and {Fermi-LAT Collaboration}},
        title = "{Fermipy: An open-source Python package for analysis of Fermi-LAT Data}",
    booktitle = {35th International Cosmic Ray Conference (ICRC2017)},
         year = 2017,
       series = {International Cosmic Ray Conference},
       volume = {301},
        month = jul,
          eid = {824},
        pages = {824},
          doi = {10.22323/1.301.0824},
archivePrefix = {arXiv},
       eprint = {1707.09551},
 primaryClass = {astro-ph.IM},
       adsurl = {https://ui.adsabs.harvard.edu/abs/2017ICRC...35..824W}
}

@ARTICLE{Zhang2020,
       author = {{Zhang}, Hai-Ming and {Xi}, Shao-Qiang and {Liu}, Ruo-Yu and {Xin}, Yu-Liang and {Liu}, Siming and {Wang}, Xiang-Yu},
        title = "{Discovery of a Spatially Extended GeV Source in the Vicinity of the TeV Halo Candidate 2HWC J1912+099: a TeV Halo or Supernova Remnant?}",
      journal = {\apj},
         year = 2020,
        month = jan,
       volume = {889},
       number = {1},
          eid = {12},
        pages = {12},
          doi = {10.3847/1538-4357/ab5af6},
archivePrefix = {arXiv},
       eprint = {1909.13185},
 primaryClass = {astro-ph.HE},
       adsurl = {https://ui.adsabs.harvard.edu/abs/2020ApJ...889...12Z}
}

@ARTICLE{Zeng2021,
       author = {{Zeng}, Houdun and {Xin}, Yuliang and {Zhang}, Shuinai and {Liu}, Siming},
        title = "{TeV Cosmic-Ray Nucleus Acceleration in Shell-type Supernova Remnants with Hard {\ensuremath{\gamma}}-Ray Spectra}",
      journal = {\apj},
         year = 2021,
        month = mar,
       volume = {910},
       number = {1},
          eid = {78},
        pages = {78},
          doi = {10.3847/1538-4357/abe37e},
archivePrefix = {arXiv},
       eprint = {2102.03465},
 primaryClass = {astro-ph.HE},
       adsurl = {https://ui.adsabs.harvard.edu/abs/2021ApJ...910...78Z}
}

@ARTICLE{Sun2022,
       author = {{Sun}, Xiao-Na and {Yang}, Rui-Zhi and {Liang}, En-Wei},
        title = "{Diffuse GeV emission in the field of HESS J1912+101 revisited}",
      journal = {\aap},
         year = 2022,
        month = mar,
       volume = {659},
          eid = {A83},
        pages = {A83},
          doi = {10.1051/0004-6361/202142394},
       adsurl = {https://ui.adsabs.harvard.edu/abs/2022A&A...659A..83S}
}

@ARTICLE{Li2023,
       author = {{Li}, Yuan and {Liu}, Siming and {He}, Yu},
        title = "{The Nature of {\ensuremath{\gamma}}-Ray Emission from HESS J1912+101}",
      journal = {\apj},
         year = 2023,
        month = aug,
       volume = {953},
       number = {1},
          eid = {100},
        pages = {100},
          doi = {10.3847/1538-4357/ace344},
       adsurl = {https://ui.adsabs.harvard.edu/abs/2023ApJ...953..100L}
}

@ARTICLE{HESS2018,
       author = {{H.~E.~S.~S. Collaboration} and {Abdalla}, H. and {Abramowski}, A. and {Aharonian}, F. and {Ait Benkhali}, F. and {Akhperjanian}, A.~G. and {Andersson}, T. and {Ang{\"u}ner}, E.~O. and {Arakawa}, M. and {Arrieta}, M. and {Aubert}, P. and {Backes}, M. and {Balzer}, A. and {Barnard}, M. and {Becherini}, Y. and {Becker Tjus}, J. and {Berge}, D. and {Bernhard}, S. and {Bernl{\"o}hr}, K. and {Blackwell}, R. and {B{\"o}ttcher}, M. and {Boisson}, C. and {Bolmont}, J. and {Bonnefoy}, S. and {Bordas}, P. and {Bregeon}, J. and {Brun}, F. and {Brun}, P. and {Bryan}, M. and {B{\"u}chele}, M. and {Bulik}, T. and {Capasso}, M. and {Carr}, J. and {Casanova}, S. and {Cerruti}, M. and {Chakraborty}, N. and {Chaves}, R.~C.~G. and {Chen}, A. and {Chevalier}, J. and {Coffaro}, M. and {Colafrancesco}, S. and {Cologna}, G. and {Condon}, B. and {Conrad}, J. and {Cui}, Y. and {Davids}, I.~D. and {Decock}, J. and {Degrange}, B. and {Deil}, C. and {Devin}, J. and {deWilt}, P. and {Dirson}, L. and {Djannati-Ata{\"\i}}, A. and {Domainko}, W. and {Donath}, A. and {Drury}, L.~O. 'C. and {Dutson}, K. and {Dyks}, J. and {Edwards}, T. and {Egberts}, K. and {Eger}, P. and {Ernenwein}, J. -P. and {Eschbach}, S. and {Farnier}, C. and {Fegan}, S. and {Fernandes}, M.~V. and {Fiasson}, A. and {Fontaine}, G. and {F{\"o}rster}, A. and {Funk}, S. and {F{\"u}{\ss}ling}, M. and {Gabici}, S. and {Gajdus}, M. and {Gallant}, Y.~A. and {Garrigoux}, T. and {Giavitto}, G. and {Giebels}, B. and {Glicenstein}, J.~F. and {Gottschall}, D. and {Goyal}, A. and {Grondin}, M. -H. and {Hahn}, J. and {Haupt}, M. and {Hawkes}, J. and {Heinzelmann}, G. and {Henri}, G. and {Hermann}, G. and {Hervet}, O. and {Hinton}, J.~A. and {Hofmann}, W. and {Hoischen}, C. and {Holch}, T.~L. and {Holler}, M. and {Horns}, D. and {Ivascenko}, A. and {Iwasaki}, H. and {Jacholkowska}, A. and {Jamrozy}, M. and {Janiak}, M. and {Jankowsky}, D. and {Jankowsky}, F. and {Jingo}, M. and {Jogler}, T. and {Jouvin}, L. and {Jung-Richardt}, I. and {Kastendieck}, M.~A. and {Katarzy{\'n}ski}, K. and {Katsuragawa}, M. and {Katz}, U. and {Kerszberg}, D. and {Khangulyan}, D. and {Kh{\'e}lifi}, B. and {King}, J. and {Klepser}, S. and {Klochkov}, D. and {Klu{\'z}niak}, W. and {Kolitzus}, D. and {Komin}, Nu. and {Kosack}, K. and {Krakau}, S. and {Kraus}, M. and {Kr{\"u}ger}, P.~P. and {Laffon}, H. and {Lamanna}, G. and {Lau}, J. and {Lees}, J. -P. and {Lefaucheur}, J. and {Lefranc}, V. and {Lemi{\`e}re}, A. and {Lemoine-Goumard}, M. and {Lenain}, J. -P. and {Leser}, E. and {Lohse}, T. and {Lorentz}, M. and {Liu}, R. and {L{\'o}pez-Coto}, R. and {Lypova}, I. and {Marandon}, V. and {Marcowith}, A. and {Mariaud}, C. and {Marx}, R. and {Maurin}, G. and {Maxted}, N. and {Mayer}, M. and {Meintjes}, P.~J. and {Meyer}, M. and {Mitchell}, A.~M.~W. and {Moderski}, R. and {Mohamed}, M. and {Mohrmann}, L. and {Mor{\r{a}}}, K. and {Moulin}, E. and {Murach}, T. and {Nakashima}, S. and {de Naurois}, M. and {Niederwanger}, F. and {Niemiec}, J. and {Oakes}, L. and {O'Brien}, P. and {Odaka}, H. and {{\"O}ttl}, S. and {Ohm}, S. and {Ostrowski}, M. and {Oya}, I. and {Padovani}, M. and {Panter}, M. and {Parsons}, R.~D. and {Pekeur}, N.~W. and {Pelletier}, G. and {Perennes}, C. and {Petrucci}, P. -O. and {Peyaud}, B. and {Piel}, Q. and {Pita}, S. and {Poon}, H. and {Prokhorov}, D. and {Prokoph}, H. and {P{\"u}hlhofer}, G. and {Punch}, M. and {Quirrenbach}, A. and {Raab}, S. and {Reimer}, A. and {Reimer}, O. and {Renaud}, M. and {de los Reyes}, R. and {Richter}, S. and {Rieger}, F. and {Romoli}, C. and {Rowell}, G. and {Rudak}, B. and {Rulten}, C.~B. and {Sahakian}, V. and {Saito}, S. and {Salek}, D. and {Sanchez}, D.~A. and {Santangelo}, A. and {Sasaki}, M. and {Schlickeiser}, R. and {Sch{\"u}ssler}, F. and {Schulz}, A.},
        title = "{A search for new supernova remnant shells in the Galactic plane with H.E.S.S.}",
      journal = {\aap},
         year = 2018,
        month = apr,
       volume = {612},
          eid = {A8},
        pages = {A8},
          doi = {10.1051/0004-6361/201730737},
archivePrefix = {arXiv},
       eprint = {1801.06020},
 primaryClass = {astro-ph.HE},
       adsurl = {https://ui.adsabs.harvard.edu/abs/2018A&A...612A...8H}
}

@ARTICLE{HESS2008,
       author = {{Aharonian}, F. and {Akhperjanian}, A.~G. and {Barres de Almeida}, U. and {Bazer-Bachi}, A.~R. and {Behera}, B. and {Beilicke}, M. and {Benbow}, W. and {Bernl{\"o}hr}, K. and {Boisson}, C. and {Bolz}, O. and {Borrel}, V. and {Braun}, I. and {Brion}, E. and {Brown}, A.~M. and {B{\"u}hler}, R. and {Bulik}, T. and {B{\"u}sching}, I. and {Boutelier}, T. and {Carrigan}, S. and {Chadwick}, P.~M. and {Chounet}, L. -M. and {Clapson}, A.~C. and {Coignet}, G. and {Cornils}, R. and {Costamante}, L. and {Dalton}, M. and {Degrange}, B. and {Dickinson}, H.~J. and {Djannati-Ata{\"\i}}, A. and {Domainko}, W. and {O'C. Drury}, L. and {Dubois}, F. and {Dubus}, G. and {Dyks}, J. and {Egberts}, K. and {Emmanoulopoulos}, D. and {Espigat}, P. and {Farnier}, C. and {Feinstein}, F. and {Fiasson}, A. and {F{\"o}rster}, A. and {Fontaine}, G. and {Funk}, Seb. and {F{\"u}{\ss}ling}, M. and {Gallant}, Y.~A. and {Giebels}, B. and {Glicenstein}, J.~F. and {Gl{\"u}ck}, B. and {Goret}, P. and {Hadjichristidis}, C. and {Hauser}, D. and {Hauser}, M. and {Heinzelmann}, G. and {Henri}, G. and {Hermann}, G. and {Hinton}, J.~A. and {Hoffmann}, A. and {Hofmann}, W. and {Holleran}, M. and {Hoppe}, S. and {Horns}, D. and {Jacholkowska}, A. and {de Jager}, O.~C. and {Jung}, I. and {Katarzy{\'n}ski}, K. and {Kendziorra}, E. and {Kerschhaggl}, M. and {Kh{\'e}lifi}, B. and {Keogh}, D. and {Komin}, Nu. and {Kosack}, K. and {Lamanna}, G. and {Latham}, I.~J. and {Lemi{\`e}re}, A. and {Lemoine-Goumard}, M. and {Lenain}, J. -P. and {Lohse}, T. and {Martin}, J.~M. and {Martineau-Huynh}, O. and {Marcowith}, A. and {Masterson}, C. and {Maurin}, D. and {Maurin}, G. and {McComb}, T.~J.~L. and {Moderski}, R. and {Moulin}, E. and {de Naurois}, M. and {Nedbal}, D. and {Nolan}, S.~J. and {Ohm}, S. and {Olive}, J. -P. and {de O{\~n}a Wilhelmi}, E. and {Orford}, K.~J. and {Osborne}, J.~L. and {Ostrowski}, M. and {Panter}, M. and {Pedaletti}, G. and {Pelletier}, G. and {Petrucci}, P. -O. and {Pita}, S. and {P{\"u}hlhofer}, G. and {Punch}, M. and {Raubenheimer}, B.~C. and {Raue}, M. and {Rayner}, S.~M. and {Reimer}, O. and {Renaud}, M. and {Ripken}, J. and {Rob}, L. and {Rolland}, L. and {Rosier-Lees}, S. and {Rowell}, G. and {Rudak}, B. and {Ruppel}, J. and {Sahakian}, V. and {Santangelo}, A. and {Schlickeiser}, R. and {Sch{\"o}ck}, F. and {Schr{\"o}der}, R. and {Schwanke}, U. and {Schwarzburg}, S. and {Schwemmer}, S. and {Shalchi}, A. and {Sol}, H. and {Spangler}, D. and {Stawarz}, {\L}. and {Steenkamp}, R. and {Stegmann}, C. and {Superina}, G. and {Tam}, P.~H. and {Tavernet}, J. -P. and {Terrier}, R. and {van Eldik}, C. and {Vasileiadis}, G. and {Venter}, C. and {Vialle}, J.~P. and {Vincent}, P. and {Vivier}, M. and {V{\"o}lk}, H.~J. and {Volpe}, F. and {Wagner}, S.~J. and {Ward}, M. and {Zdziarski}, A.~A. and {Zech}, A.},
        title = "{Discovery of very-high-energy {\ensuremath{\gamma}}-ray emission from the vicinity of PSR J1913+1011 with HESS}",
      journal = {\aap},
         year = 2008,
        month = jun,
       volume = {484},
       number = {2},
        pages = {435-440},
          doi = {10.1051/0004-6361:20078715},
archivePrefix = {arXiv},
       eprint = {0802.3841},
 primaryClass = {astro-ph},
       adsurl = {https://ui.adsabs.harvard.edu/abs/2008A&A...484..435A}
}

@ARTICLE{Umemoto2017,
       author = {{Umemoto}, Tomofumi and {Minamidani}, Tetsuhiro and {Kuno}, Nario and {Fujita}, Shinji and {Matsuo}, Mitsuhiro and {Nishimura}, Atsushi and {Torii}, Kazufumi and {Tosaki}, Tomoka and {Kohno}, Mikito and {Kuriki}, Mika and {Tsuda}, Yuya and {Hirota}, Akihiko and {Ohashi}, Satoshi and {Yamagishi}, Mitsuyoshi and {Handa}, Toshihiro and {Nakanishi}, Hiroyuki and {Omodaka}, Toshihiro and {Koide}, Nagito and {Matsumoto}, Naoko and {Onishi}, Toshikazu and {Tokuda}, Kazuki and {Seta}, Masumichi and {Kobayashi}, Yukinori and {Tachihara}, Kengo and {Sano}, Hidetoshi and {Hattori}, Yusuke and {Onodera}, Sachiko and {Oasa}, Yumiko and {Kamegai}, Kazuhisa and {Tsuboi}, Masato and {Sofue}, Yoshiaki and {Higuchi}, Aya E. and {Chibueze}, James O. and {Mizuno}, Norikazu and {Honma}, Mareki and {Muller}, Erik and {Inoue}, Tsuyoshi and {Morokuma-Matsui}, Kana and {Shinnaga}, Hiroko and {Ozawa}, Takeaki and {Takahashi}, Ryo and {Yoshiike}, Satoshi and {Costes}, Jean and {Kuwahara}, Sho},
        title = "{FOREST unbiased Galactic plane imaging survey with the Nobeyama 45 m telescope (FUGIN). I. Project overview and initial results}",
      journal = {\pasj},
         year = 2017,
        month = oct,
       volume = {69},
       number = {5},
          eid = {78},
        pages = {78},
          doi = {10.1093/pasj/psx061},
archivePrefix = {arXiv},
       eprint = {1707.05981},
 primaryClass = {astro-ph.GA},
       adsurl = {https://ui.adsabs.harvard.edu/abs/2017PASJ...69...78U}
}

@ARTICLE{Atwood2009,
       author = {{Atwood}, W.~B. and {Abdo}, A.~A. and {Ackermann}, M. and {Althouse}, W. and {Anderson}, B. and {Axelsson}, M. and {Baldini}, L. and {Ballet}, J. and {Band}, D.~L. and {Barbiellini}, G. and {Bartelt}, J. and {Bastieri}, D. and {Baughman}, B.~M. and {Bechtol}, K. and {B{\'e}d{\'e}r{\`e}de}, D. and {Bellardi}, F. and {Bellazzini}, R. and {Berenji}, B. and {Bignami}, G.~F. and {Bisello}, D. and {Bissaldi}, E. and {Blandford}, R.~D. and {Bloom}, E.~D. and {Bogart}, J.~R. and {Bonamente}, E. and {Bonnell}, J. and {Borgland}, A.~W. and {Bouvier}, A. and {Bregeon}, J. and {Brez}, A. and {Brigida}, M. and {Bruel}, P. and {Burnett}, T.~H. and {Busetto}, G. and {Caliandro}, G.~A. and {Cameron}, R.~A. and {Caraveo}, P.~A. and {Carius}, S. and {Carlson}, P. and {Casandjian}, J.~M. and {Cavazzuti}, E. and {Ceccanti}, M. and {Cecchi}, C. and {Charles}, E. and {Chekhtman}, A. and {Cheung}, C.~C. and {Chiang}, J. and {Chipaux}, R. and {Cillis}, A.~N. and {Ciprini}, S. and {Claus}, R. and {Cohen-Tanugi}, J. and {Condamoor}, S. and {Conrad}, J. and {Corbet}, R. and {Corucci}, L. and {Costamante}, L. and {Cutini}, S. and {Davis}, D.~S. and {Decotigny}, D. and {DeKlotz}, M. and {Dermer}, C.~D. and {de Angelis}, A. and {Digel}, S.~W. and {do Couto e Silva}, E. and {Drell}, P.~S. and {Dubois}, R. and {Dumora}, D. and {Edmonds}, Y. and {Fabiani}, D. and {Farnier}, C. and {Favuzzi}, C. and {Flath}, D.~L. and {Fleury}, P. and {Focke}, W.~B. and {Funk}, S. and {Fusco}, P. and {Gargano}, F. and {Gasparrini}, D. and {Gehrels}, N. and {Gentit}, F. -X. and {Germani}, S. and {Giebels}, B. and {Giglietto}, N. and {Giommi}, P. and {Giordano}, F. and {Glanzman}, T. and {Godfrey}, G. and {Grenier}, I.~A. and {Grondin}, M. -H. and {Grove}, J.~E. and {Guillemot}, L. and {Guiriec}, S. and {Haller}, G. and {Harding}, A.~K. and {Hart}, P.~A. and {Hays}, E. and {Healey}, S.~E. and {Hirayama}, M. and {Hjalmarsdotter}, L. and {Horn}, R. and {Hughes}, R.~E. and {J{\'o}hannesson}, G. and {Johansson}, G. and {Johnson}, A.~S. and {Johnson}, R.~P. and {Johnson}, T.~J. and {Johnson}, W.~N. and {Kamae}, T. and {Katagiri}, H. and {Kataoka}, J. and {Kavelaars}, A. and {Kawai}, N. and {Kelly}, H. and {Kerr}, M. and {Klamra}, W. and {Kn{\"o}dlseder}, J. and {Kocian}, M.~L. and {Komin}, N. and {Kuehn}, F. and {Kuss}, M. and {Landriu}, D. and {Latronico}, L. and {Lee}, B. and {Lee}, S. -H. and {Lemoine-Goumard}, M. and {Lionetto}, A.~M. and {Longo}, F. and {Loparco}, F. and {Lott}, B. and {Lovellette}, M.~N. and {Lubrano}, P. and {Madejski}, G.~M. and {Makeev}, A. and {Marangelli}, B. and {Massai}, M.~M. and {Mazziotta}, M.~N. and {McEnery}, J.~E. and {Menon}, N. and {Meurer}, C. and {Michelson}, P.~F. and {Minuti}, M. and {Mirizzi}, N. and {Mitthumsiri}, W. and {Mizuno}, T. and {Moiseev}, A.~A. and {Monte}, C. and {Monzani}, M.~E. and {Moretti}, E. and {Morselli}, A. and {Moskalenko}, I.~V. and {Murgia}, S. and {Nakamori}, T. and {Nishino}, S. and {Nolan}, P.~L. and {Norris}, J.~P. and {Nuss}, E. and {Ohno}, M. and {Ohsugi}, T. and {Omodei}, N. and {Orlando}, E. and {Ormes}, J.~F. and {Paccagnella}, A. and {Paneque}, D. and {Panetta}, J.~H. and {Parent}, D. and {Pearce}, M. and {Pepe}, M. and {Perazzo}, A. and {Pesce-Rollins}, M. and {Picozza}, P. and {Pieri}, L. and {Pinchera}, M. and {Piron}, F. and {Porter}, T.~A. and {Poupard}, L. and {Rain{\`o}}, S. and {Rando}, R. and {Rapposelli}, E. and {Razzano}, M. and {Reimer}, A. and {Reimer}, O. and {Reposeur}, T. and {Reyes}, L.~C. and {Ritz}, S. and {Rochester}, L.~S. and {Rodriguez}, A.~Y. and {Romani}, R.~W. and {Roth}, M. and {Russell}, J.~J. and {Ryde}, F. and {Sabatini}, S. and {Sadrozinski}, H.~F. -W. and {Sanchez}, D. and {Sander}, A. and {Sapozhnikov}, L. and {Parkinson}, P.~M. Saz and {Scargle}, J.~D. and {Schalk}, T.~L. and {Scolieri}, G.},
        title = "{The Large Area Telescope on the Fermi Gamma-Ray Space Telescope Mission}",
      journal = {\apj},
         year = 2009,
        month = jun,
       volume = {697},
       number = {2},
        pages = {1071-1102},
          doi = {10.1088/0004-637X/697/2/1071},
archivePrefix = {arXiv},
       eprint = {0902.1089},
 primaryClass = {astro-ph.IM},
       adsurl = {https://ui.adsabs.harvard.edu/abs/2009ApJ...697.1071A}
}

@ARTICLE{HESS2018_GPS,
       author = {{H.~E.~S.~S. Collaboration} and {Abdalla}, H. and {Abramowski}, A. and {Aharonian}, F. and {Ait Benkhali}, F. and {Ang{\"u}ner}, E.~O. and {Arakawa}, M. and {Arrieta}, M. and {Aubert}, P. and {Backes}, M. and {Balzer}, A. and {Barnard}, M. and {Becherini}, Y. and {Becker Tjus}, J. and {Berge}, D. and {Bernhard}, S. and {Bernl{\"o}hr}, K. and {Blackwell}, R. and {B{\"o}ttcher}, M. and {Boisson}, C. and {Bolmont}, J. and {Bonnefoy}, S. and {Bordas}, P. and {Bregeon}, J. and {Brun}, F. and {Brun}, P. and {Bryan}, M. and {B{\"u}chele}, M. and {Bulik}, T. and {Capasso}, M. and {Carrigan}, S. and {Caroff}, S. and {Carosi}, A. and {Casanova}, S. and {Cerruti}, M. and {Chakraborty}, N. and {Chaves}, R.~C.~G. and {Chen}, A. and {Chevalier}, J. and {Colafrancesco}, S. and {Condon}, B. and {Conrad}, J. and {Davids}, I.~D. and {Decock}, J. and {Deil}, C. and {Devin}, J. and {deWilt}, P. and {Dirson}, L. and {Djannati-Ata{\"\i}}, A. and {Domainko}, W. and {Donath}, A. and {Drury}, L.~O. 'C. and {Dutson}, K. and {Dyks}, J. and {Edwards}, T. and {Egberts}, K. and {Eger}, P. and {Emery}, G. and {Ernenwein}, J. -P. and {Eschbach}, S. and {Farnier}, C. and {Fegan}, S. and {Fernandes}, M.~V. and {Fiasson}, A. and {Fontaine}, G. and {F{\"o}rster}, A. and {Funk}, S. and {F{\"u}{\ss}ling}, M. and {Gabici}, S. and {Gallant}, Y.~A. and {Garrigoux}, T. and {Gast}, H. and {Gat{\'e}}, F. and {Giavitto}, G. and {Giebels}, B. and {Glawion}, D. and {Glicenstein}, J.~F. and {Gottschall}, D. and {Grondin}, M. -H. and {Hahn}, J. and {Haupt}, M. and {Hawkes}, J. and {Heinzelmann}, G. and {Henri}, G. and {Hermann}, G. and {Hinton}, J.~A. and {Hofmann}, W. and {Hoischen}, C. and {Holch}, T.~L. and {Holler}, M. and {Horns}, D. and {Ivascenko}, A. and {Iwasaki}, H. and {Jacholkowska}, A. and {Jamrozy}, M. and {Jankowsky}, D. and {Jankowsky}, F. and {Jingo}, M. and {Jouvin}, L. and {Jung-Richardt}, I. and {Kastendieck}, M.~A. and {Katarzy{\'n}ski}, K. and {Katsuragawa}, M. and {Katz}, U. and {Kerszberg}, D. and {Khangulyan}, D. and {Kh{\'e}lifi}, B. and {King}, J. and {Klepser}, S. and {Klochkov}, D. and {Klu{\'z}niak}, W. and {Komin}, Nu. and {Kosack}, K. and {Krakau}, S. and {Kraus}, M. and {Kr{\"u}ger}, P.~P. and {Laffon}, H. and {Lamanna}, G. and {Lau}, J. and {Lees}, J. -P. and {Lefaucheur}, J. and {Lemi{\`e}re}, A. and {Lemoine-Goumard}, M. and {Lenain}, J. -P. and {Leser}, E. and {Lohse}, T. and {Lorentz}, M. and {Liu}, R. and {L{\'o}pez-Coto}, R. and {Lypova}, I. and {Marandon}, V. and {Malyshev}, D. and {Marcowith}, A. and {Mariaud}, C. and {Marx}, R. and {Maurin}, G. and {Maxted}, N. and {Mayer}, M. and {Meintjes}, P.~J. and {Meyer}, M. and {Mitchell}, A.~M.~W. and {Moderski}, R. and {Mohamed}, M. and {Mohrmann}, L. and {Mor{\r{a}}}, K. and {Moulin}, E. and {Murach}, T. and {Nakashima}, S. and {de Naurois}, M. and {Ndiyavala}, H. and {Niederwanger}, F. and {Niemiec}, J. and {Oakes}, L. and {O'Brien}, P. and {Odaka}, H. and {Ohm}, S. and {Ostrowski}, M. and {Oya}, I. and {Padovani}, M. and {Panter}, M. and {Parsons}, R.~D. and {Paz Arribas}, M. and {Pekeur}, N.~W. and {Pelletier}, G. and {Perennes}, C. and {Petrucci}, P. -O. and {Peyaud}, B. and {Piel}, Q. and {Pita}, S. and {Poireau}, V. and {Poon}, H. and {Prokhorov}, D. and {Prokoph}, H. and {P{\"u}hlhofer}, G. and {Punch}, M. and {Quirrenbach}, A. and {Raab}, S. and {Rauth}, R. and {Reimer}, A. and {Reimer}, O. and {Renaud}, M. and {de los Reyes}, R. and {Rieger}, F. and {Rinchiuso}, L. and {Romoli}, C. and {Rowell}, G. and {Rudak}, B. and {Rulten}, C.~B. and {Safi-Harb}, S. and {Sahakian}, V. and {Saito}, S. and {Sanchez}, D.~A. and {Santangelo}, A. and {Sasaki}, M. and {Schandri}, M. and {Schlickeiser}, R. and {Sch{\"u}ssler}, F. and {Schulz}, A. and {Schwanke}, U. and {Schwemmer}, S.},
        title = "{The H.E.S.S. Galactic plane survey}",
      journal = {\aap},
         year = 2018,
        month = apr,
       volume = {612},
          eid = {A1},
        pages = {A1},
          doi = {10.1051/0004-6361/201732098},
archivePrefix = {arXiv},
       eprint = {1804.02432},
 primaryClass = {astro-ph.HE},
       adsurl = {https://ui.adsabs.harvard.edu/abs/2018A&A...612A...1H}
}

@ARTICLE{LHAASO_1stCat,
       author = {{Cao}, Zhen and {Aharonian}, F. and {An}, Q. and {Axikegu} and {Bai}, Y.~X. and {Bao}, Y.~W. and {Bastieri}, D. and {Bi}, X.~J. and {Bi}, Y.~J. and {Cai}, J.~T. and {Cao}, Q. and {Cao}, W.~Y. and {Cao}, Zhe and {Chang}, J. and {Chang}, J.~F. and {Chen}, A.~M. and {Chen}, E.~S. and {Chen}, Liang and {Chen}, Lin and {Chen}, Long and {Chen}, M.~J. and {Chen}, M.~L. and {Chen}, Q.~H. and {Chen}, S.~H. and {Chen}, S.~Z. and {Chen}, T.~L. and {Chen}, Y. and {Cheng}, N. and {Cheng}, Y.~D. and {Cui}, M.~Y. and {Cui}, S.~W. and {Cui}, X.~H. and {Cui}, Y.~D. and {Dai}, B.~Z. and {Dai}, H.~L. and {Dai}, Z.~G. and {Danzengluobu} and {Della Volpe}, D. and {Dong}, X.~Q. and {Duan}, K.~K. and {Fan}, J.~H. and {Fan}, Y.~Z. and {Fang}, J. and {Fang}, K. and {Feng}, C.~F. and {Feng}, L. and {Feng}, S.~H. and {Feng}, X.~T. and {Feng}, Y.~L. and {Gabici}, S. and {Gao}, B. and {Gao}, C.~D. and {Gao}, L.~Q. and {Gao}, Q. and {Gao}, W. and {Gao}, W.~K. and {Ge}, M.~M. and {Geng}, L.~S. and {Giacinti}, G. and {Gong}, G.~H. and {Gou}, Q.~B. and {Gu}, M.~H. and {Guo}, F.~L. and {Guo}, X.~L. and {Guo}, Y.~Q. and {Guo}, Y.~Y. and {Han}, Y.~A. and {He}, H.~H. and {He}, H.~N. and {He}, J.~Y. and {He}, X.~B. and {He}, Y. and {Heller}, M. and {Hor}, Y.~K. and {Hou}, B.~W. and {Hou}, C. and {Hou}, X. and {Hu}, H.~B. and {Hu}, Q. and {Hu}, S.~C. and {Huang}, D.~H. and {Huang}, T.~Q. and {Huang}, W.~J. and {Huang}, X.~T. and {Huang}, X.~Y. and {Huang}, Y. and {Huang}, Z.~C. and {Ji}, X.~L. and {Jia}, H.~Y. and {Jia}, K. and {Jiang}, K. and {Jiang}, X.~W. and {Jiang}, Z.~J. and {Jin}, M. and {Kang}, M.~M. and {Ke}, T. and {Kuleshov}, D. and {Kurinov}, K. and {Li}, B.~B. and {Li}, Cheng and {Li}, Cong and {Li}, D. and {Li}, F. and {Li}, H.~B. and {Li}, H.~C. and {Li}, H.~Y. and {Li}, J. and {Li}, Jian and {Li}, Jie and {Li}, K. and {Li}, W.~L. and {Li}, W.~L. and {Li}, X.~R. and {Li}, Xin and {Li}, Y.~Z. and {Li}, Zhe and {Li}, Zhuo and {Liang}, E.~W. and {Liang}, Y.~F. and {Lin}, S.~J. and {Liu}, B. and {Liu}, C. and {Liu}, D. and {Liu}, H. and {Liu}, H.~D. and {Liu}, J. and {Liu}, J.~L. and {Liu}, J.~Y. and {Liu}, M.~Y. and {Liu}, R.~Y. and {Liu}, S.~M. and {Liu}, W. and {Liu}, Y. and {Liu}, Y.~N. and {Lu}, R. and {Luo}, Q. and {Lv}, H.~K. and {Ma}, B.~Q. and {Ma}, L.~L. and {Ma}, X.~H. and {Mao}, J.~R. and {Min}, Z. and {Mitthumsiri}, W. and {Mu}, H.~J. and {Nan}, Y.~C. and {Neronov}, A. and {Ou}, Z.~W. and {Pang}, B.~Y. and {Pattarakijwanich}, P. and {Pei}, Z.~Y. and {Qi}, M.~Y. and {Qi}, Y.~Q. and {Qiao}, B.~Q. and {Qin}, J.~J. and {Ruffolo}, D. and {S{\'a}iz}, A. and {Semikoz}, D. and {Shao}, C.~Y. and {Shao}, L. and {Shchegolev}, O. and {Sheng}, X.~D. and {Shu}, F.~W. and {Song}, H.~C. and {Stenkin}, Yu. V. and {Stepanov}, V. and {Su}, Y. and {Sun}, Q.~N. and {Sun}, X.~N. and {Sun}, Z.~B. and {Tam}, P.~H.~T. and {Tang}, Q.~W. and {Tang}, Z.~B. and {Tian}, W.~W. and {Wang}, C. and {Wang}, C.~B. and {Wang}, G.~W. and {Wang}, H.~G. and {Wang}, H.~H. and {Wang}, J.~C. and {Wang}, K. and {Wang}, L.~P. and {Wang}, L.~Y. and {Wang}, P.~H. and {Wang}, R. and {Wang}, W. and {Wang}, X.~G. and {Wang}, X.~Y. and {Wang}, Y. and {Wang}, Y.~D. and {Wang}, Y.~J. and {Wang}, Z.~H. and {Wang}, Z.~X. and {Wang}, Zhen and {Wang}, Zheng and {Wei}, D.~M. and {Wei}, J.~J. and {Wei}, Y.~J. and {Wen}, T. and {Wu}, C.~Y. and {Wu}, H.~R.},
        title = "{The First LHAASO Catalog of Gamma-Ray Sources}",
      journal = {\apjs},
         year = 2024,
        month = mar,
       volume = {271},
       number = {1},
          eid = {25},
        pages = {25},
          doi = {10.3847/1538-4365/acfd29},
archivePrefix = {arXiv},
       eprint = {2305.17030},
 primaryClass = {astro-ph.HE},
       adsurl = {https://ui.adsabs.harvard.edu/abs/2024ApJS..271...25C}
}

@ARTICLE{LHAASO_Inst,
       author = {{Ma}, Xin-Hua and {Bi}, Yu-Jiang and {Cao}, Zhen and {Chen}, Ming-Jun and {Chen}, Song-Zhan and {Cheng}, Yao-Dong and {Gong}, Guang-Hua and {Gu}, Min-Hao and {He}, Hui-Hai and {Hou}, Chao and {Huang}, Wen-Hao and {Huang}, Xing-Tao and {Liu}, Cheng and {Shchegolev}, Oleg and {Sheng}, Xiang-Dong and {Stenkin}, Yuri and {Wu}, Chao-Yong and {Wu}, Han-Rong and {Wu}, Sha and {Xiao}, Gang and {Yao}, Zhi-Guo and {Zhang}, Shou-Shan and {Zhang}, Yi and {Zuo}, Xiong},
        title = "{Chapter 1 LHAASO Instruments and Detector technology}",
      journal = {Chinese Physics C},
         year = 2022,
        month = mar,
       volume = {46},
       number = {3},
          eid = {030001},
        pages = {030001},
          doi = {10.1088/1674-1137/ac3fa6},
       adsurl = {https://ui.adsabs.harvard.edu/abs/2022ChPhC..46c0001M}
}

@ARTICLE{Amenomori2021,
       author = {{Tibet AS{\ensuremath{\gamma}} Collaboration} and {Amenomori}, M. and {Bao}, Y.~W. and {Bi}, X.~J. and {Chen}, D. and {Chen}, T.~L. and {Chen}, W.~Y. and {Chen}, Xu and {Chen}, Y. and {Cirennima}, S.~W., Cui and {Danzengluobu}, L.~K., Ding and {Fang}, J.~H. and {Fang}, K. and {Feng}, C.~F. and {Feng}, Zhaoyang and {Feng}, Z.~Y. and {Gao}, Qi and {Gou}, Q.~B. and {Guo}, Y.~Q. and {Guo}, Y.~Y. and {He}, H.~H. and {He}, Z.~T. and {Hibino}, K. and {Hotta}, N. and {Hu}, Haibing and {Hu}, H.~B. and {Huang}, J. and {Jia}, H.~Y. and {Jiang}, L. and {Jin}, H.~B. and {Kasahara}, K. and {Katayose}, Y. and {Kato}, C. and {Kato}, S. and {Kawata}, K. and {Kihara}, W. and {Ko}, Y. and {Kozai}, M. and {Labaciren}, G.~M., Le and {Li}, A.~F. and {Li}, H.~J. and {Li}, W.~J. and {Lin}, Y.~H. and {Liu}, B. and {Liu}, C. and {Liu}, J.~S. and {Liu}, M.~Y. and {Liu}, W. and {Lou}, Y. -Q. and {Lu}, H. and {Meng}, X.~R. and {Munakata}, K. and {Nakada}, H. and {Nakamura}, Y. and {Nanjo}, H. and {Nishizawa}, M. and {Ohnishi}, M. and {Ohura}, T. and {Ozawa}, S. and {Qian}, X.~L. and {Qu}, X.~B. and {Saito}, T. and {Sakata}, M. and {Sako}, T.~K. and {Shao}, J. and {Shibata}, M. and {Shiomi}, A. and {Sugimoto}, H. and {Takano}, W. and {Takita}, M. and {Tan}, Y.~H. and {Tateyama}, N. and {Torii}, S. and {Tsuchiya}, H. and {Udo}, S. and {Wang}, H. and {Wu}, H.~R. and {Xue}, L. and {Yamamoto}, Y. and {Yang}, Z. and {Yokoe}, Y. and {Yuan}, A.~F. and {Zhai}, L.~M. and {Zhang}, H.~M. and {Zhang}, J.~L. and {Zhang}, X. and {Zhang}, X.~Y. and {Zhang}, Y. and {Zhang}, Yi and {Zhang}, Ying and {Zhao}, S.~P. and {Zhaxisangzhu}, X.~X., Zhou},
        title = "{Potential PeVatron supernova remnant G106.3+2.7 seen in the highest-energy gamma rays}",
      journal = {Nature Astronomy},
         year = 2021,
        month = jan,
       volume = {5},
        pages = {460-464},
          doi = {10.1038/s41550-020-01294-9},
archivePrefix = {arXiv},
       eprint = {2109.02898},
 primaryClass = {astro-ph.HE},
       adsurl = {https://ui.adsabs.harvard.edu/abs/2021NatAs...5..460T}
}

@ARTICLE{LHAASOJ1848-0001,
       author = {{The LHAASO Collaboration}},
        title = "{An extreme particle accelerator powered by PSR J1849-0001}",
      journal = {arXiv e-prints},
         year = 2026,
        month = mar,
          eid = {arXiv:2603.15537},
        pages = {arXiv:2603.15537},
          doi = {10.48550/arXiv.2603.15537},
archivePrefix = {arXiv},
       eprint = {2603.15537},
 primaryClass = {astro-ph.HE},
       adsurl = {https://ui.adsabs.harvard.edu/abs/2026arXiv260315537T}
}

@ARTICLE{Seta2025,
       author = {{Seta}, Amit and {McClure-Griffiths}, N.~M.},
        title = "{Magnetic fields in the multiphase interstellar medium of the Milky Way: turbulent kinetic and magnetic energy density relation}",
      journal = {\mnras},
         year = 2025,
        month = may,
       volume = {539},
       number = {2},
        pages = {1024-1039},
          doi = {10.1093/mnras/staf520},
archivePrefix = {arXiv},
       eprint = {2503.23634},
 primaryClass = {astro-ph.GA},
       adsurl = {https://ui.adsabs.harvard.edu/abs/2025MNRAS.539.1024S}
}

@ARTICLE{Temim2013,
       author = {{Temim}, Tea and {Slane}, Patrick and {Castro}, Daniel and {Plucinsky}, Paul P. and {Gelfand}, Joseph and {Dickel}, John R.},
        title = "{High-energy Emission from the Composite Supernova Remnant MSH 15-56}",
      journal = {\apj},
         year = 2013,
        month = may,
       volume = {768},
       number = {1},
          eid = {61},
        pages = {61},
          doi = {10.1088/0004-637X/768/1/61},
archivePrefix = {arXiv},
       eprint = {1303.2425},
 primaryClass = {astro-ph.HE},
       adsurl = {https://ui.adsabs.harvard.edu/abs/2013ApJ...768...61T}
}

@ARTICLE{Temim2015,
       author = {{Temim}, Tea and {Slane}, Patrick and {Kolb}, Christopher and {Blondin}, John and {Hughes}, John P. and {Bucciantini}, Niccol{\'o}},
        title = "{Late-Time Evolution of Composite Supernova Remnants: Deep Chandra Observations and Hydrodynamical Modeling of a Crushed Pulsar Wind Nebula in SNR G327.1-1.1}",
      journal = {\apj},
         year = 2015,
        month = jul,
       volume = {808},
       number = {1},
          eid = {100},
        pages = {100},
          doi = {10.1088/0004-637X/808/1/100},
archivePrefix = {arXiv},
       eprint = {1506.03069},
 primaryClass = {astro-ph.HE},
       adsurl = {https://ui.adsabs.harvard.edu/abs/2015ApJ...808..100T}
}

@ARTICLE{Gong2025,
       author = {{Gong}, Yunlu and {Zhou}, Liancheng and {Xia}, Qi and {Zhang}, Haiyun and {Fang}, Jun and {Zhang}, Li},
        title = "{Multiband Nonthermal Radiative Properties of the Pulsar Wind Nebula CTB 87}",
      journal = {\apj},
         year = 2025,
        month = mar,
       volume = {981},
       number = {1},
          eid = {7},
        pages = {7},
          doi = {10.3847/1538-4357/adae90},
archivePrefix = {arXiv},
       eprint = {2502.03011},
 primaryClass = {astro-ph.HE},
       adsurl = {https://ui.adsabs.harvard.edu/abs/2025ApJ...981....7G}
}

@ARTICLE{Foreman2013PASP,
       author = {{Foreman-Mackey}, Daniel and {Hogg}, David W. and {Lang}, Dustin and {Goodman}, Jonathan},
        title = "{emcee: The MCMC Hammer}",
      journal = {\pasp},
         year = 2013,
        month = mar,
       volume = {125},
       number = {925},
        pages = {306},
          doi = {10.1086/670067},
archivePrefix = {arXiv},
       eprint = {1202.3665},
 primaryClass = {astro-ph.IM},
       adsurl = {https://ui.adsabs.harvard.edu/abs/2013PASP..125..306F}
}

@ARTICLE{Cummings2016ApJ,
       author = {{Cummings}, A.~C. and {Stone}, E.~C. and {Heikkila}, B.~C. and {Lal}, N. and {Webber}, W.~R. and {J{\'o}hannesson}, G. and {Moskalenko}, I.~V. and {Orlando}, E. and {Porter}, T.~A.},
        title = "{Galactic Cosmic Rays in the Local Interstellar Medium: Voyager 1 Observations and Model Results}",
      journal = {\apj},
         year = 2016,
        month = nov,
       volume = {831},
       number = {1},
          eid = {18},
        pages = {18},
          doi = {10.3847/0004-637X/831/1/18},
       adsurl = {https://ui.adsabs.harvard.edu/abs/2016ApJ...831...18C}
}

@ARTICLE{Morris2002MNRAS,
       author = {{Morris}, D.~J. and {Hobbs}, G. and {Lyne}, A.~G. and {Stairs}, I.~H. and {Camilo}, F. and {Manchester}, R.~N. and {Possenti}, A. and {Bell}, J.~F. and {Kaspi}, V.~M. and {Amico}, N. D' and {McKay}, N.~P.~F. and {Crawford}, F. and {Kramer}, M.},
        title = "{The Parkes Multibeam Pulsar Survey - II. Discovery and timing of 120 pulsars}",
      journal = {\mnras},
         year = 2002,
        month = sep,
       volume = {335},
       number = {2},
        pages = {275-290},
          doi = {10.1046/j.1365-8711.2002.05551.x},
archivePrefix = {arXiv},
       eprint = {astro-ph/0204238},
 primaryClass = {astro-ph},
       adsurl = {https://ui.adsabs.harvard.edu/abs/2002MNRAS.335..275M}
}

@ARTICLE{Kassim1988ApJ,
       author = {{Kassim}, Namir E.},
        title = "{SNR Candidates from the Clark Lake Galactic Plane Survey}",
      journal = {\apjl},
         year = 1988,
        month = may,
       volume = {328},
        pages = {L55},
          doi = {10.1086/185159},
       adsurl = {https://ui.adsabs.harvard.edu/abs/1988ApJ...328L..55K}
}

@ARTICLE{Gorham1990ApJ,
       author = {{Gorham}, Peter W.},
        title = "{A Radio/Infrared/Optical Study of Candidate Supernova Remnants from the Clark Lake 30.9 MHz Galactic Plane Survey}",
      journal = {\apj},
         year = 1990,
        month = nov,
       volume = {364},
        pages = {187},
          doi = {10.1086/169401},
       adsurl = {https://ui.adsabs.harvard.edu/abs/1990ApJ...364..187G}
}

@ARTICLE{Ranasinghe2023ApJS,
       author = {{Ranasinghe}, S. and {Leahy}, D.},
        title = "{A Statistical Analysis of Galactic Radio Supernova Remnants}",
      journal = {\apjs},
         year = 2023,
        month = apr,
       volume = {265},
       number = {2},
          eid = {53},
        pages = {53},
          doi = {10.3847/1538-4365/acc1de},
archivePrefix = {arXiv},
       eprint = {2302.06593},
 primaryClass = {astro-ph.GA},
       adsurl = {https://ui.adsabs.harvard.edu/abs/2023ApJS..265...53R}
}

@ARTICLE{Gabici2007,
       author = {{Gabici}, Stefano and {Aharonian}, Felix A. and {Blasi}, Pasquale},
        title = "{Gamma rays from molecular clouds}",
      journal = {\apss},
         year = 2007,
        month = jun,
       volume = {309},
       number = {1-4},
        pages = {365-371},
          doi = {10.1007/s10509-007-9427-6},
archivePrefix = {arXiv},
       eprint = {astro-ph/0610032},
 primaryClass = {astro-ph},
       adsurl = {https://ui.adsabs.harvard.edu/abs/2007Ap&SS.309..365G}
}

@ARTICLE{DAngelo2018MNRAS,
       author = {{D'Angelo}, Marta and {Morlino}, Giovanni and {Amato}, Elena and {Blasi}, Pasquale},
        title = "{Diffuse gamma-ray emission from self-confined cosmic rays around Galactic sources}",
      journal = {\mnras},
         year = 2018,
        month = feb,
       volume = {474},
       number = {2},
        pages = {1944-1954},
          doi = {10.1093/mnras/stx2828},
archivePrefix = {arXiv},
       eprint = {1710.10937},
 primaryClass = {astro-ph.HE},
       adsurl = {https://ui.adsabs.harvard.edu/abs/2018MNRAS.474.1944D}
}

@ARTICLE{Aharonian2008A&A_W28,
       author = {{Aharonian}, F. and {Akhperjanian}, A.~G. and {Bazer-Bachi}, A.~R. and {Behera}, B. and {Beilicke}, M. and {Benbow}, W. and {Berge}, D. and {Bernl{\"o}hr}, K. and {Boisson}, C. and {Bolz}, O. and {Borrel}, V. and {Braun}, I. and {Brion}, E. and {Brown}, A.~M. and {B{\"u}hler}, R. and {Bulik}, T. and {B{\"u}sching}, I. and {Boutelier}, T. and {Carrigan}, S. and {Chadwick}, P.~M. and {Chounet}, L.-M. and {Clapson}, A.~C. and {Coignet}, G. and {Cornils}, R. and {Costamante}, L. and {Degrange}, B. and {Dickinson}, H.~J. and {Djannati-Ata{\"\i}}, A. and {Domainko}, W. and {O'C. Drury}, L. and {Dubus}, G. and {Dyks}, J. and {Egberts}, K. and {Emmanoulopoulos}, D. and {Espigat}, P. and {Farnier}, C. and {Feinstein}, F. and {Fiasson}, A. and {F{\"o}rster}, A. and {Fontaine}, G. and {Fukui}, Y. and {Funk}, Seb. and {Funk}, S. and {F{\"u}{\ss}ling}, M. and {Gallant}, Y.~A. and {Giebels}, B. and {Glicenstein}, J.~F. and {Gl{\"u}ck}, B. and {Goret}, P. and {Hadjichristidis}, C. and {Hauser}, D. and {Hauser}, M. and {Heinzelmann}, G. and {Henri}, G. and {Hermann}, G. and {Hinton}, J.~A. and {Hoffmann}, A. and {Hofmann}, W. and {Holleran}, M. and {Hoppe}, S. and {Horns}, D. and {Jacholkowska}, A. and {de Jager}, O.~C. and {Kendziorra}, E. and {Kerschhaggl}, M. and {Kh{\'e}lifi}, B. and {Komin}, Nu. and {Kosack}, K. and {Lamanna}, G. and {Latham}, I.~J. and {Le Gallou}, R. and {Lemi{\`e}re}, A. and {Lemoine-Goumard}, M. and {Lenain}, J.-P. and {Lohse}, T. and {Martin}, J.~M. and {Martineau-Huynh}, O. and {Marcowith}, A. and {Masterson}, C. and {Maurin}, G. and {McComb}, T.~J.~L. and {Moderski}, R. and {Moriguchi}, Y. and {Moulin}, E. and {de Naurois}, M. and {Nedbal}, D. and {Nolan}, S.~J. and {Olive}, J.-P. and {Orford}, K.~J. and {Osborne}, J.~L. and {Ostrowski}, M. and {Panter}, M. and {Pedaletti}, G. and {Pelletier}, G. and {Petrucci}, P.-O. and {Pita}, S. and {P{\"u}hlhofer}, G. and {Punch}, M. and {Ranchon}, S. and {Raubenheimer}, B.~C. and {Raue}, M. and {Rayner}, S.~M. and {Reimer}, O. and {Renaud}, M. and {Ripken}, J. and {Rob}, L. and {Rolland}, L. and {Rosier-Lees}, S. and {Rowell}, G. and {Rudak}, B. and {Ruppel}, J. and {Sahakian}, V. and {Santangelo}, A. and {Saug{\'e}}, L. and {Schlenker}, S. and {Schlickeiser}, R. and {Schr{\"o}der}, R. and {Schwanke}, U. and {Schwarzburg}, S. and {Schwemmer}, S. and {Shalchi}, A. and {Sol}, H. and {Spangler}, D. and {Stawarz}, {\L}. and {Steenkamp}, R. and {Stegmann}, C. and {Superina}, G. and {Takeuchi}, T. and {Tam}, P.~H. and {Tavernet}, J.-P. and {Terrier}, R. and {van Eldik}, C. and {Vasileiadis}, G. and {Venter}, C. and {Vialle}, J.~P. and {Vincent}, P. and {Vivier}, M. and {V{\"o}lk}, H.~J. and {Volpe}, F. and {Wagner}, S.~J. and {Ward}, M.},
        title = "{Discovery of very high energy gamma-ray emission coincident with molecular clouds in the W 28 (G6.4-0.1) field}",
      journal = {\aap},
         year = 2008,
        month = apr,
       volume = {481},
       number = {2},
        pages = {401-410},
          doi = {10.1051/0004-6361:20077765},
archivePrefix = {arXiv},
       eprint = {0801.3555},
 primaryClass = {astro-ph},
       adsurl = {https://ui.adsabs.harvard.edu/abs/2008A&A...481..401A}
}

@ARTICLE{Oka2025ApJ,
       author = {{Oka}, Tomohiko and {Ishizaki}, Wataru and {Mori}, Masaki and {Sano}, Hidetoshi and {Suzuki}, Hiromasa and {Tanaka}, Takaaki},
        title = "{Resolving the Origin of the Unidentified TeV Source HESS J1626-490 as a Relic of the Ancient Cosmic-Ray Factory SNR G335.2+0.1}",
      journal = {\apj},
         year = 2025,
        month = aug,
       volume = {989},
       number = {2},
          eid = {137},
        pages = {137},
          doi = {10.3847/1538-4357/adec80},
archivePrefix = {arXiv},
       eprint = {2507.04407},
 primaryClass = {astro-ph.HE},
       adsurl = {https://ui.adsabs.harvard.edu/abs/2025ApJ...989..137O}
}

@ARTICLE{HAWC2017Sci,
       author = {{Abeysekara}, A.~U. and {Albert}, A. and {Alfaro}, R. and {Alvarez}, C. and {{\'A}lvarez}, J.~D. and {Arceo}, R. and {Arteaga-Vel{\'a}zquez}, J.~C. and {Avila Rojas}, D. and {Ayala Solares}, H.~A. and {Barber}, A.~S. and {Bautista-Elivar}, N. and {Becerril}, A. and {Belmont-Moreno}, E. and {BenZvi}, S.~Y. and {Berley}, D. and {Bernal}, A. and {Braun}, J. and {Brisbois}, C. and {Caballero-Mora}, K.~S. and {Capistr{\'a}n}, T. and {Carrami{\~n}ana}, A. and {Casanova}, S. and {Castillo}, M. and {Cotti}, U. and {Cotzomi}, J. and {Couti{\~n}o de Le{\'o}n}, S. and {De Le{\'o}n}, C. and {De la Fuente}, E. and {Dingus}, B.~L. and {DuVernois}, M.~A. and {D{\'\i}az-V{\'e}lez}, J.~C. and {Ellsworth}, R.~W. and {Engel}, K. and {Enr{\'\i}quez-Rivera}, O. and {Fiorino}, D.~W. and {Fraija}, N. and {Garc{\'\i}a-Gonz{\'a}lez}, J.~A. and {Garfias}, F. and {Gerhardt}, M. and {Gonz{\'a}lez Mu{\~n}oz}, A. and {Gonz{\'a}lez}, M.~M. and {Goodman}, J.~A. and {Hampel-Arias}, Z. and {Harding}, J.~P. and {Hern{\'a}ndez}, S. and {Hern{\'a}ndez-Almada}, A. and {Hinton}, J. and {Hona}, B. and {Hui}, C.~M. and {H{\"u}ntemeyer}, P. and {Iriarte}, A. and {Jardin-Blicq}, A. and {Joshi}, V. and {Kaufmann}, S. and {Kieda}, D. and {Lara}, A. and {Lauer}, R.~J. and {Lee}, W.~H. and {Lennarz}, D. and {Vargas}, H. Le{\'o}n and {Linnemann}, J.~T. and {Longinotti}, A.~L. and {Luis Raya}, G. and {Luna-Garc{\'\i}a}, R. and {L{\'o}pez-Coto}, R. and {Malone}, K. and {Marinelli}, S.~S. and {Martinez}, O. and {Martinez-Castellanos}, I. and {Mart{\'\i}nez-Castro}, J. and {Mart{\'\i}nez-Huerta}, H. and {Matthews}, J.~A. and {Miranda-Romagnoli}, P. and {Moreno}, E. and {Mostaf{\'a}}, M. and {Nellen}, L. and {Newbold}, M. and {Nisa}, M.~U. and {Noriega-Papaqui}, R. and {Pelayo}, R. and {Pretz}, J. and {P{\'e}rez-P{\'e}rez}, E.~G. and {Ren}, Z. and {Rho}, C.~D. and {Rivi{\`e}re}, C. and {Rosa-Gonz{\'a}lez}, D. and {Rosenberg}, M. and {Ruiz-Velasco}, E. and {Salazar}, H. and {Salesa Greus}, F. and {Sandoval}, A. and {Schneider}, M. and {Schoorlemmer}, H. and {Sinnis}, G. and {Smith}, A.~J. and {Springer}, R.~W. and {Surajbali}, P. and {Taboada}, I. and {Tibolla}, O. and {Tollefson}, K. and {Torres}, I. and {Ukwatta}, T.~N. and {Vianello}, G. and {Weisgarber}, T. and {Westerhoff}, S. and {Wisher}, I.~G. and {Wood}, J. and {Yapici}, T. and {Yodh}, G. and {Younk}, P.~W. and {Zepeda}, A. and {Zhou}, H. and {Guo}, F. and {Hahn}, J. and {Li}, H. and {Zhang}, H.},
        title = "{Extended gamma-ray sources around pulsars constrain the origin of the positron flux at Earth}",
      journal = {Science},
         year = 2017,
        month = nov,
       volume = {358},
       number = {6365},
        pages = {911-914},
          doi = {10.1126/science.aan4880},
archivePrefix = {arXiv},
       eprint = {1711.06223},
 primaryClass = {astro-ph.HE},
       adsurl = {https://ui.adsabs.harvard.edu/abs/2017Sci...358..911A}
}

@ARTICLE{Crutcher2012ARA&A,
       author = {{Crutcher}, Richard M.},
        title = "{Magnetic Fields in Molecular Clouds}",
      journal = {\araa},
         year = 2012,
        month = sep,
       volume = {50},
        pages = {29-63},
          doi = {10.1146/annurev-astro-081811-125514},
       adsurl = {https://ui.adsabs.harvard.edu/abs/2012ARA&A..50...29C}
}
\bibliographystyle{aasjournalv7}



\end{document}